\documentclass[%
 reprint,
superscriptaddress,
 amsmath,amssymb,
 aps,
 prb,
 floatfix
]{revtex4-2}

\usepackage{graphicx}
\usepackage{dcolumn}
\usepackage{bm}
\usepackage[T1]{fontenc}
\usepackage[latin9]{inputenc}
\usepackage{xcolor}
\usepackage{float}
\usepackage{units}
\usepackage{textcomp}
\usepackage{amsmath}
\usepackage{amssymb}
\usepackage{graphicx}
\usepackage{mathtools}
\usepackage{physics}
\usepackage{wasysym}
\usepackage{hyperref}
\hypersetup{
    colorlinks=true,
    linkcolor=blue,
    citecolor=blue,      
    urlcolor=blue,
} 
\makeatletter

\usepackage{placeins} 
\makeatother
\usepackage{babel}

\begin{document}
\title{Magnetotransport of Sm$_2$Ir$_2$O$_7$ across the pressure-induced quantum-critical phase boundary}
\author{M.J. Coak}
\email{m.j.coak@bham.ac.uk}
\affiliation{Department of Physics, University of Warwick, Gibbet Hill Road, Coventry CV4 7AL, United Kingdom}
\affiliation{London Centre for Nanotechnology, University College London, Gordon St, London WC1H 0AH, United Kingdom}
\affiliation{School of Physics and Astronomy, University of Birmingham, Edgbaston, Birmingham B15 2TT, United Kingdom}
\author{K. G\"otze}
\affiliation{Department of Physics, University of Warwick, Gibbet Hill Road, Coventry CV4 7AL, United Kingdom}
\affiliation{Deutsches Elektronen-Synchrotron (DESY), 22607 Hamburg, Germany}
\author{T. Northam De La Fuente}
\affiliation{Department of Physics, Royal Holloway, University of London, Egham TW20 0EX, United Kingdom}
\affiliation{Institut Laue-Langevin, CS 20156, 38042 Grenoble Cedex 9, France}
\author{C. Castelnovo}
\affiliation{Cavendish Laboratory, University of Cambridge, JJ Thomson Ave, Cambridge CB3 0HE, United Kingdom}
\author{J. P. Tidey}
\affiliation{Department of Chemistry, University of Warwick, Gibbet Hill Road, Coventry CV4 7AL, United Kingdom}
\author{J. Singleton}
\affiliation{National High Magnetic Field Laboratory, Los Alamos National Laboratory, MS-E536, Los Alamos, NM 87545, United States of America}
\author{A.T. Boothroyd}
\affiliation{Department of Physics, University of Oxford, Clarendon Laboratory, Oxford, OX1 3PU, United Kingdom}
\author{D. Prabhakaran}
\affiliation{Department of Physics, University of Oxford, Clarendon Laboratory, Oxford, OX1 3PU, United Kingdom}
\author{P.A. Goddard}
\email{p.goddard@warwick.ac.uk}
\affiliation{Department of Physics, University of Warwick, Gibbet Hill Road, Coventry CV4 7AL, United Kingdom}
\date{\today}

\begin{abstract}
Rare-earth pyrochlore iridates host two interlocking magnetic sublattices of corner-sharing tetrahedra and can harbour a unique combination of frustrated moments, exotic excitations and highly correlated electrons. They are also the first systems predicted to display both topological Weyl semimetal and axion insulator phases.  We have measured the transport and magnetotransport properties of single-crystal Sm$_2$Ir$_2$O$_7$ up to and beyond the pressure-induced quantum critical point for all-in-all-out (AIAO) Ir order at $p_{{\rm c}}$ = 63~kbar previously identified by resonant X-ray scattering and close to which Weyl semimetallic behavior has been previously predicted. Our findings overturn the accepted expectation that the suppression of AIAO order should lead to metallic conduction persisting down to zero temperature. Instead, the resistivity-minimum temperature, which tracks the decrease in the AIAO ordering temperature for pressures up to 30~kbar, begins to increase under further application of pressure, pointing to the presence of a second as-yet unidentified mechanism leading to non-metallic behavior. The magnetotransport does track the suppression of Ir magnetism, however, with a strong hysteresis observed only within the AIAO phase boundary, similar to that found for Ho$_2$Ir$_2$O$_7$ and attributed to plastic deformation of Ir domains. Around $p_{{\rm c}}$ we find the emergence of a new type of electronic phase, characterized by a negative magnetoresistance with small hysteresis at the lowest temperatures, and hysteresis-free positive magnetoresistance above approximately 5~K. The temperature dependence of our low-temperature transport data are found to be best described by a model consistent with a Weyl semimetal across the entire pressure range.
\end{abstract}

\maketitle

The magnetic pyrochlore lattice, with spins arranged into a network of corner-sharing tetrahedra, has a strong history of yielding new physical states, discoveries and insights. The seminal observation of a spin-ice state in Ho$_2$Ti$_2$O$_7$ and Dy$_2$Ti$_2$O$_7$ \citep{Harris1997,Bramwell2001} and the associated magnetic monopole picture for its low-energy excitations \citep{Castelnovo2008,Morris2009,Fennell2009,Kadowaki2009} have led to lasting interest in this family of materials. The pyrochlore iridates add to the richness of these compounds by introducing a second inter-penetrating magnetic lattice with its own order, and interactions between the two. This complex magnetic behavior, their highly correlated electronic states, and predictions of topological properties, make pyrochlore iridates fascinating systems for investigating exotic quantum phenomena.

\begin{figure*}[t]
\begin{centering}
\includegraphics[width=0.9\linewidth]{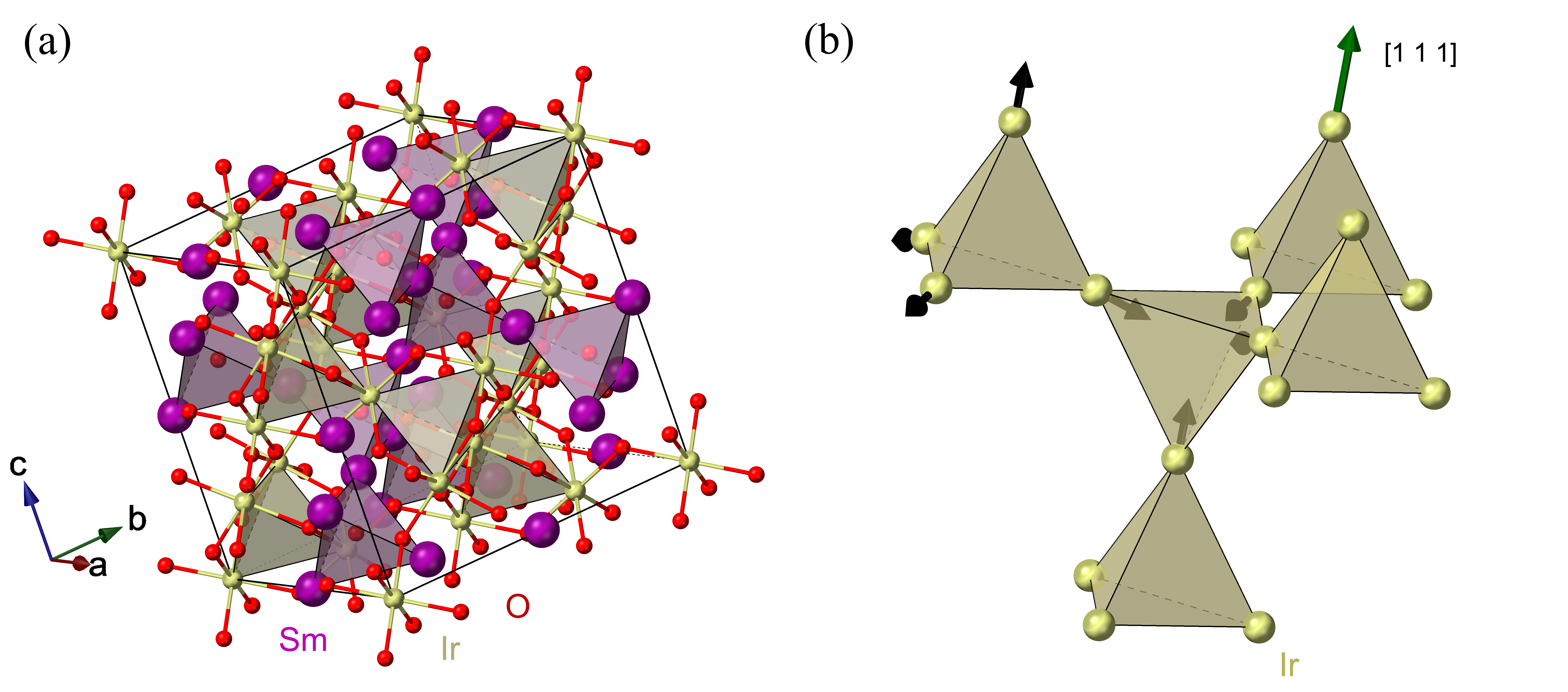}
\end{centering}
\caption{Crystal structure - (a) Ambient-pressure crystal structure of Sm$_2$Ir$_2$O$_7$, based on structural data from Refs \citep{Takatsu2014,Donnerer2016}. The crystal forms a cubic pyrochlore structure, with space group $Fd$--$3m$. The Sm atoms form a network of corner-sharing tetrahedra. Between these is a second network of corner-sharing tetrahedra of Ir atoms shown in (b). Each Ir is surrounded by an oxygen octahedron. The black arrows in (b) denote an example of the all-in-all-out Ir magnetic order. The green arrow shows the direction of one of the $\left[1\,1\,1\right]$ axes. The lattice parameter $a$ = 10.311(2)~\r{A} for Sm$_2$Ir$_2$O$_7$ at room temperature and ambient pressure.}
\label{fig:Structures}
\end{figure*}

The family of lanthanide pyrochlore iridates $Ln_2$Ir$_2$O$_7$, where $Ln$ is a magnetic trivalent lanthanide ion, provide an ideal environment for investigating fundamental magnetism and correlated-electron physics. They share the crystal structure shown in Fig~\ref{fig:Structures}(a) \textemdash{} a cubic pyrochlore structure with space group $Fd$--$3m$ \citep{Taira2001,Takatsu2014,Donnerer2016}. There are two interwoven sub-lattices of magnetic ions: a network of corner-sharing $Ln^{3+}$ tetrahedra with an oxygen site in the center, and another of corner-sharing Ir$^{4+}$ tetrahedra. Each Ir site sits in the center of an oxygen octahedron. While Pr$_2$Ir$_2$O$_7$ remains paramagnetic to low temperatures, the Ir sublattices in most (if not all) of the other members of this family undergo a transition to all-in-all-out (AIAO) antiferromagnetic long-range order \citep{Tomiyasu2012,Donnerer2016,Chun2018,Wang2020a,Jacobsen2020}, as illustrated in Fig. \ref{fig:Structures}(b). In this configuration, all the spins on one Ir tetrahedron point radially outwards, and all those on its neighboring tetrahedra point inwards, giving zero total moment.

In addition to the AIAO magnetism of the Ir ions, the rare-earth sublattice in these materials may also show interesting and correlated behavior. In Dy$_2$Ir$_2$O$_7$ and Ho$_2$Ir$_2$O$_7$, the large rare-earth moments are constrained to lie along the local $\left[1\,1\,1\right]$ directions and are coupled by both dipolar and exchange interactions, which are overall ferromagnetic~\citep{Lefrancois2017,Cathelin2020,Pearce2022}. These interactions favor the two-in-two-out (2I2O) configuration, promoting spin-ice physics, akin to that seen in the frustrated pyrochlore titanates, and the appearance of magnetic monopole quasiparticles~\citep{Bramwell2001,Castelnovo2012}. The added effect of exchange with Ir nearest neighbors means that at sufficiently low temperatures both Dy$_2$Ir$_2$O$_7$ and Ho$_2$Ir$_2$O$_7$ are expected to undergo a transition to a fragmented monopole crystal~\citep{BrooksBartlett2014, Lefrancois2017,Cathelin2020}. Experiments on Ho$_2$Ir$_2$O$_7$ at intermediate temperatures show that the magnetoresistance is sensitive to the density of magnetic monopoles and that the Ir AIAO domains can be manipulated, via the Ir--Ho coupling, using an applied magnetic field~\citep{Pearce2022}. The other members of the family have considerably smaller moments on the rare-earth site and their magnetic ground state remains the subject of debate.

The materials in this series additionally all share equivalent transport behavior \textemdash{} metallic high-temperature resistivity and a resistivity minimum $T_{\mathrm{min}}$ (described in the literature as a metal-insulator transition temperature) below which the resistivity increases with cooling. This $T_{\mathrm{min}}$ occurs at the same temperature as the Ir AIAO ordering and has been attributed to a spin-orbit-coupling assisted Mott transition \citep{Ueda2015,WitczakKrempa2012}. The exception is $Ln$ = Pr, in which the AIAO order is absent and which remains metallic at all temperatures. ARPES measurements have shown that in the paramagnetic phase the band structure exhibits quadratic-band-touching points~\cite{Kondo2015,Nakayama2016}. Perturbations to this electronic structure, such as symmetry breaking due to magnetic order on the rare-earth sites, are predicted to give rise to topologically non-trivial electronic states, including Weyl semimetals (WSMs) and accompanying Fermi arcs~\citep{Kondo2015}, and evidence for such phenomena has been reported extensively~\citep{Sakata2011,Tian2016,Ueda2015,Ueda2017,Ma2015,Ueda2018}.

It is possible to suppress the AIAO order in this family leading to what is reported in the literature as a quantum critical point, or sometimes a quantum metal-insulator transition, associated with significant quantum fluctuations~\citep{WitczakKrempa2012,Ueda2017,Savary2014,Nakatsuji2006,Ni2021a,Tian2016,Wang2020a}. There are two principal means to tune interactions in these materials. Compressing the lattice parameter, $a$, will predominately strengthen the hopping term $t$ in the Mott insulator description, while preserving local spin environments. Chemical substitution of the rare-earth site, on the other hand, mostly affects the oxygen coordinate parameter $x$. This parameter describes distortions away from a perfect octahedral environment ($x$=0.3125) around the Ir sites, towards a cubic environment around the $Ln$ sites ($x$=0.375) as its value increases (and the Ir-O-Ir bond angle decreases)~\citep{Wang2020a}. The reduction in $x$ caused by decreasing $Ln$ ionic radius as one moves from Gd $\rightarrow$ Eu $\rightarrow$ Sm $\rightarrow$ Nd reduces the local Ir Ising anisotropy, pushing down the AIAO ordering temperature \citep{Ueda2015,Wang2020a,Matsuhira2011,Zhang2017,Liu2018b}. (Contrary to the Sm magnetism, where the presence of a strong single-ion anisotropy due to crystal electric field effects is robustly established, the single-ion behavior of the Ir magnetism is less well understood. By `Ising anisotropy' of the Ir moments we refer to the fact that an AIAO order with well-defined local axes develops at low temperature and pressure. It remains to be determined whether this is purely an effect of the anisotropic direct Dzyaloshinskii-Moriya interaction combined with isotropic AFM exchange, or whether a single-ion CEF contribution plays a role). The reduction of AIAO ordering tempearture suppresses $T_{\mathrm{min}}$ accordingly and reduces the resistivity towards metallic behavior. The next compound in this progression, Pr$_2$Ir$_2$O$_7$ \citep{Nakatsuji2006,Tokiwa2014,Ueda2017}, is then metallic at all temperatures and does not show any magnetic order \citep{Nakatsuji2006,Matsuhira2011,Wang2020a}. Reduction of either $a$ or $x$ has the result of suppressing the Ir AIAO order, but through distinct mechanisms. It is intriguing to ask what differences may result from the two approaches. 

The strong sensitivity of the Ir ordering temperature, and the concomitant resistivity minimum, to the fine details of the crystal structure, as well as the many theoretical predictions of the presence of topological electronic states in the phase diagram \citep{WitczakKrempa2014}, means that controlled tuning of the lattice is expected to be particularly enlightening in these materials. Applied hydrostatic pressure provides a clean, continuous and reversible method for achieving this tuning.

There have been several pressure studies reporting the evolution of resistivity in $Ln_2$Ir$_2$O$_7$ samples. Most show a reduction in resistivity with pressure as they are driven towards metallization. Polycrystalline Nd$_2$Ir$_2$O$_7$ reproduces the metallic behavior of Pr$_2$Ir$_2$O$_7$ beyond 65~kbar of applied pressure \citep{Ueda2015,Ueda2017}. The $T_{\mathrm{min}}$ transition is continuously suppressed with pressure until it disappears at this point \citep{Liu2018b,Ueda2015,Ueda2017}. Magnetic field has also been shown to induce metallicity in Nd$_2$Ir$_2$O$_7$ by flipping the AIAO order to a 2-in-2-out state \citep{Ueda2015a}. In Gd$_2$Ir$_2$O$_7$, $T_{\mathrm{min}}$ linearly increases with applied pressure \citep{Liu2018b}. In Eu$_2$Ir$_2$O$_7$ there are reports of a slight increase of $T_{\mathrm{min}}$ under pressures up to 22~kbar in polycrystalline pellets \citep{Liu2018b} or very little change up to 120~kbar in single crystals \citep{Tafti2012a}.

The differences in this behavior across the family are likely due to the interplay of the competing changes in $x$ and $a$ lattice parameters induced in each compound. Sm$_2$Ir$_2$O$_7$, which in the $Ln$ series lies between Eu and Nd and hence between these two regimes of different pressure dependence, is therefore an interesting material for investigation.

The poly- or single-crystalline nature of the samples in this family is known to be important. Polycrystalline samples of $Ln_2$Ir$_2$O$_7$ can differ in transition temperature and form of resistivity to single-crystal samples \citep{Ishikawa2012,Tian2016,Cathelin2020}. These differences likely arise from slight variations in composition, the effects of the broken symmetry at the surface or the restricted magnetic domain size in polycrystalline samples, the random orientation of individual grains, and the unknown transport properties of inter-grain boundaries present in pressed pellets. For these reasons, conclusions derived from transport measurements on polycrystalline samples should be treated with caution. Single crystals are indispensable for true clean tests of the physics in these systems, and for direction-sensitive magnetotransport measurements, but have only recently been successfully grown for many of these materials~\citep{Cathelin2020,Pearce2022,Wang2020a, Ishikawa2012}.

Single crystals of Sm$_2$Ir$_2$O$_7$ exhibit a high-temperature metallic phase (in which resistivity drops with decreasing temperature). A transition to AIAO Ir order occurs around $T_{\mathrm{AIAO}}$ = 115~K at ambient pressure, and the associated $T_{\mathrm{min}}$ is clearly seen in transport data as a resistivity minimum below which resistivity rises with decreasing temperature~\citep{Matsuhira2011,Liu2018b}. There is some evidence from muon-spin relaxation measurements that the Sm moments undergo ordering, possibly short-range, at temperatures below about 10~K~\citep{Asih2017}, but the details remain unknown. Recently Wang \emph{et al.} showed, using resonant X-ray diffraction on single crystals, that the Ir AIAO ordering temperature is reduced to a quantum critical point (QCP) at a pressure $p_{{\rm c}}$ = 63~kbar~\citep{Wang2020a}. Above $p_{{\rm c}}$ a metallic state is expected once the magnetic transition responsible for driving insulating behavior is removed, similar to that seen in Pr$_2$Ir$_2$O$_7$ or in Nd$_2$Ir$_2$O$_7$ in applied fields or pressures \citep{Liu2018b,Ueda2015,Ueda2017}. There are also several predictions that Sm$_2$Ir$_2$O$_7$ hosts WSM states~\citep{Wan2011,WitczakKrempa2012,Ishii2015,Donnerer2016,Wang2020a}, most likely in proximity to the quantum-critical region close to where the magnetic order is suppressed \citep{Ueda2017,Wang2020a}. Initial transport measurements on polycrystalline samples have indeed shown that $T_{\mathrm{min}}$ drops gradually in Sm$_2$Ir$_2$O$_7$ as pressure is applied in the low pressure region up to 20~kbar~\citep{Liu2018b}. However, so far no measurements of electronic properties have been performed in the vicinity of the QCP, nor have any transport experiments been performed on single crystals of this material.

In this work, we report magnetotransport measurements on a high-quality single-crystal sample of Sm$_2$Ir$_2$O$_7$ in fields up to 15~T and pressures up to and beyond the 63~kbar quantum critical point. Our magnetoresistance data show that the AIAO Ir order is indeed suppressed at this pressure, in line with the X-ray measurements~\citep{Wang2020a}. However, while the resistance drops dramatically with pressure, the $T_{\mathrm{min}}$ resistivity minimum does not fall to zero temperature with the magnetic order as expected. Instead, for pressures above about 30~kbar the high-temperature metallic state is disrupted upon cooling by another process, which becomes more robust as pressure is increased further. Moreover, the magnetotransport changes significantly in the vicinity of the QCP, indicative of the emergence of a new low-temperature electronic state. Finally, a model consistent with Weyl semimetallic behavior is found to best describe the temperature dependence of the low-temperature resistivity across the entire pressure range. The temperature window over which the model effectively reproduces the data is largest at the lowest pressures.

\begin{figure}
\centering
\includegraphics[width=0.99\linewidth]{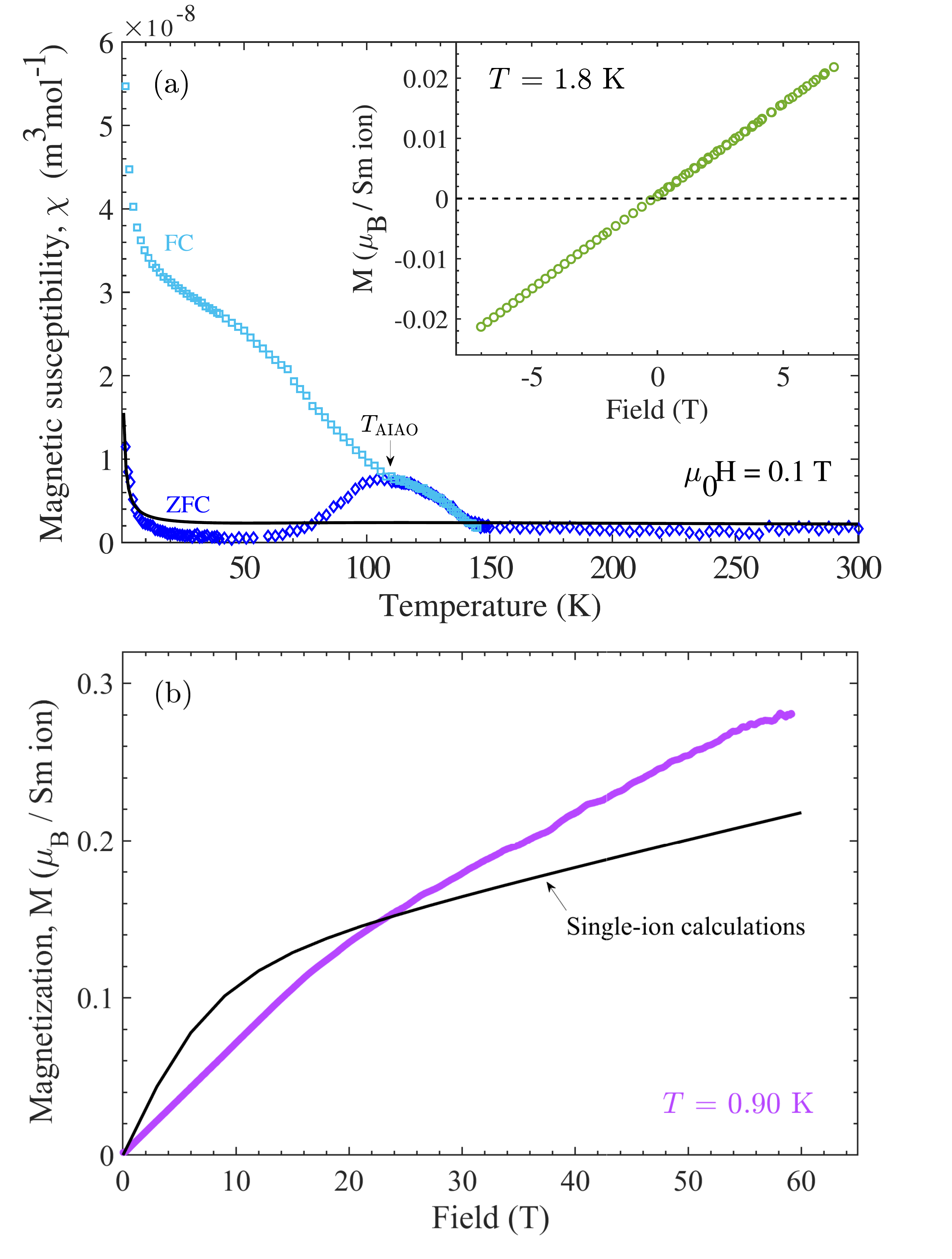}
\caption{Magnetic properties - (a) Magnetic susceptibility data (dark and light blue symbols) of a single crystal of Sm$_2$Ir$_2$O$_7$ in an applied field of 0.1~T parallel to $\left[1\,1\,1\right]$, measured as a function of temperature after cooling in zero field (ZFC) and with the field applied (FC). The Ir AIAO ordering temperature where the ZFC and FC data bifurcate, $T_{\mathrm{AIAO}}$, is marked with an arrow. Inset shows the magnetic-field dependence of magnetization in a powder sample at 1.8 K. (b) Magnetization data (purple symbols) against field at 0.9~K for a powder sample measured to 60~T in a pulsed magnet. In both panels the black line is the result of single-ion calculations of the Sm$^{3+}$ moment taking into account the crystal-field environment as described in the text.}
\label{fig:SQUIDdata}
\end{figure}

\section*{Results}

\subsection*{Magnetic properties}

The main panel of Fig. \ref{fig:SQUIDdata}(a) shows the magnetic susceptibility of Sm$_2$Ir$_2$O$_7$, measured in a $\left[1\,1\,1\right]$ magnetic field of 0.1\,T while warming after cooling in zero field (ZFC) and after cooling in the same 0.1\,T field (FC). The temperature where the FC and ZFC data bifurcate, $T_{\mathrm{AIAO}}$ =110(2)~K, coincides with the onset of the Ir AIAO ordered state, in agreement with the results of previous magnetic measurements \citep{Taira2001}, resonant X-ray diffraction \citep{Donnerer2016,Wang2020a} and muon-spin rotation \citep{Asih2017} studies. The FC susceptibility data continue to rise as temperature is reduced further. A hump is seen in the data at temperatures just below 150\,K, and an additional shoulder-like feature appears in the FC data centered at about 60--70\,K. There is no strong evidence in these data for long-range magnetic order on the Sm sublattice below 10~K, as suggested by Ref. \citep{Asih2017} from features in the muon-spin relaxation rate. Instead, our data are consistent with the presence of paramagnetic Sm moments at temperatures down to 2~K.

The inset of Fig. \ref{fig:SQUIDdata}(a) shows the field dependence of the magnetization $M(H)$ of a powder sample at low temperature. Fig. \ref{fig:SQUIDdata}(b) shows the extension of these data to high magnetic field. The high-field data were measured on a powder sample using extraction magnetometry \citep{Goddard2008} at the National High Magnetic Field Laboratory, Los Alamos, in pulsed fields up to 60~T and temperatures down to 0.9~K. As the Ir moments are ordered below 110~K, we expect that these data are dominated by the paramagnetic Sm moments. 

Initial crystalline electric field (CEF) calculations using a point-charge CEF Hamiltonian and the Stevens' operator formalism (see Supplementary Information, Supplementary Note 13), show that the Sm$^{3+}$ ground-state doublet wavefunction is approximately $\vert J, m_J \rangle = \vert 5/2, \pm 3/2 \rangle$. The magnetic moment of this approximate ground state is expected to be 0.43\,$\mu_{\rm B}$ at $T = 0$. Calculations of the single-ion properties for Sm$_2$Ir$_2$O$_7$ can be improved by employing intermediate coupling and allowing mixing of higher $J$ levels, as implemented in the SPECTRE software~\citep{SPECTRE}. This method was used in Ref.~\citep{PecanhaAntonio2019} to obtain a set of CEF parameters for Sm$^{3+}$ in Sm$_2$Ti$_2$O$_7$. Assuming the parameters therein for Sm$_2$Ti$_2$O$_7$ are a good approximation for Sm$_2$Ir$_2$O$_7$, we calculate that at $T = 0$\,K and $\mu_0H = 0.5$\,T, the moments exhibit a strong easy-axis anisotropy along the local $\left[1\,1\,1\right]$ axis with $\mu_{\parallel[111]} = 0.233\,$ $\mu_{\rm B}$ and $\mu_{\perp[111]}$ = 0.001\,$\mu_{\rm B}$ leading to four non-colinear Ising axes in each Sm tetrahedron. The moment calculated in this way is around half the value of that from the $\vert 5/2, \pm 3/2 \rangle$ doublet alone.  The discrepancy is largely due to a small admixture of terms from higher $J$ levels in the ground state, especially the $J = 7/2$ level~\cite{PecanhaAntonio2019}.

The black line in the main panel of Fig.~\ref{fig:SQUIDdata}(a) is the single-ion susceptibility for $H//[111]$ calculated using the intermediate-coupling wavefunctions. In spite of the simplicity of the single-ion approximation, which neglects Sm--Sm as well as Sm--Ir interactions, we see that it captures the leading behavior of the ZFC susceptibility. In particular, it shows that the low-temperature Curie-like tail can be accounted for entirely by the Sm sublattice. It is no surprise that this calculation is unable to reproduce the hump seen below 150~K, as this feature is likely caused by domains of weak ferromagnetism associated with the Ir AIAO order, which have been proposed in other pyrochlore iridates~\cite{Ishikawa2012,Zhu2014}. The solid black line in  Fig.~\ref{fig:SQUIDdata}(b) is the powder-averaged single-ion calculation for the magnetization. The calculations predict that elevated fields mix the CEF-split energy levels, causing the Sm moment to become less Ising-like and grow significantly beyond its low-field value, consistent with the experimental data. 

While dipolar interactions dominate the behavior of the rare-earth ions in other spin-ice materials such as holmium or dysprosium titanate, the small size of the Sm moments suggests that they can be safely disregarded in the temperature and field regime of interest in our work. The same argument however does not apply to exchange interactions, and further work beyond the scope of the present paper is needed to ascertain the effects of Sm-Sm and Sm-Ir exchange couplings to the behavior of the magnetisation in an applied field.  It is reasonable to expect the Sm ions to prefer to adopt an AIAO order at sufficiently low temperature, similar to Ho$_2$Ir$_2$O$_7$ and Dy$_2$Ir$_2$O$_7$, but they may favour a 2I2O configuration, depending on the sign and relative size of the Sm--Sm exchange interaction compared to the Sm--Ir interaction \citep{Lefrancois2017,Cathelin2020}.

\subsection*{Transport properties under applied pressure - temperature dependence of resistivity}

\begin{figure}
\centering
\includegraphics[width=0.99\linewidth]{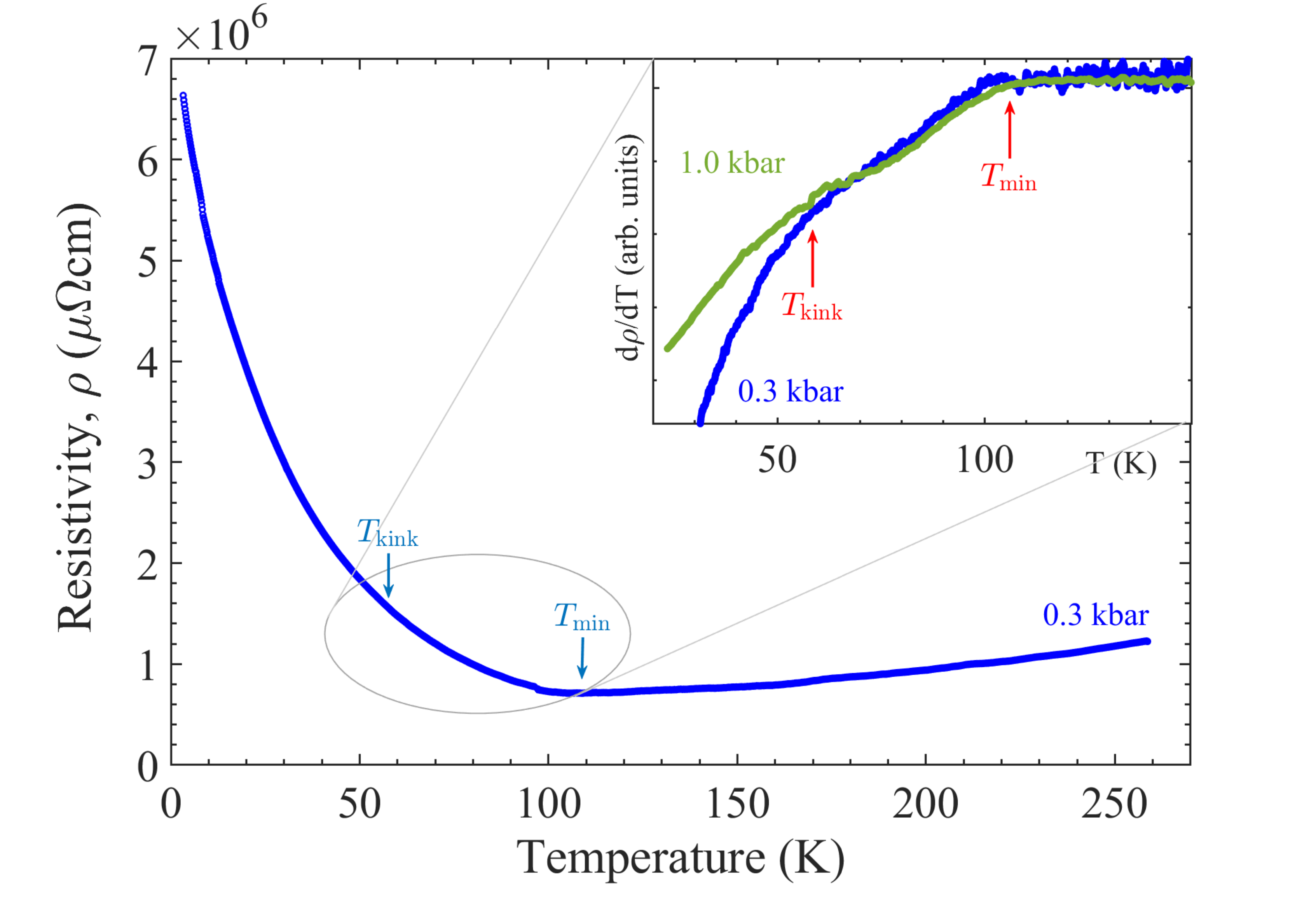}
\caption{Resistivity vs temperature at near-ambient pressure - The temperature of the resistivity minimum, $T_{\mathrm{min}}$, is marked with an arrow. The inset shows the result of differentiating the data with respect to temperature, which reveals a subtle kink or shoulder around 60~K, $T_{\mathrm{kink}}$.}
\label{fig:RvT-Ambient}
\end{figure}

Fig.~\ref{fig:RvT-Ambient} shows the temperature dependence of the resistivity of a single crystal of Sm$_2$Ir$_2$O$_7$, measured in a diamond anvil cell as discussed in the Methods section and at approximately zero applied pressure (0.3(3) kbar).

The resistivity shows metallic behavior (defined as positive 
${\rm d}\rho/{\rm d}T$) at high temperature, but undergoes a transition to a negative ${\rm d}\rho/{\rm d}T$ regime below $T_{\mathrm{min}}$~=~110(3)~K, where the resistivity exhibits a minimum. This transition is referred to as a metal-insulator transition in the literature. However, as discussed below, our data indicate that the low-temperature state cannot be described as a conventional insulator characterized by a gap at the Fermi energy. To avoid confusion, we refer to this region as `non-metallic'. The resistivity minimum occurs at the same temperature as the Ir AIAO magnetic transition seen in the magnetization, in agreement with the literature. Our resistivity data are consistent with previously reported low-pressure data on polycrystals by Liu~\emph{et al.} \citep{Liu2018b} and on pressed pellets by Matsuhira \emph{et al.} \citep{Matsuhira2011}. We note that the anomalous results of Graf~\emph{et al.} \citep{Graf2014}, also on pressed pellets, who report a much lower transition temperature are not reproduced in other studies.

\begin{figure}
\centering
\includegraphics[width=0.9\linewidth]{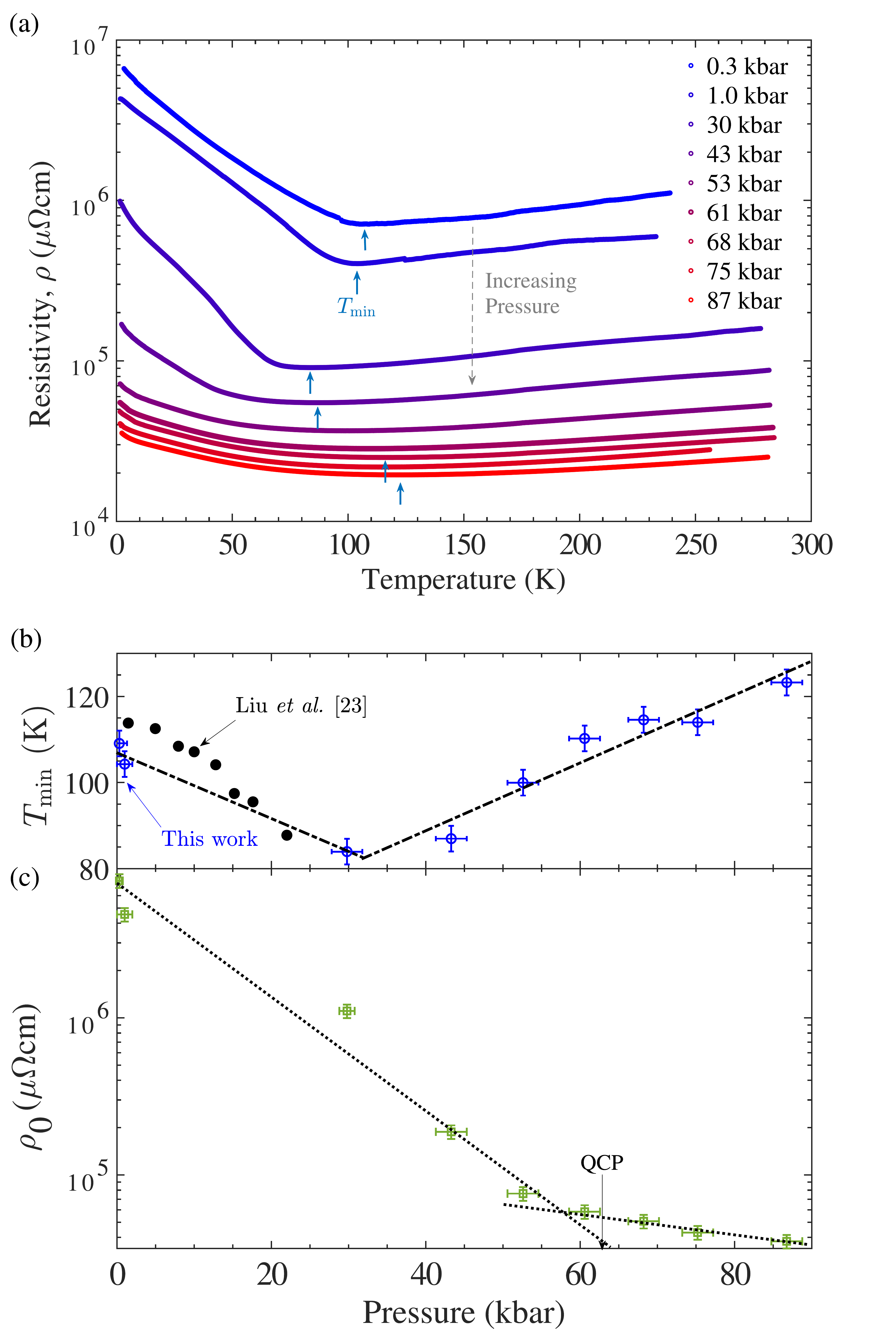}
\centering
\caption{Resistivity vs temperature at high pressures - (a) Resistivity vs temperature for increasing applied pressures, plotted on a logarithmic y-axis. The temperature position of the resistivity minimum, $T_{\mathrm{min}}$, is marked with arrows. (b) Values of $T_{\mathrm{min}}$ plotted against applied pressure, blue open circles. Data from the work of Liu \emph{et al.} \citep{Liu2018b} are overlayed as closed black circles. (c) Resistivity extrapolated (via a quadratic fit below 8\,K) to $T$ = 0~K plotted against pressure. Note that the vertical axis is logarithmic. Dotted lines are intended as guides to the eye, highlighting a change of behavior around 60~kbar, close to the Ir AIAO QCP. Error bars show the uncertainties in determining pressure (see Methods) and in the values of $T_{\mathrm{min}}$ and $\rho_0$.}
\label{fig:RvT}
\end{figure}

The temperature derivative of the resistivity reveals a subtle change of gradient or kink, which we denote $T_{\mathrm{kink}}$, at around 60\textendash{}70~K in both the 0.3(3)~kbar and 1(1)~kbar data sets (Fig.~\ref{fig:RvT-Ambient} Inset). This feature is not seen or resolved in our data for $p\geq30$~kbar, but it is seen, after differentiation, in the data of Liu~\emph{et al.} across their 0\textendash{}22~kbar pressure range, where it increases in temperature with applied pressure (see Supplementary Figure 12).

\begin{figure}
\begin{centering}
\includegraphics[width=0.9\columnwidth]{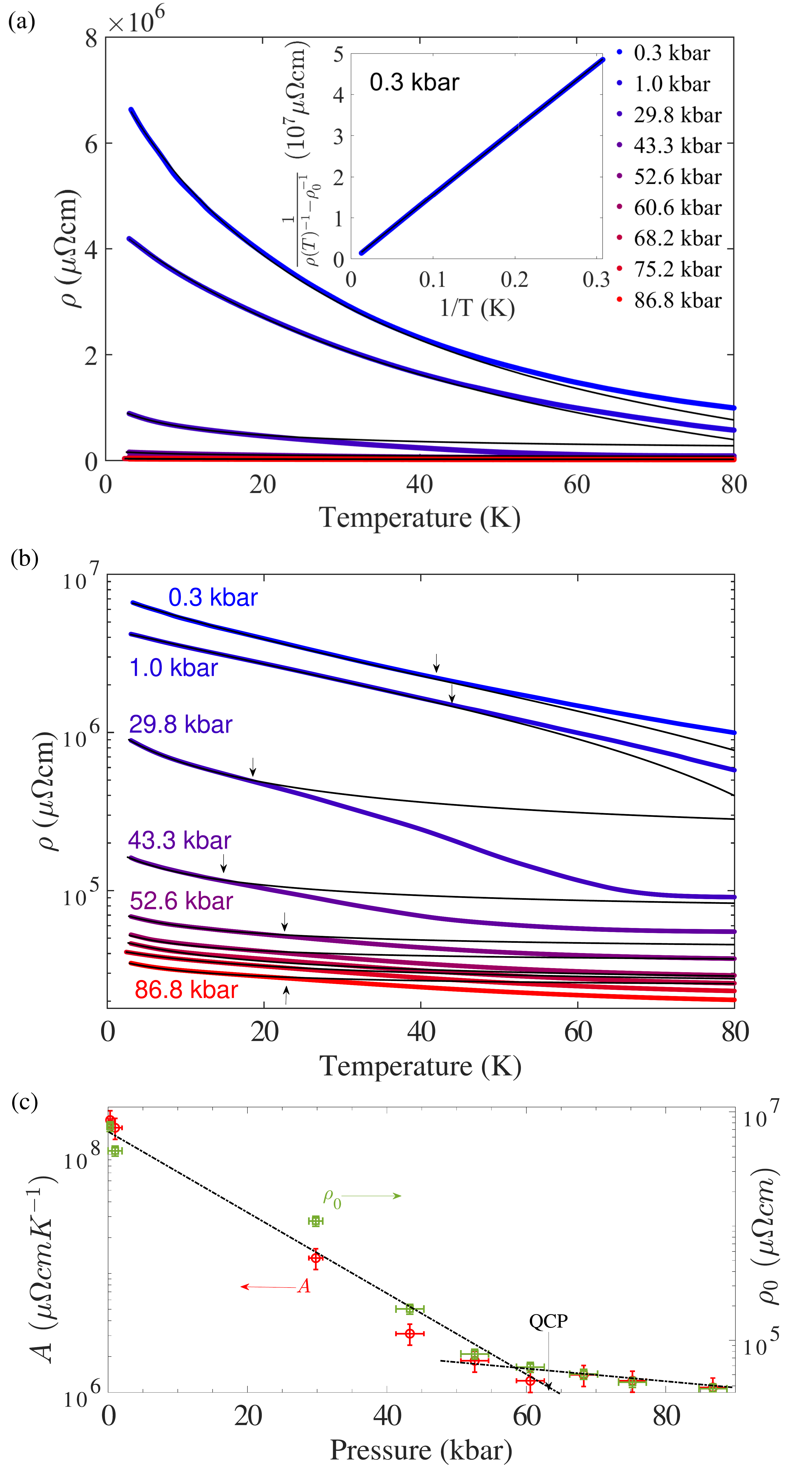}
\par\end{centering}
\caption{Fits of the low-temperature resistivity - (a) Resistivity data (points) and fits (black line) to $\rho^{-1} = \rho_0^{-1} + (A/T)^{-1}$, as described in the text. Inset shows $\rho$ with the $\rho_0$ contribution removed versus $1/T$ for $p = 0.3(3)$~kbar, yielding a straight line for $T$ up to 55~K. (b) The same as (a) plotted on a logarithmic vertical axis, to allow the higher pressure curves to be seen more clearly. Black arrows indicate the temperatures where each fit diverges from the data by 2\%. (c) Pressure dependence of the fit parameter $A$, the coefficient of the $1/T$ term (red circles). Dashed lines are guides to the eye (note that the vertical axis is logarithmic), highlighting a change of character around 60~kbar close to the QCP. 
The pressure dependence of $\rho_0$ from Fig \ref{fig:RvT}(c) is also shown overlaid as green squares for comparison. Error bars show the uncertainties in determining pressure (see Methods) and the standard error on the fit for $A$.}
\label{fig:RvT-Fits}
\end{figure}

Fig.~\ref{fig:RvT}(a) shows the resistivity of the same Sm$_2$Ir$_2$O$_7$ crystal at each pressure measured, increasing from 0.3(3)~kbar at the top, blue, to 87(2)~kbar at the bottom, red. Applying pressure up to 87~kbar continuously reduces the magnitude of the resistivity across all temperatures (by a factor of 40 at room temperature and 190 at 2~K). The absolute value of resistivity is initially comparatively high ($\sim 10^6$~$\mu\Omega$cm) in the metallic state, suggesting that this material may not be best described as a simple coherent metal even above $T_\mathrm{min}$.

A transition from metallic to non-metallic behavior upon cooling is observed at all pressures. Liu~\emph{et al.} \citep{Liu2018b} measured the pressure dependence of the resistivity of their polycrystal sample up to 22~kbar and reported a linear decrease in $T_{\mathrm{min}}$ over this range and this fits well with our results. However, just above the maximum pressure achieved by Liu~\emph{et al.}, our data show a change in behavior: the value of $T_{\mathrm{min}}$ begins to increase again, in a linear trend, from approximately 30~kbar, as shown in Fig. \ref{fig:RvT}(b). This resistivity minimum is sharp and well defined up to 30(2)~kbar, but becomes markedly broader and smoother at higher pressures, possibly indicative of a change from a phase transition to a crossover. Also at this point, the rise in resistivity for $T<T_{\mathrm{min}}$ becomes less pronounced than that observed at lower pressures.

The nature of the electronic state below $T_{\mathrm{min}}$ in this family remains the subject of debate --- several studies have compared the form of the low-temperature resistivity to various models of localized transport to attempt to shed light on this matter, see e.g. Refs \citep{Telang2019, LaBarre2019, Liu2020, Liu2018b, Han2015, Telang2021}. In the present case, the functional form of the resistivity temperature dependence in the non-metallic phase below $T_{\mathrm{min}}$ proves challenging to describe with common simple models. To start, a simple Arrhenius-type activated exponential behavior falls very far from describing the data, which are much shallower in slope than an exponential. It is clear from inspection of the curves in Fig. \ref{fig:RvT}(a) that although the magnitude of the resistivity increases dramatically at low temperatures, it does not appear to diverge at zero temperature. Extrapolating the curves to $T=0$ results in a finite $y$-axis intercept $\rho_{0}$ for each pressure. We show the result of doing just this in Fig. \ref{fig:RvT}(c), with a quadratic function fit to the data below 8~K and extrapolated below their lower limit. The extrapolated $\rho_{0}$ value is seen to drop off rapidly with applied pressure up to around 60\,kbar, the location of the AIAO QCP, after which it decreases at a much slower rate.

As discussed by Savary \emph{et al.} \citep{Savary2014}, a theoretical treatment of the transport properties in the pyrochlore iridates is complex and challenging; no clear predictions have been published at this time and there is as yet no experimental consensus. In polycrystalline Sm$_2$Ir$_2$O$_7$~\citep{Liu2018b}, and in Gd$_2$Ir$_2$O$_7$ \citep{Han2015}, the low-temperature resistivity has been described with a $\rho\propto\mathrm{exp}\left(T_{0}/T\right)^{1/4}$ variable-range hopping expression \textemdash{} an activated insulating behavior typical of systems with a high degree of local disorder \citep{Mott1990,Mott1969,Hill1976,Keuls1997,Polyakov1993}. Telang and Singh, in their study of the effect of Bi doping in Sm$_2$Ir$_2$O$_7$ pressed powders~\citep{Telang2021}, describe the undoped sample as obeying a $1/T$ temperature dependence typifying WSM behavior. This then reportedly evolves through a $T^{1/4}$ power law dependence tentatively attributed to quantum critical effects as doping suppresses the resistivity, to a $\mathrm{ln}(T)$ region interpreted as signifying quadratic-band-touching. In polycrystalline Y$_2$Ir$_2$O$_7$, a $T^{-4}$ model \citep{LaBarre2019} and a $T^{-3/2}$ model \citep{Liu2020} were fit to the transport data, and attributed to a WSM state with thermally screened charged impurities, and with the presence of impurities and electron--electron correlations, respectively. Fits to all these models, and more, are shown in the Supplementary Information, Supplementary Notes 1-6, to compare their relative merits with respect to our single-crystal data. With the exception of the $T^{1/4}$ function, all of these models diverge at zero temperature, so it is unsurprising that they are not able to describe our data beyond highly restricted temperature ranges. (We point out that a $\rho(T)\propto{}T^{1/3}$ temperature dependence, shown in Supplementary Figure 3, was found to give a markedly better fit to our data than any of these models previously employed in the literature. However, there seems to be no clear physical basis for this functional form and we just note its ability to fit the data).

The simplest physical model that can reproduce the behavior of our resistivity data is a combination of two parallel resistance channels, one which diverges and accounts for the non-metallic nature of the material and a second term which leads to a finite resistivity at zero temperature. The latter term could be any function that varies slowly compared to the diverging term, including $\rho\propto T$ or $T^2$, but for simplicity we choose a temperature-independent value, set to be equal to the $\rho_{0}$ values determined from data extrapolation (shown in Fig.~\ref{fig:RvT}(c)). This term implies a metallic channel with impurity scattering is present in the sample that could potentially arise due to surface states or conduction along magnetic domain-wall boundaries, as suggested to be present in Nd$_2$Ir$_2$O$_7$~\citep{Ueda2015a}. (The strong, monotonic pressure dependence of $\rho_0$ rules out that it arises from an experimental artefact, such as an electrical short).

The best fits (see Supplementary Notes 3-5) to the data at all pressures were found using the model $\rho^{-1} = \rho_0^{-1} + (A/T)^{-1}$, where $A$ is a (pressure dependent) constant, and are shown in Fig.~\ref{fig:RvT-Fits}(a) and (b). As mentioned earlier, a $\rho\propto1/T$ divergent term has previously been predicted for WSMs~\citep{Hosur2012} and it describes our data particularly well at low pressures as can be seen in the inset to Fig.~\ref{fig:RvT-Fits}(a). This shows the 0.3(3)~kbar $\rho(T)$ data lying on a straight line for $2 < T < 55$K when plotted versus $1/T$ after the $\rho_0^{-1}$ contribution has been subtracted. The temperature range over which the model effectively reproduces the data is largest at the two lowest pressures, farthest from the QCP. This is indicated by the arrows in Fig.~\ref{fig:RvT-Fits}(b), which show the upper limit of the region of validity of the fit (pragmatically set here at 2\% divergence between fit and data). The coefficient $A$ of the $1/T$ term also varies with pressure, as shown in Fig.~\ref{fig:RvT-Fits}(c). Similar to $\rho_{0}$, $A$ is dramatically suppressed with initial pressurization, by two orders of magnitude up to the vicinity of the QCP at $p_{c}$, where a change of behavior is seen, and further reductions are much more gradual. Indeed, the pressure dependences of $\rho_{0}$ and $A$ are found to agree strikingly well, see Fig.~\ref{fig:RvT-Fits}(c), suggesting that the two conduction channels may be connected in some way.

In Sm$_2$Ir$_2$O$_7$, the lattice constant is known to decrease linearly with pressure across the whole range measured here, while the oxygen-coordinate parameter $x$ changes only a little \citep{Wang2020a}. The swift drop on increasing pressure up to 60\,kbar in the parameters $\rho_0$ and $A$ that quantify the non-metallic behavior at low temperatures indicates that the degree of localization in the AIAO Ir-ordered phase is highly sensitive to reductions in the lattice constant and the concomitant enhancement of the hopping parameter, which is consistent with the ground state being underpinned by electronic correlations. By contrast, in the state that emerges above the QCP, the localization of electrons is much reduced as compared to the behavior seen under ambient conditions, and further changes in pressure have only a small effect. The relative decoupling of this state from changes in the lattice constant suggests that correlations play a smaller role in the electronic ground state in this pressure regime.

\subsection*{Transport properties under applied pressure - hysteresis in magnetoresistance}

\begin{figure}
\centering
\includegraphics[width=1\linewidth]{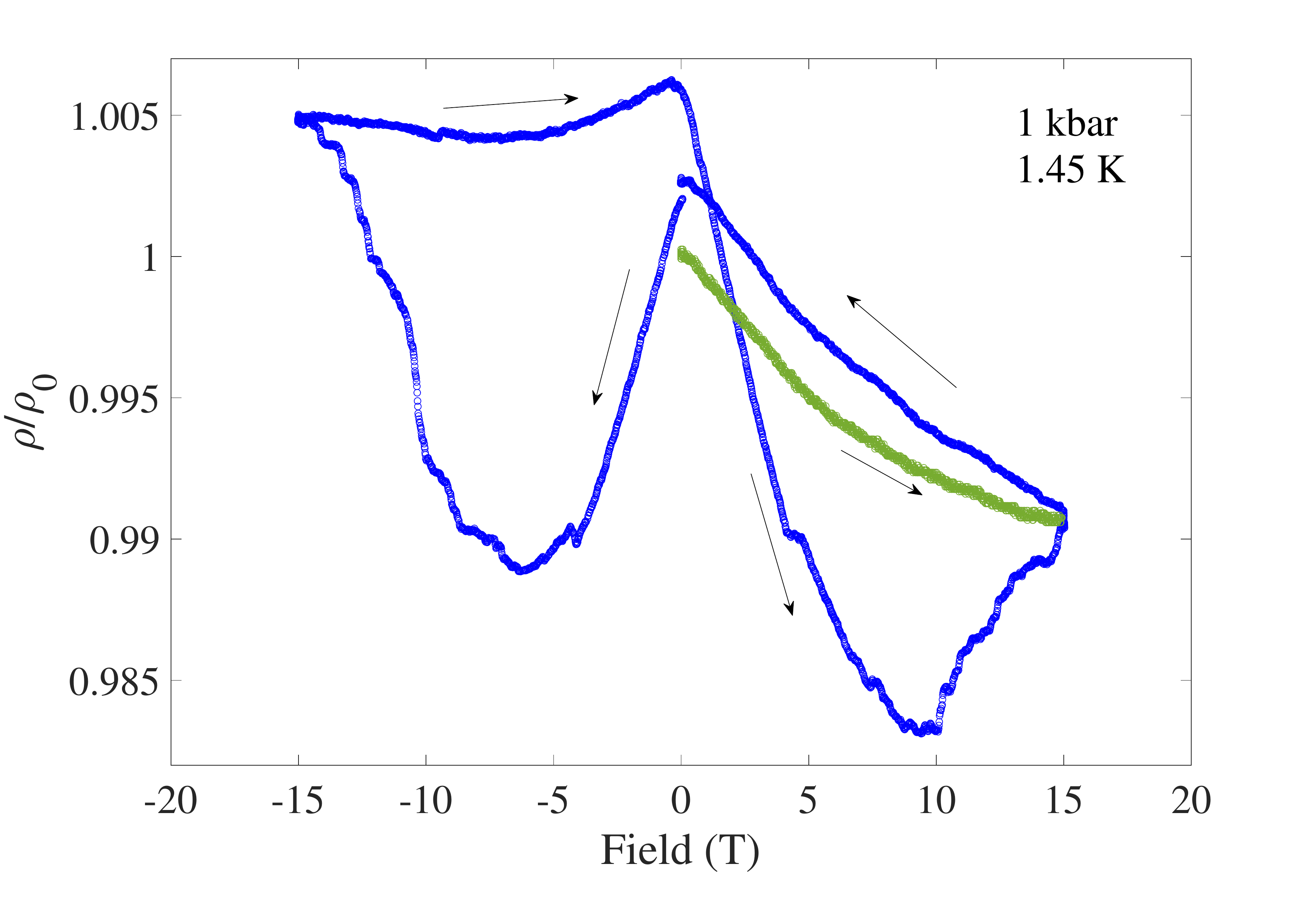}
\caption{Hysteresis in MR - Magnetoresistance at $T=$1.45(5)~K and near-ambient pressure, showing a clear hysteresis loop. The field was applied along the approximate $\left[1\,1\,1\right]$ direction, perpendicular to the plane of measurement. Data are normalized to their zero-field values.The green points denote the first application of the field up to +15~T after cooling the sample in zero field, then arrows around the subsequent blue data points show the evolution of the resistivity as the field is swept to -15~T and then back to +15~T again, at a rate of 1~Tmin$^{-1}$.}
\label{fig:MR1-ambient-lowT}
\end{figure}

\begin{figure}
\centering
\includegraphics[width=1\linewidth]{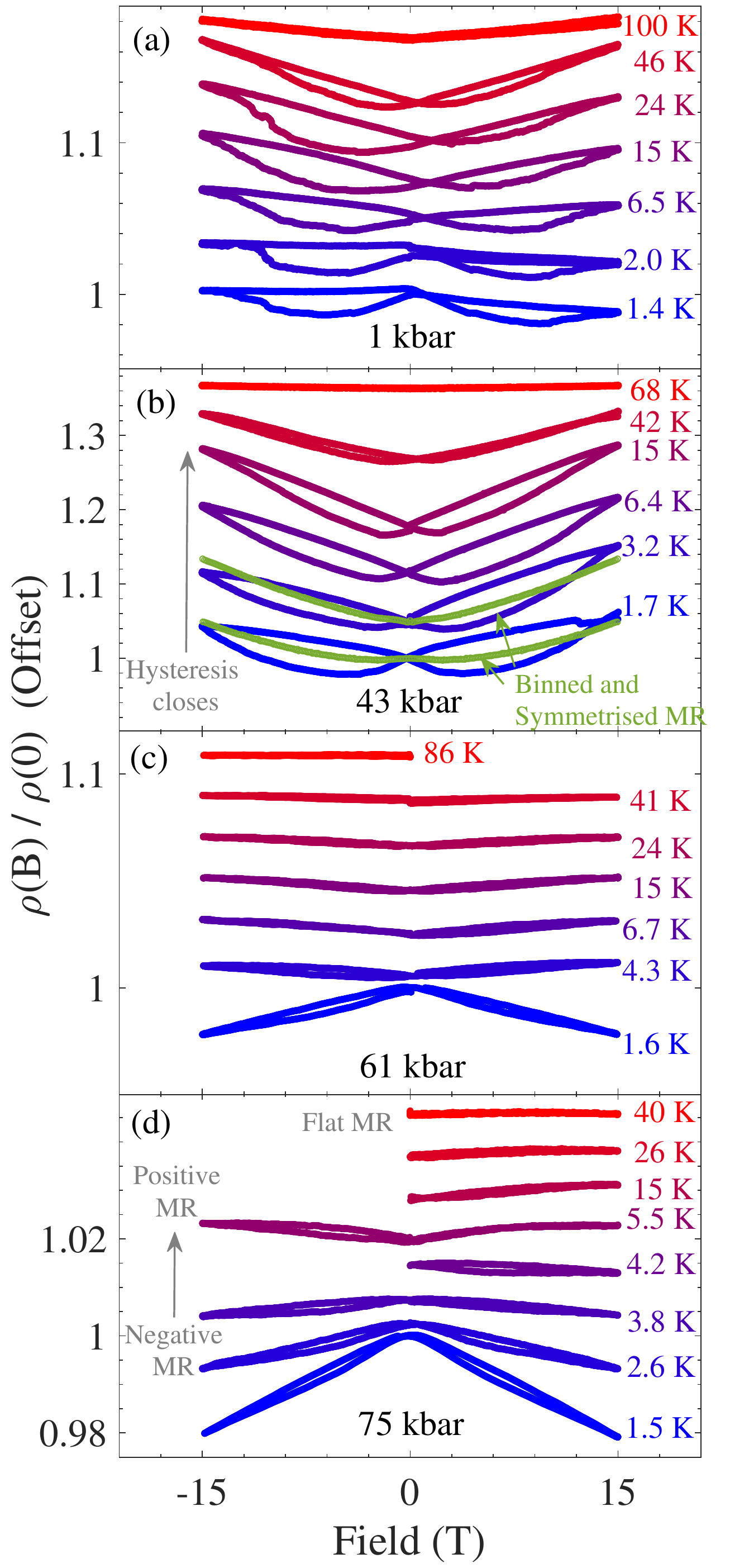}
\caption{Evolution of MR with pressure - Magnetoresistance ratio $\rho(B)$/$\rho(B=0)$ at fixed temperatures for pressures of (a) 1(1)~kbar, (b) 43(2)~kbar (c) 61(2)~kbar and (d) 75(2)~kbar. The field was applied along the approximate $\left[1\,1\,1\right]$ direction, perpendicular to the plane of measurement. By 61~kbar the form of the hysteresis loop is changed, and the magnetoresistance is negative at low temperatures. Green points in (b) illustrate examples of binning and symmetrizing the data to obtain effective single-valued $\rho(B)$ data, separating out the hysteresis effects, as displayed for all data in Fig. \ref{fig:MR-Binned}. The data all have value 1 at $B=0$, and are offset vertically for visibility.}
\label{fig:MR2}
\end{figure}

At each pressure, we measured the magnetic-field dependence of the resistivity at various temperatures down to 1.5~K. The field was varied between $\pm15$~T in the direction perpendicular to the plane of the resistivity measurement, which is along the diamond anvil cell axis and close to the $\left[1\,1\,1\right]$ direction. Fig.~\ref{fig:MR1-ambient-lowT} shows that the low-temperature $\left[1\,1\,1\right]$ magnetoresistance exhibits a marked and reproducible hysteresis, which persists for pressures below about 61~kbar. Also shown is the first field sweep performed on cooling in zero field from high temperatures (green line) which lies between the up and down sweeps of subsequent field sweeps. The strong hysteresis remains as temperature is increased up to about 50~K, but closes gradually at higher temperatures, and before the AIAO-ordering transition is reached, see Fig.~\ref{fig:MR2}. As can be seen in Fig.~\ref{fig:MR2} and the Supplementary Note 9 (data for all pressures) and Supplementary Figure 18 (an interpolated colormap plot of hysteresis magnitude across the entire temperature-pressure range) of the Supplementary Information, this trend continues at elevated pressures until about 61~kbar, at which point the size of the hysteresis reduces dramatically, disappearing close to the AIAO QCP.

This hysteretic behavior is highly reminiscent of that observed in Ho$_2$Ir$_2$O$_7$ (see Supplementary Figure 17 for a comparison of Fig.~\ref{fig:MR1-ambient-lowT} with Ho$_2$Ir$_2$O$_7$ and Nd$_2$Ir$_2$O$_7$ data). In that case, a magnetic field close to the $\left[1\,1\,1\right]$ direction acts to align the large Ho moments in a 3I1O/1I3O monopole-crystal arrangement which, via the Ho-Ir coupling, energetically favors one Ir AIAO domain over the other and alters the domain ratio in the sample. On sweeping down, the domain ratio remains constant until a negative field is applied, whereupon the second domain is favored and begins to grow preferentially. In Sm$_2$Ir$_2$O$_7$, the Sm ions have a significantly smaller moment than the Ho ions in Ho$_2$Ir$_2$O$_7$ (=~10$\mu_{\rm B}$). If the same effect is responsible for the hysteresis in both materials, this would explain why the magnitude of the hysteresis is lower in Sm$_2$Ir$_2$O$_7$ and also why the hysteresis is seen not to close even at the highest fields reached. At these temperatures, a field of about 4~T is sufficient to close the hysteresis in Ho$_2$Ir$_2$O$_7$ and achieve a steady 70:30\% domain ratio (a value determined by the presence of strong pinning sites). By contrast, in Sm$_2$Ir$_2$O$_7$, the field required to saturate the moments and fully close the hysteresis loops is far greater than the 15~T achieved in these measurements; there are no high-field regions where the resistance becomes single-valued. Indeed, from the ratio of the moment sizes in the two cases we would expect the field required to close the hysteresis to be roughly ten times larger in Sm$_2$Ir$_2$O$_7$ than in Ho$_2$Ir$_2$O$_7$. This also explains why the hysteresis loops shown by Pearce \emph{et al.} appear smoother than found here for Sm$_2$Ir$_2$O$_7$ \textendash{} the data in Fig.~\ref{fig:MR1-ambient-lowT} access only the low-field region of the hysteresis loop, which likely extends out towards 100~T.

\begin{figure}
\centering
\includegraphics[width=0.99\linewidth]{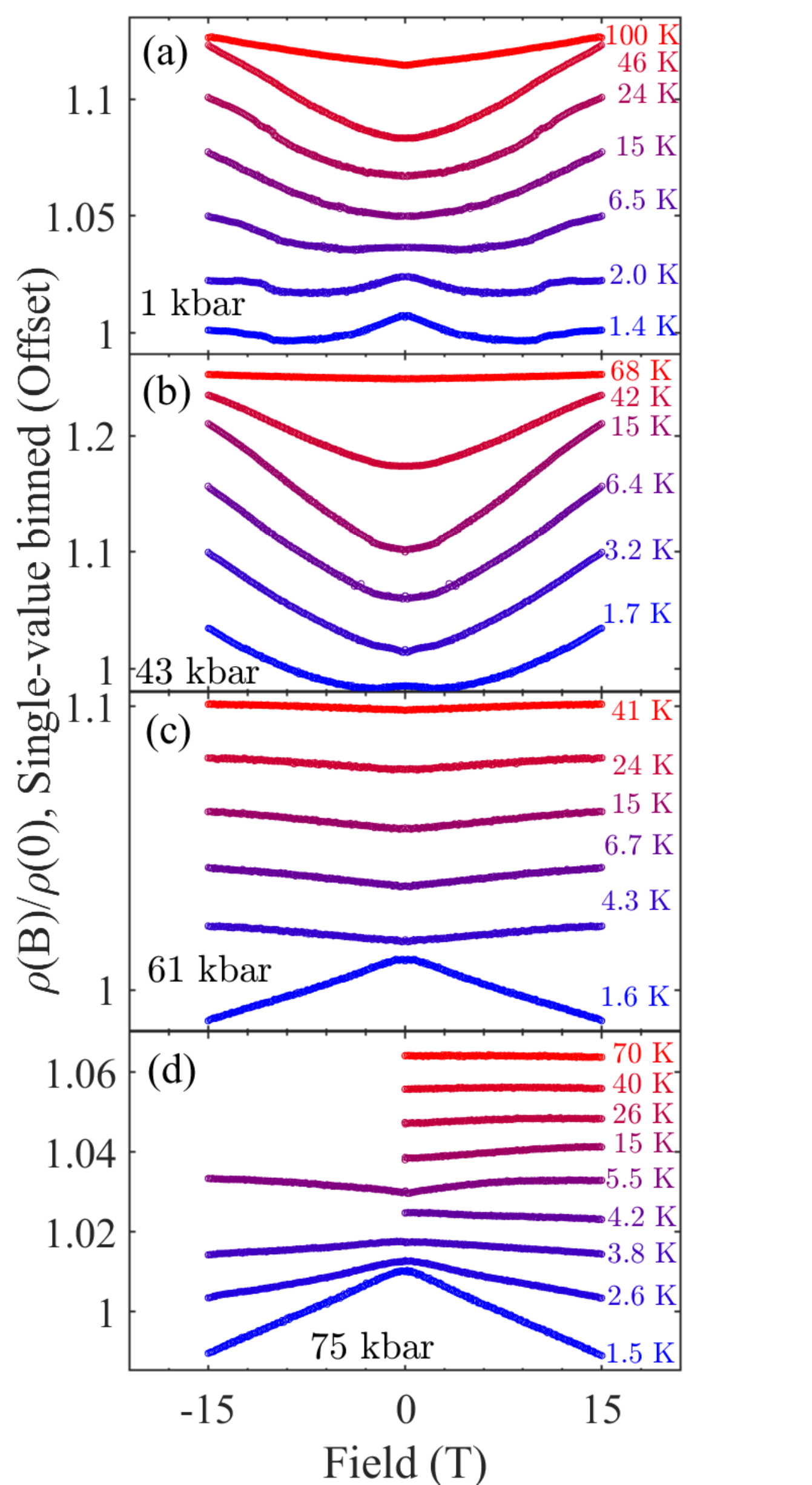}
\caption{Single-valued MR at high pressures - Background magnetoresistance ratio $\rho(B)$/$\rho(B=0)$ at fixed temperatures for pressures of (a) 1(1)~kbar, (b) 43(2)~kbar (c) 61(2)~kbar and (d) 75(2)~kbar. By 61~kbar the MR is markedly negative at low temperature. These data are extracted from Fig. \ref{fig:MR2} by binning and symmetrizing the data to remove $\rho_{xy}$ components, and averaging up and down sweeps to separate out the hysteresis effects. The data all have value 1 at $B=0$, and are offset vertically for visibility.}
\label{fig:MR-Binned}
\end{figure}

The sensitivity of the magnetoresistance in Ho$_2$Ir$_2$O$_7$ to changes in Ir-domain ratio arises because the monopole density varies differently with field in the two Ir-domain types, which in turn affects the monopole-quasiparticle scattering rate via two coupling mechanisms linking: (i) electronic spin and magnetic monopole charges; and (ii) electronic charge and the electric dipoles associated with monopoles. The validity of the magnetic monopole description in Sm$_2$Ir$_2$O$_7$ has not yet been definitively established and so it is not fully clear if the hysteresis has the same origin as in Ho$_2$Ir$_2$O$_7$; however we can estimate that mechanism (i) should be smaller for the low-moment Sm ions. On the other hand, mechanism (ii), which is driven by Sm\textendash{}Sm interactions, could be of a similar size in both cases, as we expect the Sm exchange energy to be similar to or stronger than the equivalent in Ho$_2$Ir$_2$O$_7$.

An alternative, related effect is proposed to account for the hysteresis observed in Nd$_2$Ir$_2$O$_7$, which also has small rare-earth moments. Ueda \emph{et al.} suggest that the current is preferentially carried through the Ir-domain walls and so the resistance responds to changes in the size and distribution of these as the field is swept~\citep{Ueda2015a}. Which of these mechanisms is responsible for the effects seen in Sm$_2$Ir$_2$O$_7$ remains an interesting open question, but what is clear is that the large hysteresis is characteristic of the AIAO phase and disappears as the QCP is approached. For pressures above $p_{{\rm c}}$, some hysteresis is still seen in the magnetoresistance at the lowest temperatures (see Fig.~\ref{fig:MR2}(c,d)), but it is significantly smaller than that observed within the AIAO phase. Furthermore, the background form of the magnetoresistance undergoes a marked change at this pressure.

\subsection*{Transport properties under applied pressure - background magnetoresistance}

In order to highlight the background form of the magnetoresistance of the sample separately from the hysteresis effects, one can average the multivalued resistivity at the same field value, to find an effective single-valued $\rho(H)$ for each pressure and temperature.

The low-temperature magnetoresistance data at all pressures show an asymmetry between positive and negative fields (Fig.~\ref{fig:MR1-ambient-lowT})---an overall linear slope. Recent work has shown a strong, but non-hysteretic asymmetry in the low-temperature magnetoresistance of Eu$_2$Ir$_2$O$_7$ after field-cooling, caused by the formation of single domains of AIAO Ir order which remain robust up to 9~T \citep{Fujita2015,Xu2020b}. This is contrary to the situations in Nd$_2$Ir$_2$O$_7$ and Ho$_2$Ir$_2$O$_7$ which both show switching of domains at low fields and strong hysteresis in magnetoresistance that, as mentioned above, are attributed to coupling between the Ir ions and the Ising-like rare-earth ions \citep{Ueda2015a,Pearce2022}. As we have shown, the Sm ions in the present material are also Ising-like and the MR is strongly hysteretic with only a weak asymmetry, and so we expect Sm$_2$Ir$_2$O$_7$ to behave similarly to those materials. Instead, we attribute the asymmetry to a component of the Hall voltage mixing into the measured longitudinal resistivity. The magnitude of such a Hall signal is inversely proportional to the carrier density (for a simple single band at least). Here, the strongest slope is seen at low pressures where resistivity is high. To reliably extract absolute values of carrier density, a Hall-bar geometry on a lamellar sample with accurately placed contacts would be required; the small and irregularly shaped sample in our restrictive diamond anvil cell measurements precludes this treatment. To remove the Hall asymmetry, the data were symmetrized by averaging negative and positive field traces. 

Examples of the resulting $\rho(H)$ plots are shown in Fig.~\ref{fig:MR2}(b) in green and in Fig.~\ref{fig:MR-Binned} (data for all pressures are shown in Supplementary Notes 7-9). With these data we can systematically describe the form and sign of the $\left[1\,1\,1\right]$ magnetoresistance separately from the large hysteresis seen at low pressures. As can be seen from Fig.~\ref{fig:MR-Binned}, the low-temperature, low-pressure background magnetoresistance first drops with field before rising as field is increased. A negative component is expected in the low-temperature magnetoresistance of materials with locally fluctuating or disordered moments, such as a paramagnet or a Kondo system. As the field aligns the spins, spin scattering is reduced and resistance drops. In this material at low pressure and temperature, only the Sm$^{3+}$ moments are free to align with the field, as indicated by magnetometry measurements. The behavior of the magnetoresistance in this region is reminiscent of that seen in Ho$_2$Ir$_2$O$_7$ under an applied $\left[1\,1\,1\right]$ field \citep{Pearce2022}. In that case, there is also a negative magnetoresistance at low fields, caused by the alignment of the Ho$^{3+}$ moments, followed by change to positive magnetoresistance as the effects of enhanced monopole scattering starts to dominate. As temperature is raised the magnetoresistance flattens out in both materials in similar way.

As pressure is increased, the background magnetoresistance retains the overall shape it has at low pressures, until the critical pressure for complete suppression of the Ir AIAO is approached. Here, the low temperature magnetoresistance becomes strongly negative across the whole field range (see Fig.~\ref{fig:MR-Binned}) with a small amount of hysteresis (Fig.~\ref{fig:MR2}). As temperature is raised, the negative magnetoresistance diminishes, becomes weakly positive and then flattens out. The narrow hysteresis closes and disappears at the same temperature the MR changes sign. This form of the magnetoresistance continues for all pressures measured above 61(2) kbar. 

There are several potential causes for positive magnetoresistive behavior. Simplest is the case of orbital motion of charge carriers in non-magnetic or antiferromagnetic (semi)metals with closed Fermi surfaces. Tafti \emph{et al.} attribute the positive magnetoresistance in high-pressure Eu$_2$Ir$_2$O$_7$ to this effect \citep{Tafti2012a}.

\begin{figure*}
\centering
\includegraphics[width=1.0\linewidth]{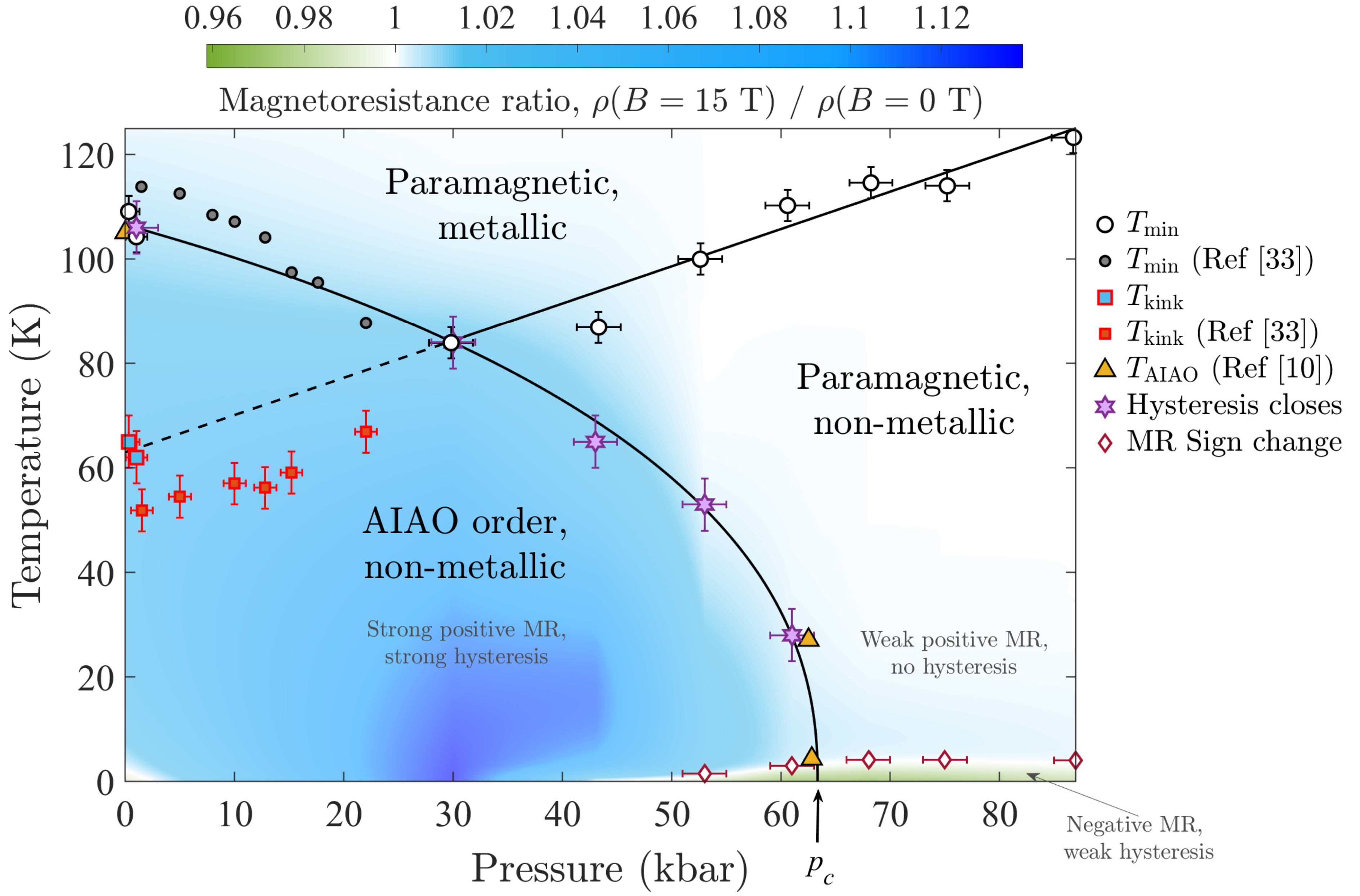}
\caption{Pressure--temperature phase diagram for Sm$_2$Ir$_2$O$_7$ - The background colormap displays interpolated values of the magnetoresistance (MR) ratio and is indicative of the size of the MR. Temperatures of the resistivity minimum $T_{\mathrm{min}}$ are plotted as open circles. Previous data from the literature up to 20~kbar \citep{Liu2018b} are overlaid as closed circles Above about 30\,kbar $T_{\mathrm{min}}$ shows a linear increase with pressure (solid line), which can be extrapolated back to ambient pressure (dotted line) where it coincides with the kink feature observed in the gradient of $\rho(\mathrm{T})$ at the lowest pressures. These are plotted (open squares) along with similar features found in the data of Liu \emph{et al.}~\citep{Liu2018b} up to 20\,kbar (closed squares). Triangles denote the Ir AIAO ordered phase observed by Wang \emph{et al.}~\citep{Wang2020a} and star symbols represent the temperatures at which the magnitude of the hysteresis in the MR extrapolates to zero. The curved solid line is a guide to the eye through these points. The green shaded region highlights a regime around the QCP with a negative background magnetoresistance. The diamond symbols denote the temperatures where the magnetoresistance switches from negative to positive, which bound this region. Error bars show the uncertainties in determining pressure (see Methods) and estimated uncertainty for their corresponding temperature values.}
\label{fig:PhaseDiagram}
\end{figure*}

\section*{Discussion}

Figure \ref{fig:PhaseDiagram} shows the pressure-temperature phase diagram of both the magnetic and transport properties of Sm$_2$Ir$_2$O$_7$ constructed by pulling together all of the features seen in our measurements, as well as prior results in the literature. The background colormap displays the ratio of the single-valued magnetoresistance at 15~T and zero field, to distinguish regions of positive (blue) and negative (green) MR. Regions where the MR is flat, showing no variation in field, are shown in white on this scale. (See Supplementary Figure 16 and Supplementary Note 12). The purple stars represent the temperatures at which the hysteresis in MR closes (see Supplementary Figure 18) and are seen to follow closely the AIAO magnetic transitions at $T_{\mathrm{AIAO}}$ (orange triangles) found in the resonant X-ray measurements of Wang \emph{et al.}~\citep{Wang2020a}. The curved solid line is a guide to the eye that connects these points.

The measured transport properties are clearly consistent with the pressure-induced suppression of the AIAO phase as manifested in three specific trends in the data. \emph{First}, the crossover from metallic to non-metallic conduction, marked by the resistivity minimum at $T$ = $T_{\mathrm{min}}$, as reported by Liu \emph{et al.} \citep{Liu2018b} (closed circles) and our low-pressure data (open circles), closely tracks the decrease in $T_{\mathrm{AIAO}}$ at all pressures up to 30~kbar. This is consistent with the interpretation of $T_{\mathrm{min}}$ as a Mott transition associated with the Ir ordering at low pressures \citep{Ueda2015,WitczakKrempa2012,Wang2020a}. \emph{Second}, the magnitude of the resistivity continuously decreases as pressure is raised (Fig. \ref{fig:RvT}), and the extrapolated finite $\rho_{0}$ values drop rapidly with pressure up to 60(4)~kbar, which is where AIAO Ir order is completely quashed. As mentioned earlier, the AIAO magnetic order is driven by the combination of $U/t$ (Coulomb repulsion/hopping strength) and local Ising anisotropy \citep{Wang2020a}. As pressure is applied, the lattice parameter decreases significantly, the $5d$ bandwidth and hopping strength increases, and $U/t$ is suppressed \citep{Wang2020a}. The combined result of pressure then is to stymy the AIAO ordering temperature and enhance conductivity.
\emph{Third}, the marked hysteresis seen in the magnetoresistance for magnetic field applied close to the $\left[1\,1\,1\right]$ direction is present only within the AIAO ordered phase. This makes sense if, as discussed earlier, the hysteresis has its origin in the plastic deformation of the Ir domains via coupling with the Sm$^{3+}$ moments, equivalent to that reported in Ho$_2$Ir$_2$O$_7$ or Nd$_2$Ir$_2$O$_7$. Once the AIAO order is suppressed, the domains are no longer present and the large magnetoresistance hysteresis disappears. All these magnetotransport features affirm the existence of the same quantum critical point at $p_{{\rm c}}$ = 63~kbar suggested by Wang \emph{et al.} from their X-ray data.

Despite these connections between the Ir AIAO order and the measured transport properties, an important and clear result of our experiments is that Sm$_2$Ir$_2$O$_7$ remains metallic at high temperatures and non-metallic at low temperatures below a finite $T_{\mathrm{min}}$ for the entire pressure range measured. In particular, the transition marked by the resistivity minimum at $T_{\mathrm{min}}$ is not suppressed to zero temperature along with the Ir AIAO order to yield metallic behavior at all temperatures, as was naively anticipated from the coincidence of the two transitions at low pressures and from the pressure dependence of $T_{\mathrm{min}}$ in Nd$_2$Ir$_2$O$_7$ \citep{Liu2018b,Ueda2015,Ueda2017}.

As mentioned, $T_{\mathrm{min}}$ initially decreases, following the phase line of the Ir AIAO order observed by Wang \emph{et al.}~\citep{Wang2020a}. However, above 30\,kbar this trend reverses and $T_{\mathrm{min}}$ increases linearly up to the highest pressures measured. It seems that some additional mechanism is causing a resurgence of localization of the charge carriers once the pressure pushes the AIAO transition below a certain temperature. This new localization mechanism then strengthens with further application of pressure. Interestingly, an extrapolation of the high-pressure linear trend of $T_{\mathrm{min}}(p)$ back to low pressures, (dashed black line) coincides with the position of the subtle change in $d\rho/dT$ observed in our sample at $T_{\mathrm{kink}}$. This feature is also present in the low-pressure data of Liu \emph{et al.} (closed squares, see also Supplementary Figure 12) \textemdash{} albeit at a slightly lower temperature than in our data \textemdash{} and is seen to increase linearly with pressure at a gradient equal to the high-pressure dependence of $T_{\mathrm{min}}$. We note that a similar feature is observed in the resistivity of Eu$_2$Ir$_2$O$_7$, where it also appears in the heat capacity \citep{Ishikawa2012}. There it is shown to be most clear in the purest crystals with smallest deviations from Eu and Ir stoichiometric values, suggesting in this case that it is an intrinsic effect that gets smeared out in lower quality samples. Possible origins for this kink feature include low-temperature structural and/or magnetic variations, and we note again that a shoulder-like feature appears in the field-cooled susceptibility data close to $T_{\rm kink}$ (see Figure~\ref{fig:SQUIDdata}).

Whatever the cause, it is apparent that $T_{\mathrm{min}}$ and $T_{\mathrm{AIAO}}$ separate for pressures above 30\,kbar, and the Ir AIAO order ceases to be the driver of the non-metallic state when the temperature scale associated with the new, as yet unidentified, mechanism begins to dominate. This explains why the material remains non-metallic once the AIAO order is suppressed at the $p_{{\rm c}}$ = 63\,kbar QCP. For pressures between 30 and 63~kbar, the Ir ordering still takes place below $T_{\mathrm{AIAO}}$. The main markers of the AIAO ground state in the transport in this intermediate pressure region are the large magnetoresistive hysteresis discussed earlier and the strong negative pressure dependence of the extrapolated $\rho_{0}$ value.

A new form of magnetotransport emerges around the QCP. The large hysteresis and competition between negative and positive magnetoresistance components, seen at low temperatures in the AIAO phase, now changes to a region of weakly hysteretic but strongly negative magnetoresistance that extends from just below $p_{{\rm c}}$ to beyond the highest pressure measured (green region in Fig.~\ref{fig:PhaseDiagram}). The negative magnetoresistance evolves slowly to a weakly positive response as the temperature is raised above about 5~K, disappearing altogether at elevated temperatures. At around the same temperature the weak hysteresis closes gradually (as can be seen more clearly in Supplementary Figures 18 and 15). The open diamond symbols in  Fig.~\ref{fig:PhaseDiagram} illustrate the temperatures where the MR is zero (see Supplementary Figures 16 and 19). According to Wang \emph{et al.}, while the long-range Ir AIAO order is suppressed above $p_{{\rm c}}$, the axial anisotropy is preserved, leaving behind weakly correlated or paramagnetic Ising-like moments \citep{Wang2020a}. Any low-temperature magnetic order on the Sm sublattice is also predicted to be lost once the Ir order is gone. Alignment of these fluctuating spins in an applied magnetic field could account for the negative magnetoresistance observed above the QCP. The positive magnetoresistance at higher temperatures is consistent with orbital effects arising from small Fermi-surface pockets, such as might be expected for a disordered semimetal and, as mentioned above, a similar origin was proposed for the high-pressure magnetoresistance seen in Eu$_2$Ir$_2$O$_7$ \citep{Tafti2012a}.

The precise origin of these new electronic properties remains to be determined. Several theoretical proposals have been made for the electronic phases that might exist in the vicinity of the AIAO QCP \citep{Wan2011,WitczakKrempa2012,Savary2014,Ishii2015,Yamaji2016}. These include a WSM phase expected to be located in a relatively narrow region of Coulomb $U$ between the AIAO Mott insulator and the paramagnetic-metal phase (previously believed to occur close to where AIAO order is suppressed) \citep{Ueda2016,Wang2020a,Ueda2020}, and tentatively reported in the results of Telang and Singh \citep{Telang2021}. However, as Wang \emph{et al.} point out from their estimates of the shift in hopping $t$ required to realize the QCP in Sm$_2$Ir$_2$O$_7$, this theoretically predicted narrow region where the WSM exists could span the entire pressure range measured \citep{Wang2020a}. Our analysis of the single-crystal transport data finds them to be well described by a $1/T$ functional form, which is indeed consistent with models of WSMs \citep{Hosur2012}, albeit with the inclusion of a parallel slowly-varying (relative to $1/T$) conduction channel, possibly arising from surface or domain-wall conduction. The temperature range over which this $1/T$ dependence is valid is found to be largest at the lowest pressure, away from the QCP. 

Another theoretical prediction is that the QCP in Sm$_2$Ir$_2$O$_7$ separates the WSM from a non-Fermi-liquid semimetal phase distinguished by quadratic-band touching (QBT) \citep{Savary2014}. Indeed, ARPES measurements suggest that this is the ambient pressure phase exhibited by Pr$_2$Ir$_2$O$_7$ at low temperatures \citep{Kondo2015} and by Nd$_2$Ir$_2$O$_7$ in its metallic state at elevated temperatures \citep{Nakayama2016}. It should also be pointed out that optical conductivity spectra for Sm$_2$Ir$_2$O$_7$ point towards a semimetal nature \citep{Ueda2016}, and are distinct from the finite gap seen in these data for Eu$_2$Ir$_2$O$_7$. Recent work on Pr$_2$Ir$_2$O$_7$ has additionally reported topological Weyl semimetal states under the application of magnetic field \citep{Ueda2022a}. A recent paper reporting transport measurements on Sm$_2$Ir$_2$O$_7$ speculates that the WSM is present at ambient pressure and that doping more than 5\% of Bi on the Sm site leads to a transformation to the QBT phase, although the evidence in favor of this interpretation is limited \citep{Telang2021}. Our data also do not provide evidence for QBT behavior emerging under pressure; however, we highlight again that our experiments reveal that the paramagnetic metallic phase is not realized once AIAO order is destroyed and instead another, as yet mysterious, electronic phase emerges. 

Our data share certain similarities with work done by Tafti \emph{et al.} on Eu$_2$Ir$_2$O$_7$, neighbor to Sm$_2$Ir$_2$O$_7$ in the chemical series of this family \citep{Tafti2012a}. While the $T_{\mathrm{min}}$ in that material does not appear to shift very much at all with applied pressure, the authors do find that the resistivity drops significantly as pressure is increased. In the low-pressure region, they can fit their $\rho(T)$ data to an activated model at low temperatures (which does not work for Sm$_2$Ir$_2$O$_7$). The size of the insulating gap in Eu$_2$Ir$_2$O$_7$ drops swiftly towards zero at 80~kbar and then remains low and slowly varying upon further increase in pressure, reminiscent of the $A$ coefficient we extract from the $1/T$ fits for our Sm$_2$Ir$_2$O$_7$ data. The authors describe their high-pressure, low-temperature state as a novel metal with resistivity that increases on cooling. In contrast to our results, Tafti \emph{et al.} find that the magnetoresistance of Eu$_2$Ir$_2$O$_7$ is positive down to the lowest temperatures in the high pressure region \citep{Tafti2012a}. It should be noted that the AIAO phase transition has not yet been tracked with pressure in the Eu material. Tafti \emph{et al.} also find that the resistivity at high temperatures and low pressures in Eu$_2$Ir$_2$O$_7$ increases on cooling. This is in contrast to the more metallic behavior (i.e. decreasing resistivity on cooling) observed at ambient pressure in single-crystal measurements of other pyrochlore iridates $Ln_2$Ir$_2$O$_7$, including $Ln$ = Pr \citep{Nakatsuji2006}, Nd \citep{Ueda2015a}, Sm (our data), Ho \citep{Pearce2022} and even in other studies of the same Eu material considered by Tafti \emph{et al.} \citep{Ishikawa2012}, as well as in some pressed-powder measurements, e.g. Pr \citep{Matsuhira2011}, Nd \citep{Matsuhira2011, Ueda2014, Ueda2015} and  Sm \citep{Graf2014}. We suggest that a metallic temperature dependence above $T_{\mathrm{min}}$ is the true intrinsic behavior of the pyrochlore iridates in this region of the phase diagram.

We point out that another origin has been proposed for the metal insulator transition seen in the pyrochlore osmate Cd$_2$Os$_2$O$_7$. This material is a well-compensated, low-carrier density semimetal in its high-temperature paramagnetic phase. At 227~K, the AIAO magnetic transition on the Os sublattice is suggested to shift the electron and hole bands by a relatively small, but sufficient amount to cause a Lifshitz-type transition in this finely balanced semimetal, opening a small band gap \citep{Hiroi2015,Shinaoka2012}. The transport below the transition is governed by the opening gap, a number of in-gap impurity states caused by oxygen non-stoichiometry, and the presence of conduction channels in magnetic domain walls (as already mentioned above in the context of Nd$_2$Ir$_2$O$_7$). It is interesting to draw parallels with our material, which also shows non-activated behavior and a non-diverging resistivity. Additional Hall effect measurements would be required to distinguish between this Lifshitz mechanism and the Mott transition more commonly attributed to the pyrochlore iridates. Such measurements are planned but require a further refinement of the synthesis to obtain larger high-quality single crystals than are currently available.

In summary, our data show a significant change in the electronic properties of Sm$_2$Ir$_2$O$_7$ as the AIAO Ir magnetic order is gradually destroyed by pressure. Although the metallic-to-non-metallic crossover at $T = T_{\mathrm{min}}$ is not suppressed in tandem with the magnetic transition, the magnitude of the resistivity, the extrapolated finite $T = 0$ resistivity, and the coefficient of the $1/T$ resistivity component all drop swiftly as the QCP is approached, then switch over to a much reduced rate of reduction with further increasing pressure. The $1/T$ temperature dependence, which is predicted for a Weyl semimetal, fits the data up to 55~K at 0.3~kbar, but this range is reduced at higher pressures. While it is not currently possible from these data alone to precisely identify the electronic state that emerges beyond the QCP, we have measured both transport and magnetotransport data across the entire phase diagram in order to provide information to guide theoretical studies. Our results also act as a basis for future experimental efforts, with optical conductivity and angle-dependent magnetotransport particularly well suited to shedding light on the possible Weyl semimetal and anisotropic, quadratic-band-touching phases.

\FloatBarrier

\section*{Methods}

\subsection*{Crystal growth}
Single crystals of Sm$_2$Ir$_2$O$_7$ were grown by the flux method, using KF as the flux, as described in Ref. \citep{Millican2007}. They were characterized and verified to be of good crystalline quality by X-ray diffraction.

\subsection*{High pressure transport measurements}
Magnetotransport data were carried out in a diamond anvil cell \citep{Dunstan1989,Dunstan1989a} with non-magnetic BeCu body and hardened BeCu gasket. The culet of the diamonds used was 1~mm and the gasket hole was drilled to 0.5~mm. The cell was pressurized at room temperature for each pressure point, then measured on a home-built probe loaded into a Oxford Instruments 17~T cryomagnet fitted with a variable temperature insert. A single crystal was prepared for resistivity measurements in the standard 4-wire geometry with gold wires bonded to the crystal with DuPont 6838 silver-loaded epoxy. Glycerol was chosen as the pressure transmitting medium due to its favorable compressibility and hydrostatic properties. Across the pressure range of interest, glycerol is known to perform well in comparison to other choices of media \citep{Tateiwa2009}. Making diamond anvil cell measurements in pressures close to 100~kbar inevitably brings some risk of a degree of non-hydrostaticity; nevertheless we are confident that extrinsic effects of the pressure-transmitting medium can be ruled out as an explanation of all the main features in our data. The pressure was measured via the fluorescence spectrum of a chip of ruby placed within the pressure region \citep{Mao1986}. Typical pressure uncertainties, as denoted on values given in the text are estimated from two factors. Firstly, there is a gradient or distribution of pressure within the sample volume, which we quantify via the broadness of the ruby fluorescence peak and its fitting uncertainty. Secondly, previous calibrations on identical cells have shown pressure losses upon cooling, due to thermal expansion. The resulting uncertainty from these two factors we determine as $\pm$1~kbar below 10~kbar and $\pm$2~kbar above, noting that the pressure will always be above 0~kbar. The magnetic field was applied along the cell axis, a direction close to the crystallographic $\left[1\,1\,1\right]$. The crystal was sat on the diamond anvil faces and the current applied in the plane of the anvil faces, perpendicular to the field.

\subsection*{Magnetization measurements}
 Low-field DC magnetization measurements were carried out on a powder sample in a Quantum Design MPMS SQUID magnetometer system. The sample was then additionally measured in up to 60~T of applied field via extraction-coil magnetometry at the NHMFL, Los Alamos. As this technique does not return absolute values of the magnetization, these were obtained via calibrating the low-field slope up to 7~T with the MPMS data at 1.8~K. There is no discernible difference between the latter and the pulsed field data at 1.5~K and at 0.9~K shown in the manuscript.

\section*{Data availability statement}
Data presented in this paper are made available online at wrap.warwick.ac.uk/182307.

\acknowledgments{We thank T. Orton and P. Ruddy at the University of Warwick for technical assistance. This project has received funding from the European Research Council (ERC) under the European Union\textquoteright s Horizon 2020 research and innovation programme (grant agreement No. 681260).  We acknowledge the Engineering and Physical Sciences Research Council (EPSRC), UK and the Oxford-ShanghaiTech collaboration project for financial support. This work was supported by EPSRC grants No. EP/P034616/1, No. EP/V062654/1 and No. EP/N034872/1. A portion of this work was performed at the National High Magnetic Field Laboratory (NHMFL), which is supported by National Science Foundation Cooperative Agreement No. DMR-1644779 and the Department of Energy (DOE). J. S. acknowledges support from the DOE BES program ``Science at 100 T'', which permitted the design and construction of the specialized equipment used in the high-field studies.

\section*{Competing Interests}
The authors declare no competing interests.

\section*{Author Contributions}
PAG and DP conceived the experiments. DP grew and characterised the crystal. MJC prepared the pressure cell. MJC and KG ran the magnetotransport and SQUID measurements. JPT carried out a structural analysis. JS carried out the pulsed-field experiments. TNDLF, CC and ATB conceived and performed the crystal-field calculations. MJC analysed the data. MJC and PAG prepared the manuscript with input from all coauthors. PAG, DP, ATB and CC supervised the project.


\begin{thebibliography}{65}%
\makeatletter
\providecommand \@ifxundefined [1]{%
 \@ifx{#1\undefined}
}%
\providecommand \@ifnum [1]{%
 \ifnum #1\expandafter \@firstoftwo
 \else \expandafter \@secondoftwo
 \fi
}%
\providecommand \@ifx [1]{%
 \ifx #1\expandafter \@firstoftwo
 \else \expandafter \@secondoftwo
 \fi
}%
\providecommand \natexlab [1]{#1}%
\providecommand \enquote  [1]{``#1''}%
\providecommand \bibnamefont  [1]{#1}%
\providecommand \bibfnamefont [1]{#1}%
\providecommand \citenamefont [1]{#1}%
\providecommand \href@noop [0]{\@secondoftwo}%
\providecommand \href [0]{\begingroup \@sanitize@url \@href}%
\providecommand \@href[1]{\@@startlink{#1}\@@href}%
\providecommand \@@href[1]{\endgroup#1\@@endlink}%
\providecommand \@@startlink[1]{}%
\providecommand \@@endlink[0]{}%
\providecommand \Eprint [0]{\href }%
\providecommand \doibase [0]{http://dx.doi.org/}%
\providecommand \bibinfo  [0]{\@secondoftwo}%
\providecommand \bibfield  [0]{\@secondoftwo}%
\providecommand \translation [1]{[#1]}%
\let\auto@bib@innerbib\@empty  

\bibitem {Harris1997}
  \bibfield  {author} {\bibinfo {author} {\bibfnamefont {M.~J.}\ \bibnamefont
  {Harris}}, \bibinfo {author} {\bibfnamefont {S.~T.}\ \bibnamefont
  {Bramwell}}, \bibinfo {author} {\bibfnamefont {D.~F.}\ \bibnamefont
  {McMorrow}}, \bibinfo {author} {\bibfnamefont {T.}~\bibnamefont {Zeiske}}, \
  and\ \bibinfo {author} {\bibfnamefont {K.~W.}\ \bibnamefont {Godfrey}}, }
Geometrical Frustration in the Ferromagnetic Pyrochlore {Ho}$_2${Ti}$_2${O}$_7$, 
\href {\doibase 10.1103/physrevlett.79.2554} {\bibfield  {journal} {\bibinfo
   {journal} {Phys. Rev. Lett.}\ }\textbf {\bibinfo {volume} {79}},\
  \bibinfo {pages} {2554} (\bibinfo {year} {1997})}


\bibitem {Bramwell2001}
  \bibfield  {author} {\bibinfo {author} {\bibfnamefont {S.~T.}\ \bibnamefont
  {Bramwell}}\ and\ \bibinfo {author} {\bibfnamefont {M.~J.~P.}\ \bibnamefont
  {Gingras}}, }
Spin Ice State in Frustrated Magnetic Pyrochlore Materials, 
\href {\doibase 10.1126/science.1064761} {\bibfield  {journal}
  {\bibinfo  {journal} {Science}\ }\textbf {\bibinfo {volume} {294}},\ \bibinfo
  {pages} {1495} (\bibinfo {year} {2001})}


\bibitem{Castelnovo2008}
  \bibfield  {author} {\bibinfo {author} {\bibfnamefont {C.}~\bibnamefont
  {Castelnovo}}, \bibinfo {author} {\bibfnamefont {R.}~\bibnamefont
  {Moessner}} and \bibinfo {author} {\bibfnamefont {S.~L.}\ \bibnamefont
  {Sondhi}}, }
Magnetic monopoles in spin ice, 
\href {\doibase 10.1038/nature06433} {\bibfield  {journal}
  {\bibinfo  {journal} {Nature}\ }\textbf {\bibinfo {volume} {451}},\ \bibinfo
  {pages} {42} (\bibinfo {year} {2008})}


\bibitem {Morris2009}
  \bibfield  {author} {\bibinfo {author} {\bibfnamefont {D.~J.~P.}\
  \bibnamefont {Morris}} \emph{et al.}, }
Dirac Strings and Magnetic Monopoles in the Spin Ice {Dy}$_2${Ti}$_2${O}$_7$, 
\href {\doibase 10.1126/science.1178868}
  {\bibfield  {journal} {\bibinfo  {journal} {Science}\ }\textbf {\bibinfo
  {volume} {326}},\ \bibinfo {pages} {411} (\bibinfo {year}
  {2009})}


\bibitem {Fennell2009}
  \bibfield  {author} {\bibinfo {author} {\bibfnamefont {T.}~\bibnamefont
  {Fennell}} \emph{et al.}, }
Magnetic Coulomb Phase in the Spin Ice {Ho}$_2${Ti}$_2${O}$_7$, 
\href {\doibase 10.1126/science.1177582} {\bibfield  {journal} {\bibinfo  {journal}
  {Science}\ }\textbf {\bibinfo {volume} {326}},\ \bibinfo {pages} {415}
  (\bibinfo {year} {2009})}


\bibitem {Kadowaki2009}
  \bibfield  {author} {\bibinfo {author} {\bibfnamefont {H.}~\bibnamefont
  {Kadowaki}} \emph{et al.}, }
Observation of Magnetic Monopoles in Spin Ice, 
\href {\doibase 10.1143/jpsj.78.103706} {\bibfield  {journal}
  {\bibinfo  {journal} {J. Phys.Soc. Japan} }\textbf
  {\bibinfo {volume} {78}},\ \bibinfo {pages} {103706} (\bibinfo {year}
  {2009})}


\bibitem {Takatsu2014}
  \bibfield  {author} {\bibinfo {author} {\bibfnamefont {H.}~\bibnamefont
  {Takatsu}}, \bibinfo {author} {\bibfnamefont {K.}~\bibnamefont {Watanabe}},
  \bibinfo {author} {\bibfnamefont {K.}~\bibnamefont {Goto}} and \bibinfo
  {author} {\bibfnamefont {H.}~\bibnamefont {Kadowaki}}, }
Comparative study of low-temperature x-ray diffraction experiments on $R$$_2${Ir}$_2${O}$_7$($R$={Nd, Eu, and Pr}), 
\href {\doibase
  10.1103/physrevb.90.235110} {\bibfield  {journal} {\bibinfo  {journal}
  {Phys. Rev. B}\ }\textbf {\bibinfo {volume} {90}},\ \bibinfo {pages}
  {235110} (\bibinfo {year} {2014})}


\bibitem {Donnerer2016}
  \bibfield  {author} {\bibinfo {author} {\bibfnamefont {C.}~\bibnamefont
  {Donnerer}} \emph{et al.}, }
All-in{\textendash}all-Out Magnetic Order and Propagating Spin Waves in {Sm}$_2${Ir}$_2${O}$_7$, 
\href {\doibase 10.1103/physrevlett.117.037201}
  {\bibfield  {journal} {\bibinfo  {journal} {Phys. Rev. Lett.}\
  }\textbf {\bibinfo {volume} {117}},\ \bibinfo {pages} {037201} (\bibinfo
  {year} {2016})}


\bibitem {Taira2001}
  \bibfield  {author} {\bibinfo {author} {\bibfnamefont {N.}~\bibnamefont
  {Taira}}, \bibinfo {author} {\bibfnamefont {M.}~\bibnamefont {Wakeshima}} 
  and \bibinfo {author} {\bibfnamefont {Y.}~\bibnamefont {Hinatsu}}, }
Magnetic properties of iridium pyrochlores {R}$_2${Ir}$_2${O}$_7$ ({R} = {Y}, {Sm}, {Eu} and {Lu}), 
\href{\doibase 10.1088/0953-8984/13/23/312} {\bibfield  {journal} {\bibinfo
  {journal} {J. Phys.: Cond. Mat.}\ }\textbf {\bibinfo {volume}
  {13}},\ \bibinfo {pages} {5527} (\bibinfo {year} {2001})}


\bibitem {Wang2020a}
  \bibfield  {author} {\bibinfo {author} {\bibfnamefont {Y.}~\bibnamefont
  {Wang}}, \bibinfo {author} {\bibfnamefont {T.~F.}\ \bibnamefont {Rosenbaum}},
  \bibinfo {author} {\bibfnamefont {D.}~\bibnamefont {Prabhakaran}}, \bibinfo
  {author} {\bibfnamefont {A.~T.}\ \bibnamefont {Boothroyd}}  and \bibinfo
  {author} {\bibfnamefont {Y.}~\bibnamefont {Feng}}, }
Approaching the quantum critical point in a highly correlated all-in{\textendash}all-out antiferromagnet, 
\href {\doibase 10.1103/physrevb.101.220404} {\bibfield  {journal} {\bibinfo  {journal}
  {Phys. Rev. B}\ }\textbf {\bibinfo {volume} {101}},\ \bibinfo {pages}
  {220404(R)} (\bibinfo {year} {2020})}


\bibitem {Tomiyasu2012}
  \bibfield  {author} {\bibinfo {author} {\bibfnamefont {K.}~\bibnamefont
  {Tomiyasu}} \emph{et al.}, }
Emergence of Magnetic Long-range Order in Frustrated Pyrochlore {Nd}$_2${Ir}$_2${O}$_7$ with Metal{\textendash}Insulator Transition, \href {\doibase 10.1143/jpsj.81.034709}
  {\bibfield  {journal} {\bibinfo  {journal} {J. Phys. Soc. Japan}\ }\textbf {\bibinfo {volume} {81}},\ \bibinfo {pages} {034709}
  (\bibinfo {year} {2012})}


\bibitem {Chun2018}
  \bibfield  {author} {\bibinfo {author} {\bibfnamefont {S.~H.}\ \bibnamefont
  {Chun}} \emph{et al.},\ }
Magnetic Excitations across the Metal-Insulator Transition in the Pyrochlore Iridate {Eu}$_2${Ir}$_2${O}$_7$, 
\href {\doibase 10.1103/physrevlett.120.177203} {\bibfield  {journal} {\bibinfo
  {journal} {Phys. Rev. Lett.}\ }\textbf {\bibinfo {volume} {120}},\
  \bibinfo {pages} {177203} (\bibinfo {year} {2018})}
  

\bibitem {Jacobsen2020}
  \bibfield  {author} {\bibinfo {author} {\bibfnamefont {H.}\ \bibnamefont
  {Jacobsen}} \emph{et al.}, }
Strong quantum fluctuations from competition between magnetic phases in a pyrochlore iridate, 
\href {\doibase 10.1103/PhysRevB.101.104404} {\bibfield  {journal} {\bibinfo
  {journal} {Phys. Rev. B}\ }\textbf {\bibinfo {volume} {101}},\
  \bibinfo {pages} {104404} (\bibinfo {year} {2020})}


\bibitem {Pearce2022}
  \bibfield  {author} {\bibinfo {author} {\bibfnamefont {M.~J.}\ \bibnamefont
  {Pearce}} \emph{et al.}, }
Magnetic monopole density and antiferromagnetic domain control in spin-ice iridates, 
\href {\doibase 10.1038/s41467-022-27964-y} {\bibfield  {journal} {\bibinfo  {journal}
  {Nat. Commun.}\ }\textbf {\bibinfo {volume} {13}},\ \bibinfo {pages}
  {444} (\bibinfo {year} {2022})}


\bibitem {Lefrancois2017}
  \bibfield  {author} {\bibinfo {author} {\bibfnamefont {E.}~\bibnamefont
  {Lefran{\c{c}}ois}} \emph{et al.}, }
Fragmentation in spin ice from magnetic charge injection, 
\href {\doibase 10.1038/s41467-017-00277-1} {\bibfield
  {journal} {\bibinfo  {journal} {Nat. Commun.}\ }\textbf {\bibinfo
  {volume} {8}},\ \bibinfo {pages} {209} (\bibinfo {year} {2017})}


\bibitem {Cathelin2020}
  \bibfield  {author} {\bibinfo {author} {\bibfnamefont {V.}~\bibnamefont
  {Cathelin}} \emph{et al.}, }
Fragmented monopole crystal, dimer entropy, and {C}oulomb interactions in {Dy}$_2${Ir}$_2${O}$_7$, 
\href {\doibase 10.1103/physrevresearch.2.032073} {\bibfield  {journal} {\bibinfo
  {journal} {Phys. Rev. Res.}\ }\textbf {\bibinfo {volume} {2}},\
  \bibinfo {pages} {032073(R)} (\bibinfo {year} {2020})}


\bibitem {Castelnovo2012}
  \bibfield  {author} {\bibinfo {author} {\bibfnamefont {C.}~\bibnamefont
  {Castelnovo}}, \bibinfo {author} {\bibfnamefont {R.}~\bibnamefont
  {Moessner}} and \bibinfo {author} {\bibfnamefont {S.}~\bibnamefont
  {Sondhi}}, }
Spin Ice, Fractionalization, and Topological Order, 
\href {\doibase 10.1146/annurev-conmatphys-020911-125058}
  {\bibfield  {journal} {\bibinfo  {journal} {Ann. Rev. Cond. Mat. Physics}\ }\textbf {\bibinfo {volume} {3}},\ \bibinfo {pages} {35} (\bibinfo
  {year} {2012})}


\bibitem {BrooksBartlett2014}
  \bibfield  {author} {\bibinfo {author} {\bibfnamefont {M.~E.}\ \bibnamefont
  {Brooks-Bartlett}}, \bibinfo {author} {\bibfnamefont {S.~T.}\ \bibnamefont
  {Banks}}, \bibinfo {author} {\bibfnamefont {L.~D.~C.}\ \bibnamefont
  {Jaubert}}, \bibinfo {author} {\bibfnamefont {A.}~\bibnamefont
  {Harman-Clarke}} and \bibinfo {author} {\bibfnamefont {P.~C.~W.}\
  \bibnamefont {Holdsworth}}, }
Magnetic-Moment Fragmentation and Monopole Crystallization, 
\href {\doibase 10.1103/physrevx.4.011007}
  {\bibfield  {journal} {\bibinfo  {journal} {Phys. Rev. X}\ }\textbf
  {\bibinfo {volume} {4}},\ \bibinfo {pages} {011007} (\bibinfo {year}
  {2014})}


\bibitem {Ueda2015}
  \bibfield  {author} {\bibinfo {author} {\bibfnamefont {K.}~\bibnamefont
  {Ueda}}, \bibinfo {author} {\bibfnamefont {J.}~\bibnamefont {Fujioka}},
  \bibinfo {author} {\bibfnamefont {C.}~\bibnamefont {Terakura}} and
  \bibinfo {author} {\bibfnamefont {Y.}~\bibnamefont {Tokura}}, }
Pressure and magnetic field effects on metal-insulator transitions of bulk and domain wall states in pyrochlore iridates, 
\href {\doibase 10.1103/physrevb.92.121110} {\bibfield  {journal} {\bibinfo
  {journal} {Phys. Rev. B}\ }\textbf {\bibinfo {volume} {92}},\ \bibinfo
  {pages} {121110(R)} (\bibinfo {year} {2015}{\natexlab{a}})}


\bibitem {WitczakKrempa2012}
  \bibfield  {author} {\bibinfo {author} {\bibfnamefont {W.}~\bibnamefont
  {Witczak-Krempa}}\ and\ \bibinfo {author} {\bibfnamefont {Y.~B.}\
  \bibnamefont {Kim}}, }
Topological and magnetic phases of interacting electrons in the pyrochlore iridates, 
\href {\doibase 10.1103/physrevb.85.045124} {\bibfield
   {journal} {\bibinfo  {journal} {Phys. Rev. B}\ }\textbf {\bibinfo
  {volume} {85}},\ \bibinfo {pages} {045124} (\bibinfo {year}
  {2012})}


\bibitem {Kondo2015}
  \bibfield  {author} {\bibinfo {author} {\bibfnamefont {T.}~\bibnamefont
  {Kondo}} \emph{et al.}, }
Quadratic {F}ermi node in a {3D} strongly correlated semimetal, 
\href {\doibase 10.1038/ncomms10042} {\bibfield  {journal} {\bibinfo  {journal}
  {Nat. Commun.}\ }\textbf {\bibinfo {volume} {6}},\ \bibinfo {pages}
  {10042} (\bibinfo {year} {2015})}


\bibitem {Nakayama2016}
  \bibfield  {author} {\bibinfo {author} {\bibfnamefont {M.}~\bibnamefont
  {Nakayama}} \emph{et al.}, }
Slater to {M}ott Crossover in the Metal to Insulator Transition of {Nd}$_2${Ir}$_2${O}$_7$,
\href {\doibase 10.1103/physrevlett.117.056403} {\bibfield  {journal} {\bibinfo
  {journal} {Phys. Rev. Lett.}\ }\textbf {\bibinfo {volume} {117}},\
  \bibinfo {pages} {056403} (\bibinfo {year} {2016})}


\bibitem {Ueda2017}
  \bibfield  {author} {\bibinfo {author} {\bibfnamefont {K.}~\bibnamefont
  {Ueda}} \emph{et al.}, }
Magnetic-field induced multiple topological phases in pyrochlore iridates with {M}ott criticality, 
\href {\doibase 10.1038/ncomms15515}
  {\bibfield  {journal} {\bibinfo  {journal} {Nat. Commun.}\ }\textbf
  {\bibinfo {volume} {8}},\ \bibinfo {pages} {15515} (\bibinfo {year}
  {2017})}


\bibitem {Sakata2011}
  \bibfield  {author} {\bibinfo {author} {\bibfnamefont {M.}\ \bibnamefont
  {Sakata}} \emph{et al.}, }
Suppression of metal-insulator transition at high pressure and pressure-induced magnetic ordering in pyrochlore oxide {Nd}$_2${Ir}$_2${O}$_7$, 
\href {\doibase 10.1103/physrevb.83.041102} {\bibfield
  {journal} {\bibinfo  {journal} {Phys. Rev. B}\ }\textbf {\bibinfo
  {volume} {83}},\ \bibinfo {pages} {041102} (\bibinfo {year}
  {2011})}
  

\bibitem {Tian2016}
  \bibfield  {author} {\bibinfo {author} {\bibfnamefont {Z.}~\bibnamefont
  {Tian}} \emph{et al.}, }
Field-induced quantum metal{\textendash}insulator transition in the pyrochlore iridate {Nd}$_2${Ir}$_2${O}$_7$, 
\href {\doibase 10.1038/nphys3567}
  {\bibfield  {journal} {\bibinfo  {journal} {Nat. Phys.}\ }\textbf
  {\bibinfo {volume} {12}},\ \bibinfo {pages} {134} (\bibinfo {year}
  {2016})}


  \bibitem {Ma2015}
  \bibfield  {author} {\bibinfo {author} {\bibfnamefont {E.~Y.}~\bibnamefont
  {Ma}} \emph{et al.}, }
Mobile metallic domain walls in an all-in-all-out magnetic insulator, 
\href {\doibase 10.1126/science.aac8289}
  {\bibfield  {journal} {\bibinfo  {journal} {Science}\ }\textbf
  {\bibinfo {volume} {350}},\ \bibinfo {pages} {538} (\bibinfo {year}
  {2015})}

  
  \bibitem {Ueda2018}
  \bibfield  {author} {\bibinfo {author} {\bibfnamefont {K.}~\bibnamefont
  {Ueda}}, \bibinfo {author} {\bibfnamefont {R.}~\bibnamefont {Kaneko}},
  \bibinfo {author} {\bibfnamefont {H.}~\bibnamefont {Ishizuka}}, \bibinfo
  {author} {\bibfnamefont {J.}~\bibnamefont {Fujioka}}, \bibinfo {author}
  {\bibfnamefont {N.}\ \bibnamefont {Nagaosa}} \ and\ \bibinfo {author} {\bibfnamefont
  {Y.}~\bibnamefont {Tokura}}, }
Spontaneous {H}all effect in the {W}eyl semimetal candidate of all-in all-out pyrochlore iridate, 
\href {\doibase 10.1038/s41467-018-05530-9}
  {\bibfield  {journal} {\bibinfo  {journal} {Nat. Commun.}\ }\textbf
  {\bibinfo {volume} {9}},\ \bibinfo {pages} {3032} (\bibinfo {year}
  {2018})}


\bibitem{Nakatsuji2006}
  \bibfield  {author} {\bibinfo {author} {\bibfnamefont {S.}~\bibnamefont
  {Nakatsuji}} \emph{et al.}, }
Metallic Spin-Liquid Behavior of the Geometrically Frustrated {K}ondo Lattice {Pr}$_2${Ir}$_2${O}$_7$, 
\href {\doibase 10.1103/physrevlett.96.087204}
  {\bibfield  {journal} {\bibinfo  {journal} {Phys. Rev. Lett.}\
  }\textbf {\bibinfo {volume} {96}},\ \bibinfo {pages} {087204} (\bibinfo
  {year} {2006})}


\bibitem {Ni2021a}
  \bibfield  {author} {\bibinfo {author} {\bibfnamefont {J.M.}~\bibnamefont
  {Ni}}, \bibinfo {author} {\bibfnamefont {Y.Y.}\ \bibnamefont
  {Huang}}, \bibinfo {author} {\bibfnamefont {E.J.}~\bibnamefont {Cheng}}, \bibinfo {author} {\bibfnamefont {Y.J.}~\bibnamefont {Yu}}, \bibinfo {author} {\bibfnamefont {B.L.}~\bibnamefont {Pan}}, \bibinfo {author} {\bibfnamefont {Q.}~\bibnamefont {Li}}, \bibinfo {author} {\bibfnamefont {L.M.}~\bibnamefont {Xu}}, \bibinfo {author} {\bibfnamefont {Z.M.}~\bibnamefont {Tian}},
  \ and\ \bibinfo {author} {\bibfnamefont {S.Y.}~\bibnamefont {Li}}, }
Giant isotropic magneto-thermal conductivity of metallic spin liquid candidate {Pr}$_2${Ir}$_2${O}$_7$ with quantum criticality, 
\href {\doibase 10.1038/s41467-020-20562-w} {\bibfield  {journal} {\bibinfo  {journal}
  {Nat. Commun.}\ }\textbf {\bibinfo {volume} {12}},\ \bibinfo {pages}
  {307 } (\bibinfo {year} {2021})}


\bibitem {Savary2014}
  \bibfield  {author} {\bibinfo {author} {\bibfnamefont {L.}~\bibnamefont
  {Savary}}, \bibinfo {author} {\bibfnamefont {E.-G.}\ \bibnamefont {Moon}}, \
  and\ \bibinfo {author} {\bibfnamefont {L.}~\bibnamefont {Balents}}, }
New Type of Quantum Criticality in the Pyrochlore Iridates, 
\href {\doibase 10.1103/physrevx.4.041027} {\bibfield  {journal} {\bibinfo
  {journal} {Phys. Rev. X}\ }\textbf {\bibinfo {volume} {4}},\ \bibinfo
  {pages} {041027} (\bibinfo {year} {2014})}


\bibitem {Matsuhira2011}
  \bibfield  {author} {\bibinfo {author} {\bibfnamefont {K.}~\bibnamefont
  {Matsuhira}}, \bibinfo {author} {\bibfnamefont {M.}~\bibnamefont
  {Wakeshima}}, \bibinfo {author} {\bibfnamefont {Y.}~\bibnamefont {Hinatsu}} and \bibinfo {author} {\bibfnamefont {S.}~\bibnamefont {Takagi}}, }
Metal{\textendash}Insulator Transitions in Pyrochlore Oxides $Ln$$_2${Ir}$_2${O}$_7$, 
\href {\doibase 10.1143/jpsj.80.094701} {\bibfield  {journal} {\bibinfo  {journal}
  {J. Phys. Soc. Jpn.}\ }\textbf {\bibinfo {volume} {80}},\ \bibinfo {pages}
  {094701} (\bibinfo {year} {2011})}


\bibitem {Zhang2017}
  \bibfield  {author} {\bibinfo {author} {\bibfnamefont {H.}~\bibnamefont
  {Zhang}}, \bibinfo {author} {\bibfnamefont {K.}~\bibnamefont {Haule}} and
  \bibinfo {author} {\bibfnamefont {D.}~\bibnamefont {Vanderbilt}}, }
Metal-Insulator Transition and Topological Properties of Pyrochlore Iridates, 
\href {\doibase 10.1103/physrevlett.118.026404} {\bibfield  {journal} {\bibinfo
  {journal} {Phys. Rev. Lett.}\ }\textbf {\bibinfo {volume} {118}},\
  \bibinfo {pages} {026404} (\bibinfo {year} {2017})}


\bibitem {Liu2018b}
  \bibfield  {author} {\bibinfo {author} {\bibfnamefont {W.}~\bibnamefont
  {Liu}}, \bibinfo {author} {\bibfnamefont {H.}~\bibnamefont {Han}}, \bibinfo
  {author} {\bibfnamefont {L.}~\bibnamefont {Ma}}, \bibinfo {author}
  {\bibfnamefont {L.}~\bibnamefont {Pi}}, \bibinfo {author} {\bibfnamefont
  {L.}~\bibnamefont {Zhang}} and \bibinfo {author} {\bibfnamefont
  {Y.}~\bibnamefont {Zhang}}, }
Different pressure effects in {A}$_2${Ir}$_2${O}$_7$ ({A} = {Gd, Eu, and Sm}), 
\href {\doibase 10.1016/j.jallcom.2018.01.077}
  {\bibfield  {journal} {\bibinfo  {journal} {J. Alloys and Compounds}\
  }\textbf {\bibinfo {volume} {741}},\ \bibinfo {pages} {182} (\bibinfo {year}
  {2018})}


\bibitem {Tokiwa2014}
  \bibfield  {author} {\bibinfo {author} {\bibfnamefont {Y.}~\bibnamefont
  {Tokiwa}}, \bibinfo {author} {\bibfnamefont {J.~J.}\ \bibnamefont
  {Ishikawa}}, \bibinfo {author} {\bibfnamefont {S.}~\bibnamefont {Nakatsuji}} and \bibinfo {author} {\bibfnamefont {P.}~\bibnamefont {Gegenwart}}, }
Quantum criticality in a metallic spin liquid, 
\href {\doibase 10.1038/nmat3900} {\bibfield  {journal} {\bibinfo  {journal}
  {Nat. Mat.s}\ }\textbf {\bibinfo {volume} {13}},\ \bibinfo {pages}
  {356} (\bibinfo {year} {2014})}


\bibitem {WitczakKrempa2014}
  \bibfield  {author} {\bibinfo {author} {\bibfnamefont {W.}~\bibnamefont
  {Witczak-Krempa}}, \bibinfo {author} {\bibfnamefont {G.}~\bibnamefont
  {Chen}}, \bibinfo {author} {\bibfnamefont {Y.~B.}\ \bibnamefont {Kim}}
 and \bibinfo {author} {\bibfnamefont {L.}~\bibnamefont {Balents}}, }
Correlated Quantum Phenomena in the Strong Spin-Orbit Regime, 
\href {\doibase 10.1146/annurev-conmatphys-020911-125138} {\bibfield  {journal}
  {\bibinfo  {journal} {Ann. Rev. Cond. Mat. Physics}\ }\textbf
  {\bibinfo {volume} {5}},\ \bibinfo {pages} {57} (\bibinfo {year}
  {2014})}


\bibitem {Ueda2015a}
  \bibfield  {author} {\bibinfo {author} {\bibfnamefont {K.}~\bibnamefont
  {Ueda}} \emph{et al.}, }
Magnetic Field-Induced Insulator-Semimetal Transition in a Pyrochlore {Nd}$_2${Ir}$_2${O}$_7$, 
\href {\doibase 10.1103/physrevlett.115.056402} {\bibfield  {journal} {\bibinfo  {journal}
  {Phys. Rev. Lett.}\ }\textbf {\bibinfo {volume} {115}},\ \bibinfo
  {pages} {056402} (\bibinfo {year} {2015}{\natexlab{b}})}


\bibitem {Tafti2012a}
  \bibfield  {author} {\bibinfo {author} {\bibfnamefont {F.~F.}\ \bibnamefont
  {Tafti}}, \bibinfo {author} {\bibfnamefont {J.~J.}\ \bibnamefont {Ishikawa}},
  \bibinfo {author} {\bibfnamefont {A.}~\bibnamefont {McCollam}}, \bibinfo
  {author} {\bibfnamefont {S.}~\bibnamefont {Nakatsuji}} and \bibinfo
  {author} {\bibfnamefont {S.~R.}\ \bibnamefont {Julian}}, }
Pressure-tuned insulator to metal transition in {Eu}$_2${Ir}$_2${O}$_7$, 
\href {\doibase 10.1103/PhysRevB.85.205104} {\bibfield  {journal} {\bibinfo  {journal}
  {Phys. Rev. B}\ }\textbf {\bibinfo {volume} {85}},\ \bibinfo {pages}
  {205104} (\bibinfo {year} {2012})}


\bibitem {Ishikawa2012}
  \bibfield  {author} {\bibinfo {author} {\bibfnamefont {J.~J.}\ \bibnamefont
  {Ishikawa}}, \bibinfo {author} {\bibfnamefont {E.~C.~T.}\ \bibnamefont
  {O'Farrell}} and \bibinfo {author} {\bibfnamefont {S.}~\bibnamefont
  {Nakatsuji}}, }
Continuous transition between antiferromagnetic insulator and paramagnetic metal in the pyrochlore iridate {Eu}$_2${Ir}$_2${O}$_7$, 
\href {\doibase 10.1103/physrevb.85.245109} {\bibfield
  {journal} {\bibinfo  {journal} {Phys. Rev. B}\ }\textbf {\bibinfo
  {volume} {85}},\ \bibinfo {pages} {245109} (\bibinfo {year}
  {2012})}
  

\bibitem {Asih2017}
  \bibfield  {author} {\bibinfo {author} {\bibfnamefont {R.}~\bibnamefont
  {Asih}} \emph{et al.}, }
Magnetic Moments and Ordered States in Pyrochlore Iridates {Nd}$_2${Ir}$_2${O}$_7$ and {Sm}$_2${Ir}$_2${O}$_7$ Studied by Muon-Spin Relaxation, 
\href {\doibase 10.7566/jpsj.86.024705}
  {\bibfield  {journal} {\bibinfo  {journal} {J. Phys. Soc. Jpn.}\ }\textbf {\bibinfo {volume} {86}},\ \bibinfo {pages} {024705}
  (\bibinfo {year} {2017})}


\bibitem {Wan2011}
  \bibfield  {author} {\bibinfo {author} {\bibfnamefont {X.}~\bibnamefont
  {Wan}}, \bibinfo {author} {\bibfnamefont {A.~M.}\ \bibnamefont {Turner}},
  \bibinfo {author} {\bibfnamefont {A.}~\bibnamefont {Vishwanath}} and
  \bibinfo {author} {\bibfnamefont {S.~Y.}\ \bibnamefont {Savrasov}}, }
Topological semimetal and {F}ermi-arc surface states in the electronic structure of pyrochlore iridates, 
\href {\doibase 10.1103/physrevb.83.205101} {\bibfield  {journal} {\bibinfo
  {journal} {Phys. Rev. B}\ }\textbf {\bibinfo {volume} {83}},\ \bibinfo
  {pages} {205101} (\bibinfo {year} {2011})}


\bibitem {Ishii2015}
  \bibfield  {author} {\bibinfo {author} {\bibfnamefont {F.}~\bibnamefont
  {Ishii}}, \bibinfo {author} {\bibfnamefont {Y.~P.}\ \bibnamefont {Mizuta}},
  \bibinfo {author} {\bibfnamefont {T.}~\bibnamefont {Kato}}, \bibinfo {author}
  {\bibfnamefont {T.}~\bibnamefont {Ozaki}}, \bibinfo {author} {\bibfnamefont
  {H.}~\bibnamefont {Weng}} and \bibinfo {author} {\bibfnamefont
  {S.}~\bibnamefont {Onoda}}, }
First-Principles Study on Cubic Pyrochlore Iridates {Y}$_2${Ir}$_2${O}$_7$ and {Pr}$_2${Ir}$_2${O}$_7$, 
\href {\doibase 10.7566/jpsj.84.073703}
  {\bibfield  {journal} {\bibinfo  {journal} {J. Phys. Soc. Jpn.}\ }\textbf {\bibinfo {volume} {84}},\ \bibinfo {pages} {073703}
  (\bibinfo {year} {2015})}


\bibitem {Goddard2008}
  \bibfield  {author} {\bibinfo {author} {\bibfnamefont {P.~A.}\ \bibnamefont
  {Goddard}} \emph{et al.}, }
Experimentally determining the exchange parameters of quasi-two-dimensional {H}eisenberg magnets, 
\href {\doibase 10.1088/1367-2630/10/8/083025}
  {\bibfield  {journal} {\bibinfo  {journal} {New J. Phys.}\ }\textbf
  {\bibinfo {volume} {10}},\ \bibinfo {pages} {083025} (\bibinfo {year}
  {2008})}


\bibitem {SPECTRE}
  \bibinfo {author} {A. T. Boothroyd}, SPECTRE: A
  Program for Calculating Spectroscopic Properties of Rare Earth Ions in
  Crystals (1990--2014)

  
\bibitem {PecanhaAntonio2019}
  \bibfield  {author} {\bibinfo {author} {\bibfnamefont {V.}~\bibnamefont
  {Pe{\c{c}}anha-Antonio}} \emph{et al.}, }
Intermultiplet transitions and magnetic long-range order in {Sm}-based pyrochlores, 
\href {\doibase 10.1103/physrevb.99.134415} {\bibfield  {journal} {\bibinfo  {journal}
  {Phys. Rev. B}\ }\textbf {\bibinfo {volume} {99}},\ \bibinfo {pages}
  {134415} (\bibinfo {year} {2019})}


\bibitem {Zhu2014}
  \bibfield  {author} {\bibinfo {author} {\bibfnamefont {W.~K.}\ \bibnamefont
  {Zhu}}, \bibinfo {author} {\bibfnamefont {M.}~\bibnamefont {Wang}}, \bibinfo
  {author} {\bibfnamefont {B.}~\bibnamefont {Seradjeh}}, \bibinfo {author}
  {\bibfnamefont {F.}~\bibnamefont {Yang}} and \bibinfo {author}
  {\bibfnamefont {S.~X.}\ \bibnamefont {Zhang}}, }
Enhanced weak ferromagnetism and conductivity in hole-doped pyrochlore iridate {Y}$_2${Ir}$_2${O}$_7$, 
\href {\doibase 10.1103/physrevb.90.054419} {\bibfield  {journal} {\bibinfo  {journal}
  {Phys. Rev. B}\ }\textbf {\bibinfo {volume} {90}},\ \bibinfo {pages}
  {054419} (\bibinfo {year} {2014})}


\bibitem {Graf2014}
  \bibfield  {author} {\bibinfo {author} {\bibfnamefont {M.~J.}\ \bibnamefont
  {Graf}} \emph{et al.}, }
Magnetism and magnetic order in the pyrochlore iridates in the insulator-to-metal crossover region, 
\href {\doibase 10.1088/1742-6596/551/1/012020} {\bibfield
  {journal} {\bibinfo  {journal} {J. Phys.: Conference Series}\
  }\textbf {\bibinfo {volume} {551}},\ \bibinfo {pages} {012020} (\bibinfo
  {year} {2014})}


\bibitem {Telang2019}
  \bibfield  {author} {\bibinfo {author} {\bibfnamefont {P.}~\bibnamefont
  {Telang}}, \bibinfo {author} {\bibfnamefont {K.}~\bibnamefont {Mishra}},
  \bibinfo {author} {\bibfnamefont {G.}~\bibnamefont {Prando}}, \bibinfo
  {author} {\bibfnamefont {A.K.}~\bibnamefont {Sood}} and \bibinfo {author}
  {\bibfnamefont {S.}~\bibnamefont {Singh}},\ }
Anomalous lattice contraction and emergent electronic phases in {Bi}-doped {Eu}$_2${Ir}$_2${O}$_7$, 
\href {\doibase 10.1103/physrevb.99.201112} {\bibfield  {journal} {\bibinfo  {journal}
  {Phys. Rev. B}\ }\textbf {\bibinfo {volume} {99}},\ \bibinfo {pages}
  {201112(R)} (\bibinfo {year} {2019})}


\bibitem {LaBarre2019}
  \bibfield  {author} {\bibinfo {author} {\bibfnamefont {P.~G.}\ \bibnamefont
  {LaBarre}}, \bibinfo {author} {\bibfnamefont {L.}~\bibnamefont {Dong}},
  \bibinfo {author} {\bibfnamefont {J.}~\bibnamefont {Trinh}}, \bibinfo
  {author} {\bibfnamefont {T.}~\bibnamefont {Siegrist}} and \bibinfo
  {author} {\bibfnamefont {A.~P.}\ \bibnamefont {Ramirez}}, }
Evidence for undoped {W}eyl semimetal charge transport in {Y}$_2${Ir}$_2${O}$_7$, 
\href {\doibase 10.1088/1361-648x/ab4aaf} {\bibfield  {journal} {\bibinfo  {journal} {J. Phys.: Cond. Mat.}\ }\textbf {\bibinfo {volume} {32}},\ \bibinfo {pages} {02LT01} (\bibinfo {year} {2019})}


\bibitem {Liu2020}
  \bibfield  {author} {\bibinfo {author} {\bibfnamefont {X.}~\bibnamefont
  {Liu}} \emph{et al.}, }
In-situ fabrication and transport properties of (111) {Y}$_2${Ir}$_2${O}$_7$ epitaxial thin film, 
\href {\doibase 10.1063/5.0019876} {\bibfield  {journal} {\bibinfo  {journal} {App. Phys. Lett.}\ }\textbf {\bibinfo {volume} {117}},\ \bibinfo {pages} {041903} (\bibinfo {year} {2020})}


\bibitem {Han2015}
  \bibfield  {author} {\bibinfo {author} {\bibfnamefont {H.}~\bibnamefont
  {Han}} \emph{et al.}, }
Electron paramagnetic resonance study of the $f${\textendash}$d$ interaction in pyrochlore iridate {Gd}$_2${Ir}$_2${O}$_7$, 
\href {\doibase 10.1080/14786435.2015.1086033}
  {\bibfield  {journal} {\bibinfo  {journal} {Philosophical Magazine}\ }\textbf
  {\bibinfo {volume} {95}},\ \bibinfo {pages} {3014} (\bibinfo {year}
  {2015})}

\bibitem {Telang2021}
  \bibfield  {author} {\bibinfo {author} {\bibfnamefont {P.}~\bibnamefont
  {Telang}}\ and\ \bibinfo {author} {\bibfnamefont {S.}~\bibnamefont {Singh}}, }
Protracting the {W}eyl phase by a giant negative lattice expansion in {Bi} doped {Sm}$_2${Ir}$_2${O}$_7$, 
\href@noop {} {\bibfield  {journal} {\bibinfo  {journal} {ArXiv}
  \bibinfo {pages} {2106.03512}} (\bibinfo {year} {2021})},\ \Eprint
  {http://arxiv.org/abs/2106.03512v1} {2106.03512v1} 

\bibitem {Mott1990}
  \bibfield  {author} {\bibinfo {author} {\bibfnamefont {N.}~\bibnamefont
  {Mott}}, }
On metal-insulator transitions, 
\href {\doibase 10.1016/0022-4596(90)90201-8} {\bibfield
  {journal} {\bibinfo  {journal} {J. Solid State Chemistry}\ }\textbf
  {\bibinfo {volume} {88}},\ \bibinfo {pages} {5} (\bibinfo {year}
  {1990})}


\bibitem {Mott1969}
  \bibfield  {author} {\bibinfo {author} {\bibfnamefont {N.}~\bibnamefont
  {Mott}}, }
Charge transport in non-crystalline semiconductors, 
\ \href {\doibase 10.1016/b978-0-08-015543-2.50005-x} {\emph
  {\bibinfo {booktitle} {Festk\"{o}rper Probleme {IX}}}}\ (\bibinfo
  {publisher} {Elsevier},\ \bibinfo {year} {1969})\ pp.\ \bibinfo {pages}
  {22--45}


\bibitem {Hill1976}
  \bibfield  {author} {\bibinfo {author} {\bibfnamefont {R.}~\bibnamefont
  {Hill}}, }
Variable-range hopping, 
\href {\doibase 10.1002/pssa.2210340223} {\bibfield  {journal}
  {\bibinfo  {journal} {Physica Status Solidi (a)}\ }\textbf {\bibinfo {volume}
  {34}},\ \bibinfo {pages} {601} (\bibinfo {year} {1976})}


\bibitem {Keuls1997}
  \bibfield  {author} {\bibinfo {author} {\bibfnamefont {F. W.}\ \bibnamefont
  {Van Keuls}}, \bibinfo {author} {\bibfnamefont {X. L.}~\bibnamefont {Hu}}, \bibinfo
  {author} {\bibfnamefont {H. W.}~\bibnamefont {Jiang}} and \bibinfo {author}
  {\bibfnamefont {A. J.}~\bibnamefont {Dahm}}, }
Screening of the {C}oulomb interaction in two-dimensional variable-range hopping, 
\href {\doibase 10.1103/physrevb.56.1161} {\bibfield  {journal} {\bibinfo  {journal}
  {Phys. Rev. B}\ }\textbf {\bibinfo {volume} {56}},\ \bibinfo {pages}
  {1161} (\bibinfo {year} {1997})}


\bibitem {Polyakov1993}
  \bibfield  {author} {\bibinfo {author} {\bibfnamefont {D.~G.}\ \bibnamefont
  {Polyakov}} and \bibinfo {author} {\bibfnamefont {B.~I.}\ \bibnamefont
  {Shklovskii}}, }
Conductivity-peak broadening in the quantum Hall regime, 
\href {\doibase 10.1103/physrevb.48.11167} {\bibfield
  {journal} {\bibinfo  {journal} {Phys. Rev. B}\ }\textbf {\bibinfo
  {volume} {48}},\ \bibinfo {pages} {11167} (\bibinfo {year}
  {1993})}


\bibitem {Hosur2012}
  \bibfield  {author} {\bibinfo {author} {\bibfnamefont {P.}~\bibnamefont
  {Hosur}}, \bibinfo {author} {\bibfnamefont {S.~A.}\ \bibnamefont
  {Parameswaran}} and \bibinfo {author} {\bibfnamefont {A.}~\bibnamefont
  {Vishwanath}}, }
Charge Transport in {W}eyl Semimetals, 
\href {\doibase 10.1103/physrevlett.108.046602} {\bibfield
  {journal} {\bibinfo  {journal} {Phys. Rev. Lett.}\ }\textbf {\bibinfo
  {volume} {108}},\ \bibinfo {pages} {046602} (\bibinfo {year}
  {2012})}

\bibitem {Fujita2015}
  \bibfield  {author} {\bibinfo {author} {\bibfnamefont {T.C.}~\bibnamefont
  {Fujita}}, \bibinfo {author} {\bibfnamefont {Y.}~\bibnamefont {Kozuka}}, \bibinfo {author} {\bibfnamefont {M.}~\bibnamefont {Uchida}}, \bibinfo {author} {\bibfnamefont {A.}~\bibnamefont {Tsukazaki}}, \bibinfo {author} {\bibfnamefont {T.}~\bibnamefont {Arima}} and
  \bibinfo {author} {\bibfnamefont {M.}~\bibnamefont {Kawasaki}}, }
Odd-parity magnetoresistance in pyrochlore iridate thin films with broken time-reversal symmetry, 
\href {\doibase 10.1038/srep09711} {\bibfield  {journal} {\bibinfo  {journal} {Scientific Reports}\ }\textbf {\bibinfo {volume} {5}},\ \bibinfo
  {pages} {9711} (\bibinfo {year} {2015})}


\bibitem {Xu2020b}
  \bibfield  {author} {\bibinfo {author} {\bibfnamefont {L.}~\bibnamefont
  {Xu}}, \bibinfo {author} {\bibfnamefont {G.}~\bibnamefont {Gong}}, \bibinfo {author} {\bibfnamefont {C.}~\bibnamefont {Zhao}}, \bibinfo {author} {\bibfnamefont {X.}~\bibnamefont {Song}}, \bibinfo {author} {\bibfnamefont {S.}~\bibnamefont {Yuan}} and
  \bibinfo {author} {\bibfnamefont {Z.}~\bibnamefont {Tian}}, }
Asymmetric Magnetization Reversal Behaviors Driven by Exchange Coupling between All-in-All-out Magnetic Domains and Domain Walls in a {Eu}$_2${Ir}$_2${O}$_7$ Single Crystal, 
\href {\doibase 10.1021/acs.jpcc.0c06020} {\bibfield  {journal} {\bibinfo  {journal} {J. Phys. Chem. C}\ }\textbf {\bibinfo {volume} {124}},\ \bibinfo {pages} {22656} (\bibinfo {year} {2020})}


\bibitem {Yamaji2016}
  \bibfield  {author} {\bibinfo {author} {\bibfnamefont {Y.}~\bibnamefont
  {Yamaji}}\ and\ \bibinfo {author} {\bibfnamefont {M.}~\bibnamefont {Imada}}, }
Modulated helical metals at magnetic domain walls of pyrochlore iridium oxides, 
\href {\doibase 10.1103/physrevb.93.195146} {\bibfield  {journal} {\bibinfo
  {journal} {Phys. Rev. B} }\textbf {\bibinfo {volume} {93}},\ \bibinfo
  {pages} {195146} (\bibinfo {year} {2016})}


\bibitem {Ueda2016}
  \bibfield  {author} {\bibinfo {author} {\bibfnamefont {K.}~\bibnamefont
  {Ueda}}, \bibinfo {author} {\bibfnamefont {J.}~\bibnamefont {Fujioka}}
  and \bibinfo {author} {\bibfnamefont {Y.}~\bibnamefont {Tokura}}, }
Variation of optical conductivity spectra in the course of bandwidth-controlled metal-insulator transitions in pyrochlore iridates, 
\href {\doibase 10.1103/physrevb.93.245120} {\bibfield  {journal} {\bibinfo
  {journal} {Phys. Rev. B}\ }\textbf {\bibinfo {volume} {93}},\ \bibinfo
  {pages} {245120} (\bibinfo {year} {2016})}


\bibitem {Ueda2020}
  \bibfield  {author} {\bibinfo {author} {\bibfnamefont {K.}~\bibnamefont
  {Ueda}}, \bibinfo {author} {\bibfnamefont {H.}~\bibnamefont {Fukuda}},
  \bibinfo {author} {\bibfnamefont {R.}~\bibnamefont {Kaneko}}, \bibinfo
  {author} {\bibfnamefont {J.}~\bibnamefont {Fujioka}} and \bibinfo
  {author} {\bibfnamefont {Y.}~\bibnamefont {Tokura}}, }
Evolution of possible {W}eyl semimetal states across the Mott transition in pyrochlore iridates induced by hole doping, 
\href {\doibase 10.1103/physrevb.102.245131} {\bibfield  {journal} {\bibinfo {journal}
  {Phys. Rev. B}\ }\textbf {\bibinfo {volume} {102}},\ \bibinfo {pages}
  {245131} (\bibinfo {year} {2020})}


\bibitem {Ueda2022a}
  \bibfield  {author} {\bibinfo {author} {\bibfnamefont {K.}~\bibnamefont
  {Ueda}} \emph{et al.}, }
Experimental signatures of a versatile {W}eyl semimetal in a pyrochlore iridate with spin-ice-like magnetic orders, 
\href {\doibase 10.1029/jb091ib05p04673} {\bibfield  {journal} {\bibinfo  {journal} {Phys. Rev. B}\ }\textbf {\bibinfo {volume} {105}},\ \bibinfo {pages} {L161102} (\bibinfo {year} {2022})}


\bibitem {Ueda2014}
  \bibfield  {author} {\bibinfo {author} {\bibfnamefont {K.}~\bibnamefont
  {Ueda}} \emph{et al.}, }
Anomalous domain-wall conductance in pyrochlore-type {Nd}$_2${Ir}$_2${O}$_7$ on the verge of the metal-insulator transition, 
\href {\doibase 10.1103/physrevb.89.075127} {\bibfield  {journal} {\bibinfo  {journal} {Phys. Rev. B}\ }\textbf {\bibinfo {volume} {89}},\ \bibinfo {pages} {075127} (\bibinfo {year} {2014})}


\bibitem {Hiroi2015}
  \bibfield  {author} {\bibinfo {author} {\bibfnamefont {Z.}~\bibnamefont
  {Hiroi}},  \bibinfo {author} {\bibfnamefont {J.}~\bibnamefont {Yamaura}}, \bibinfo {author} {\bibfnamefont {T.}~\bibnamefont {Hirose}}, \bibinfo {author} {\bibfnamefont {I.}~\bibnamefont {Nagashima}}\ and\
  \bibinfo {author} {\bibfnamefont {Y.}~\bibnamefont {Okamoto}}, }
Lifshitz metal{\textendash}insulator transition induced by the all-in/all-out magnetic order in the pyrochlore oxide {Cd}$_2${Os}$_2${O}$_7$, 
\href {\doibase 10.1063/1.4907734} {\bibfield  {journal} {\bibinfo  {journal} {{APL} Materials}\ }\textbf {\bibinfo {volume} {3}},\ \bibinfo
  {pages} {041501} (\bibinfo {year} {2015})}


\bibitem {Shinaoka2012}
  \bibfield  {author} {\bibinfo {author} {\bibfnamefont {H.}~\bibnamefont
  {Shinaoka}},  \bibinfo {author} {\bibfnamefont {T.}~\bibnamefont {Miyake}}\ and\
  \bibinfo {author} {\bibfnamefont {S.}~\bibnamefont {Ishibashi}}, }
Noncollinear Magnetism and Spin-Orbit Coupling in Pyrochlore Oxide {Cd}$_2${Os}$_2${O}$_7$, 
\href {\doibase 10.1103/PhysRevLett.108.247204} {\bibfield  {journal} {\bibinfo  {journal} {Phys. Rev. Lett.}\ }\textbf {\bibinfo {volume} {108}},\ \bibinfo {pages} {247204} (\bibinfo {year} {2012})}


\bibitem {Millican2007}
  \bibfield  {author} {\bibinfo {author} {\bibfnamefont {J.~N.}\ \bibnamefont
  {Millican}}, \bibinfo {author} {\bibfnamefont {R.~T.}\ \bibnamefont
  {Macaluso}}, \bibinfo {author} {\bibfnamefont {S.}~\bibnamefont {Nakatsuji}},
  \bibinfo {author} {\bibfnamefont {Y.}~\bibnamefont {Machida}}, \bibinfo
  {author} {\bibfnamefont {Y.}~\bibnamefont {Maeno}} and \bibinfo {author}
  {\bibfnamefont {J.~Y.}\ \bibnamefont {Chan}}, }
Crystal growth and structure of $R$$_2${Ir}$_2${O}$_7$ ($R$={Pr, Eu}) using molten {KF}, 
\href {\doibase 10.1016/j.materresbull.2006.08.011} {\bibfield  {journal} {\bibinfo
  {journal} {Materials Research Bulletin}\ }\textbf {\bibinfo {volume} {42}},\
  \bibinfo {pages} {928} (\bibinfo {year} {2007})}


\bibitem {Dunstan1989}
  \bibfield  {author} {\bibinfo {author} {\bibfnamefont {D.}~\bibnamefont
  {Dunstan}} and \bibinfo {author} {\bibfnamefont {I.}~\bibnamefont
  {Spain}}, }
Technology of diamond anvil high-pressure cells: I. Principles, design and construction, 
\href {\doibase 10.1088/0022-3735/22/11/004} {\bibfield {journal} {\bibinfo  {journal} {J. Phys. E: Sci. Instrum.}\ }\textbf
  {\bibinfo {volume} {22}}, \bibinfo {pages} {913} (\bibinfo {year} {1989})}


\bibitem {Dunstan1989a}
  \bibfield  {author} {\bibinfo {author} {\bibfnamefont {I.}~\bibnamefont
  {Spain}} and \bibinfo {author} {\bibfnamefont {D.}~\bibnamefont
  {Dunstan}}, }
The technology of diamond anvil high-pressure cells: {II}. Operation and use, 
\href {\doibase 10.1088/0022-3735/22/11/005} {\bibfield
  {journal} {\bibinfo  {journal} {J. Phys. E: Sci. Instrum.}\ }\textbf
  {\bibinfo {volume} {22}},\ \bibinfo {pages} {923} (\bibinfo {year}
  {1989})}


\bibitem {Tateiwa2009}
  \bibfield  {author} {\bibinfo {author} {\bibfnamefont {N.}~\bibnamefont
  {Tateiwa}}\ and\ \bibinfo {author} {\bibfnamefont {Y.}~\bibnamefont {Haga}}, }
Evaluations of pressure-transmitting media for cryogenic experiments with diamond anvil cell, 
\href {\doibase 10.1063/1.3265992} {\bibfield  {journal} {\bibinfo
  {journal} {Rev. Sci. Instr.}\ }\textbf {\bibinfo {volume} {80}},\ \bibinfo
  {pages} {123901} (\bibinfo {year} {2009})}


\bibitem {Mao1986}
  \bibfield  {author} {\bibinfo {author} {\bibfnamefont {H.}~\bibnamefont
  {Mao}}, \bibinfo {author} {\bibfnamefont {J.}~\bibnamefont {Xu}} and
  \bibinfo {author} {\bibfnamefont {P.}~\bibnamefont {Bell}}, }
Calibration of the ruby pressure gauge to 800 {kbar} under quasi-hydrostatic conditions, 
\href {\doibase 10.1029/jb091ib05p04673} {\bibfield  {journal} {\bibinfo  {journal} {J. Geophys. Res.}\ }\textbf {\bibinfo {volume} {91}},\ \bibinfo {pages} {4673} (\bibinfo {year} {1986})}


\end{thebibliography}
\end{document}



\title{Magnetotransport of Sm$_2$Ir$_2$O$_7$ across the pressure-induced
quantum-critical phase boundary - Supplementary Information}

\author{M.J. Coak}
\email{m.j.coak@bham.ac.uk}
\affiliation{Department of Physics, University of Warwick, Gibbet Hill Road, Coventry CV4 7AL, United Kingdom}
\affiliation{London Centre for Nanotechnology, University College London, Gordon St, London WC1H 0AH, United Kingdom}
\affiliation{School of Physics and Astronomy, University of Birmingham, Edgbaston, Birmingham B15 2TT, United Kingdom}
\author{K. G\"otze}
\affiliation{Department of Physics, University of Warwick, Gibbet Hill Road, Coventry CV4 7AL, United Kingdom}
\affiliation{Deutsches Elektronen-Synchrotron (DESY), 22607 Hamburg, Germany}
\author{T. Northam De La Fuente}
\affiliation{Department of Physics, Royal Holloway, University of London, Egham TW20 0EX, United Kingdom}
\affiliation{Institut Laue-Langevin, CS 20156, 38042 Grenoble Cedex 9, France}
\author{C. Castelnovo}
\affiliation{Cavendish Laboratory, University of Cambridge, JJ Thomson Ave, Cambridge CB3 0HE, United Kingdom}
\author{J. P. Tidey}
\affiliation{Department of Chemistry, University of Warwick, Gibbet Hill Road, Coventry CV4 7AL, United Kingdom}
\author{J. Singleton}
\affiliation{National High Magnetic Field Laboratory, Los Alamos National Laboratory, MS-E536, Los Alamos, NM 87545, United States of America}
\author{A.T. Boothroyd}
\affiliation{Department of Physics, University of Oxford, Clarendon Laboratory, Oxford, OX1 3PU, United Kingdom}
\author{D. Prabhakaran}
\affiliation{Department of Physics, University of Oxford, Clarendon Laboratory, Oxford, OX1 3PU, United Kingdom}
\author{P.A. Goddard}
\email{p.goddard@warwick.ac.uk}
\affiliation{Department of Physics, University of Warwick, Gibbet Hill Road, Coventry CV4 7AL, United Kingdom}
\date{\today}
\maketitle

\onecolumngrid

\renewcommand{\thesection}{\arabic{section}}
\renewcommand{\thesubsection}{\Alph{subsection}}
\renewcommand{\figurename}{{\bfseries{Supplementary Figure}}}
\renewcommand{\thefigure}{{\bfseries{\arabic{figure}}}}

\makeatletter
\renewcommand{\p@subsection}{}
\renewcommand{\p@subsubsection}{}
\makeatother

\newpage

\subsection*{Supplementary Note 1}
\emph{Fitting of low-temperature resistivity} - Supplementary Notes 2 - 6 give additional detail on the analysis of the resistivity temperature dependence below $T_{\mathrm{min}}$ to that presented in the main body of the paper. As discussed in the main text, we find our data to not be readily described by most of the models previously employed in the pyrochlore iridates. We find a $1/T$ temperature dependence to be the best model to describe the observed low-temperature resistivity in Sm$_2$Ir$_2$O$_7$. As our data do not diverge at low temperatures, all gapped or diverging functions fail to match its behavior --- we therefore introduce a parallel conduction channel of a constant $\rho_0$ to the models for better agreement.
\\
\\
The model most commonly used to describe the resistivity of the pyrochlore iridates is the variable-range-hopping (VRH) model. Supplementary Note 2 gives fits and parameters for this model, with and without an additional $\rho_0$ conducting channel, for comparison with previous publications. We conclude however that this model does not describe the data well over any appreciable temperature range. 
\\
\\
Supplementary Note 3 shows comparisons of the best-performing fits to the data, including the $1/T$ employed for the analysis in the main body of the paper, which gives the best agreement with data while also being physically plausible --- it has been discussed in previous work as representative of a Weyl semimetal \cite{Hosur2012}. VRH fits from Supplementary Note 3 are shown on the same figures for comparison and perform markedly worse at describing the temperature dependence of the resistivity.
\\
\\
Additional detail to the best-performing fits of Supplementary Note 3, as well as fits applied to these systems in the literature which do not describe the data to the same level are presented for completeness and ease of comparison in Supplementary Note 4.
\\
\\
Supplementary Note 5 presents additional technical details and plots of the fits presented in the main text.
\\
\\
Finally for the fitting of low temperature resistivity, Supplementary Note 6 shows the details of how the subtle feature in resistivity at $T_{\mathrm{kink}}$ was identified and its pressure-dependent temperature values determined.

\subsection*{Supplementary Note 2}
\begin{figure}
\begin{centering}
\includegraphics[width=0.5\columnwidth]{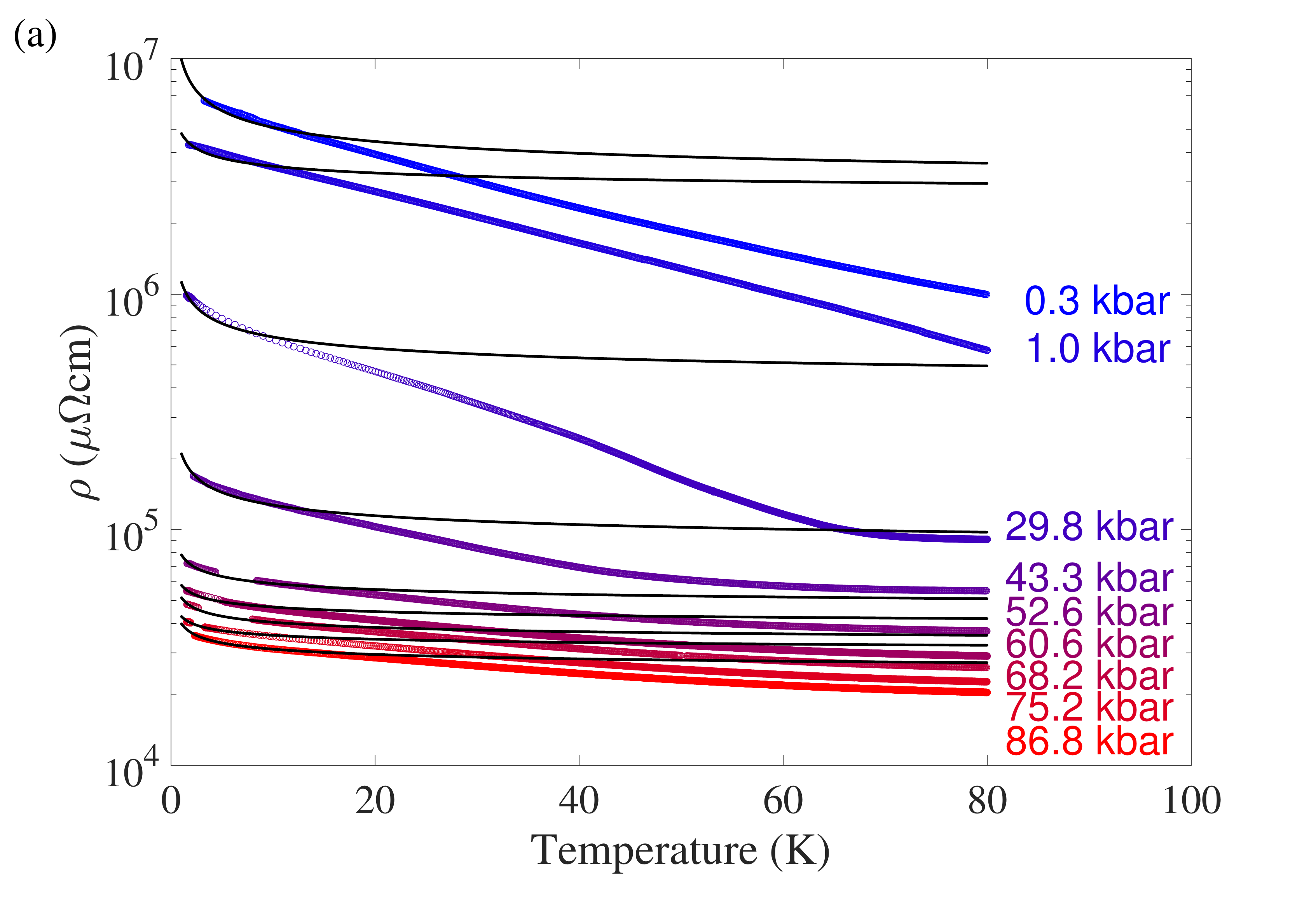}\hspace{0.01\columnwidth}\includegraphics[width=0.5\columnwidth]{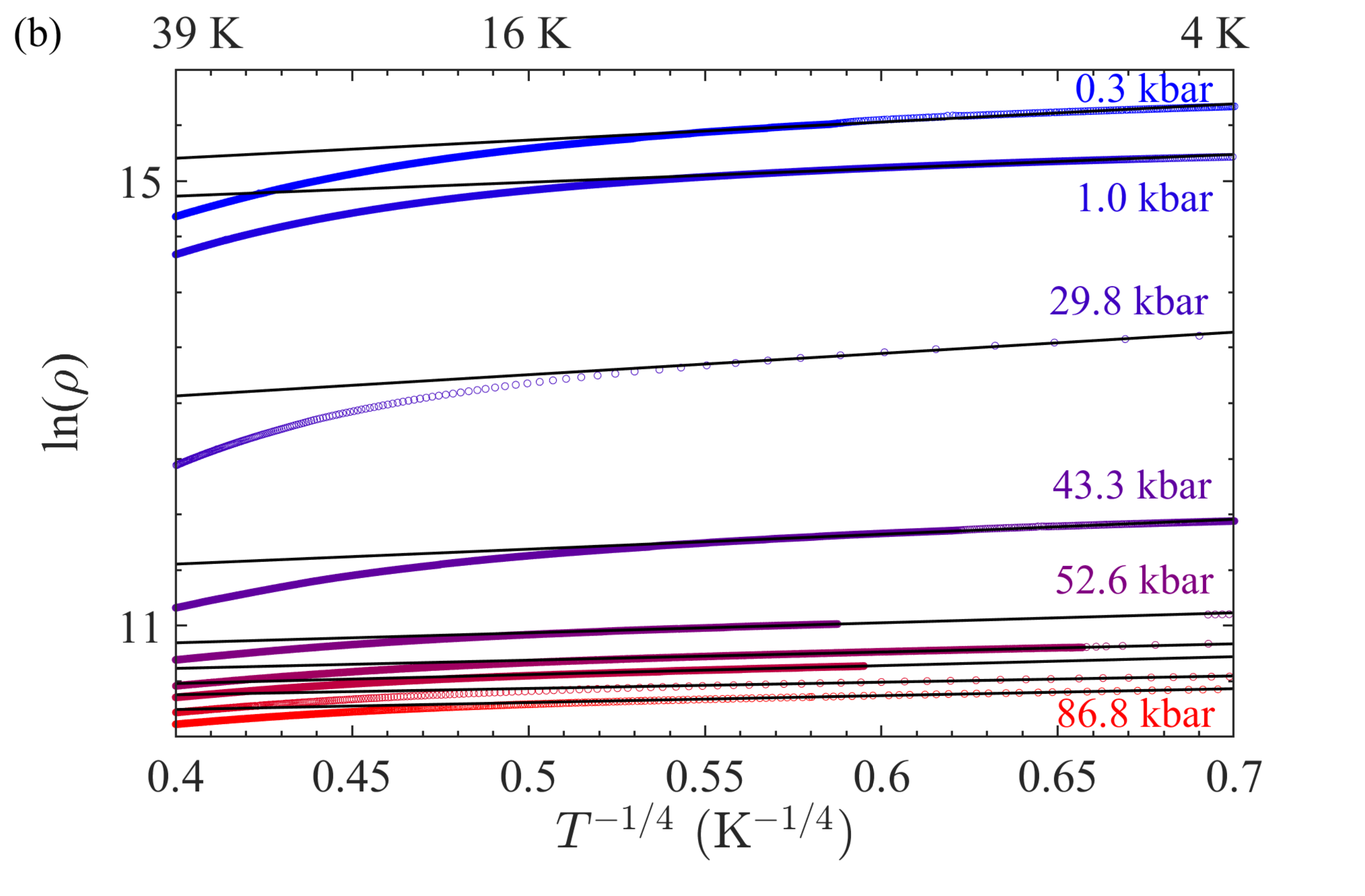}\hspace{0.01\columnwidth}\includegraphics[width=0.5\columnwidth]{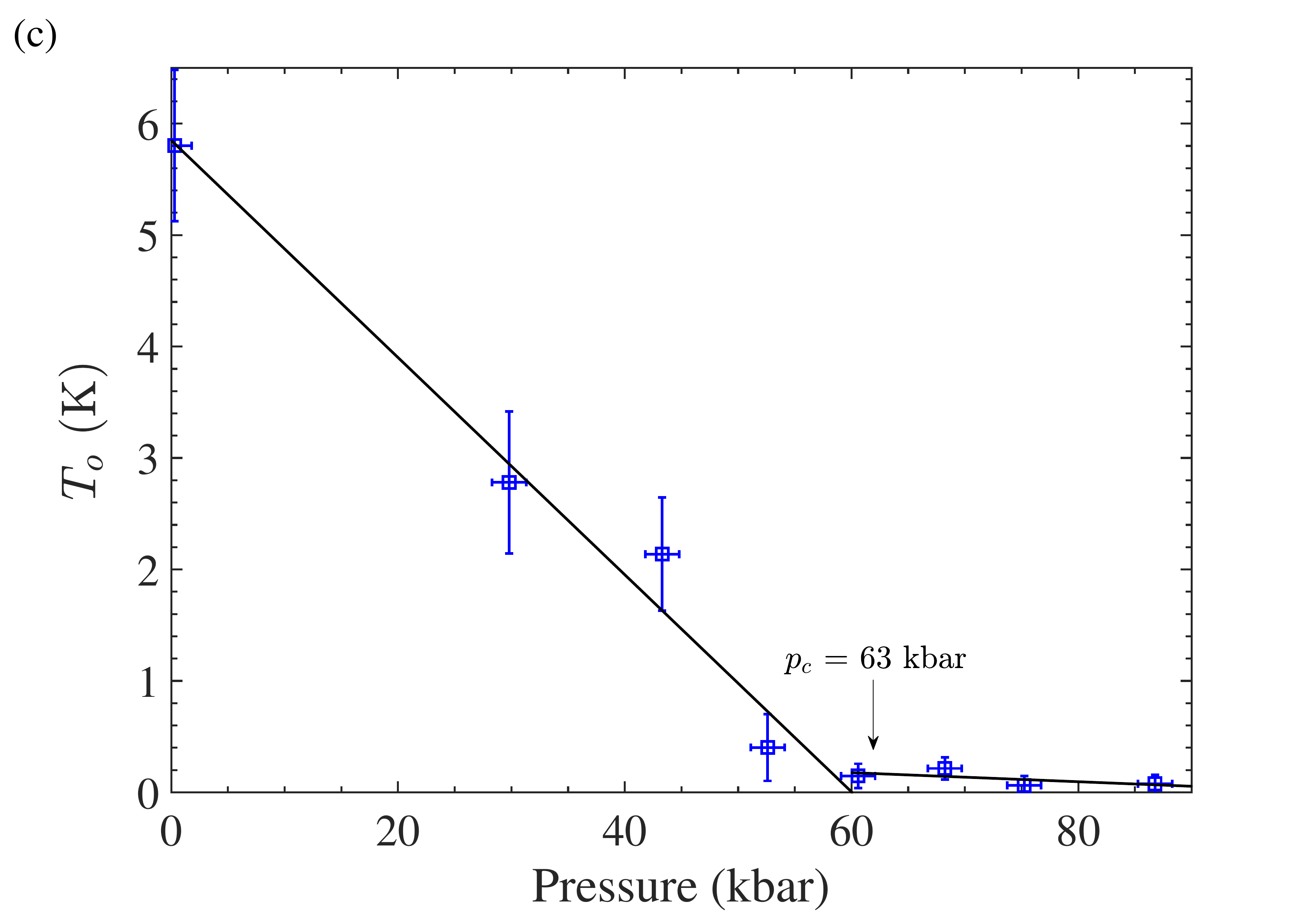}
\par\end{centering}
\centering{}\caption{\label{fig:VRHFits-1} Fits to the variable-range-hopping gapped transport model $\rho\propto \mathrm{e}{}^{(T_{0}/T)^{1/4}}$ as discussed in the main text. (a) The data for each pressure, plotted on a logarithmic scale, with fits to the model shown as solid black lines. (b) $\ln(\rho)$ vs $T^{-1/4}$, which would yield straight lines for a variable-range-hopping transport regime, and corresponding linear fits. These fit the form of the data only below around 10~K. (c) the characteristic activation energies  $T_0$ extracted from these fits. These show a steady reduction, in line with the resistivity magnitude and slope suppression seen, as pressure is increased, followed by a decrease in slope to a much shallower pressure dependence above 60~kbar.}
\end{figure}

\begin{figure}
\begin{centering}
\includegraphics[width=0.5\columnwidth]{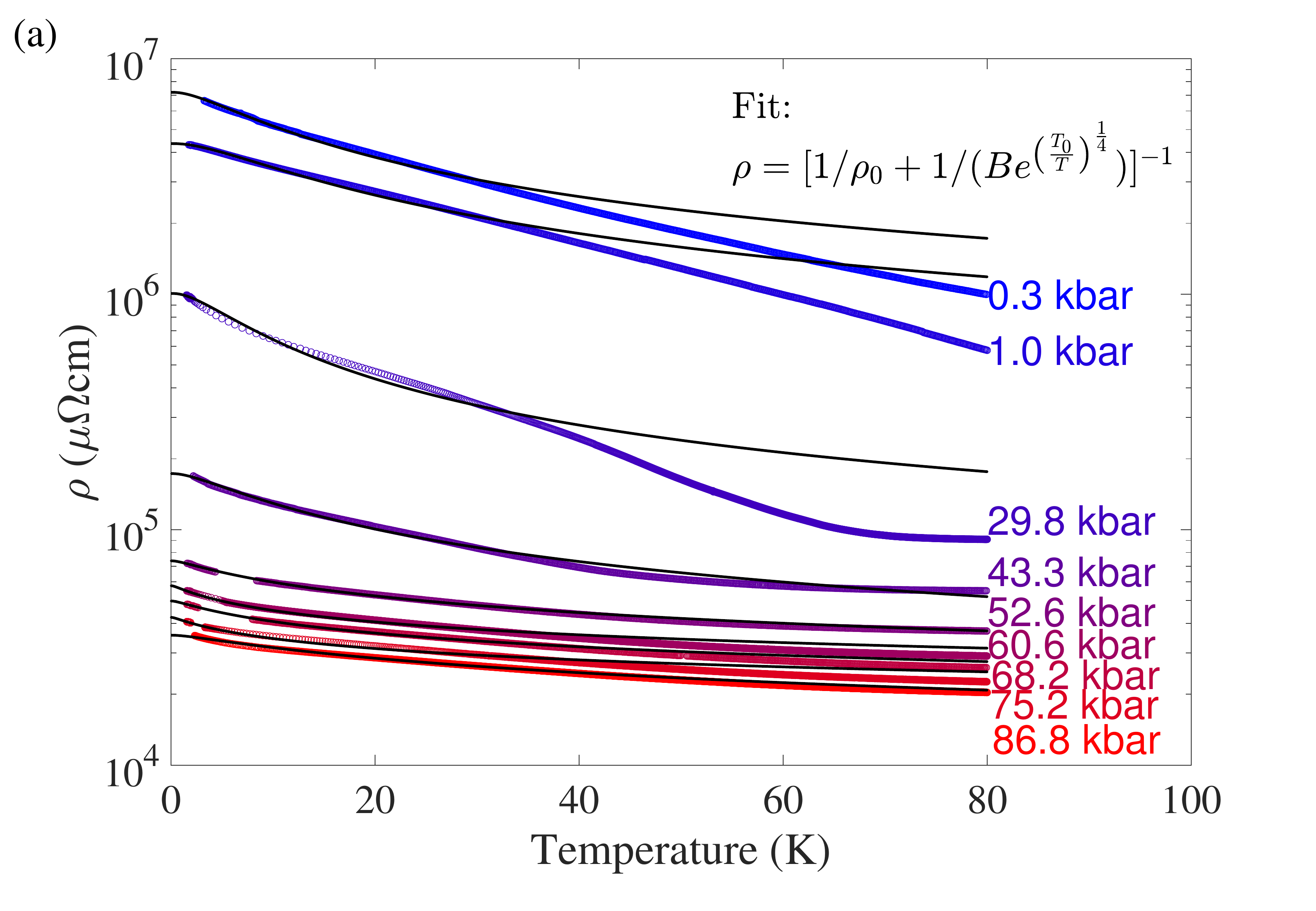}\hspace{0.01\columnwidth}\includegraphics[width=0.5\columnwidth]{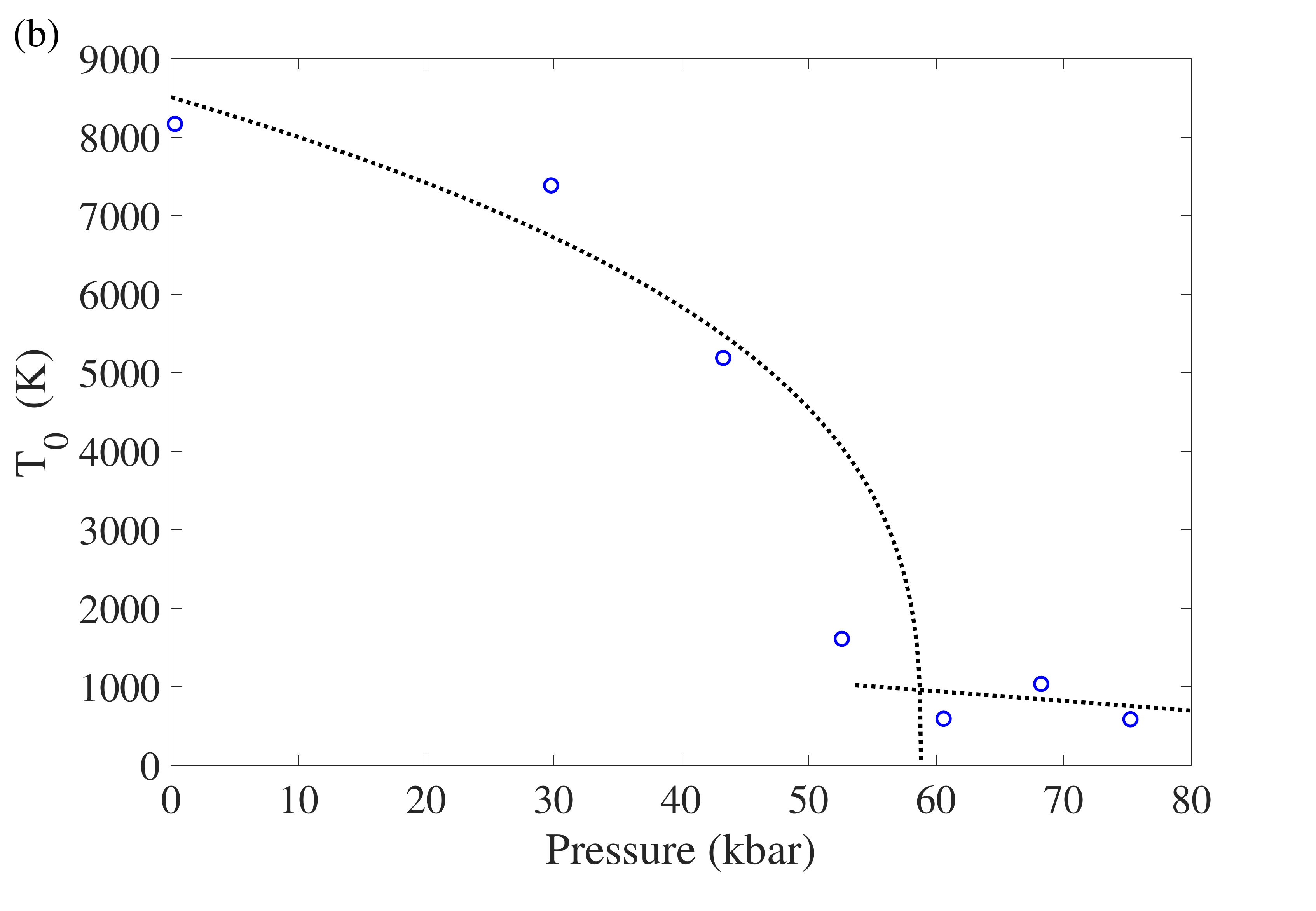}
\par\end{centering}
\centering{}\caption{\label{fig:VRHFits-2} (a) Fits to a VRH model $\rho\propto \mathrm{e}{}^{(T_{0}/T)^{1/4}}$
in parallel with a constant resistance term $\rho_0$. This is able to fit the data more convincingly than the unaltered VRH function alone and does not diverge at zero temperature. The $\rho_0$ fit parameter is the extrapolated zero-temperature resistance as plotted in the main text. (b) The resulting $T_{0}$ values, which are of significantly higher magnitude than those shown in Fig.~\ref{fig:VRHFits-1}, similarly exhibit a marked suppression with initial pressurization followed by a much shallower pressure dependence beyond $p_{c}$.}
\end{figure}

\emph{Fits to variable-range-hopping model} - Here we show fits of our low-temperature resistivity data to the variable-range-hopping model $\rho\propto \mathrm{e}{}^{(T_{0}/T)^{1/4}}$ frequently used to describe the pyrochlore iridates in the literature~\cite{Liu2018b,Han2015}. Supplementary Figure \ref{fig:VRHFits-1} shows these fits to the resistivity at each pressure and the resulting values of $T_{0}$ plotted against pressure. The VRH function does not yield a convincing fit of the data over a reasonable temperature range, in part due to the data not diverging at very low temperatures. To account for this, Supplementary Figure \ref{fig:VRHFits-2} shows fits to VRH resistance in parallel with a constant (pressure-dependent) fixed resistance channel, which fits the data much better and over a wider range. Nevertheless, the agreement with the data is significantly worse than the fits to the alternative models shown in Section Supplementary Note 3 and used in the main text.

\FloatBarrier{}

\FloatBarrier{}

\clearpage{}

\subsection*{Supplementary Note 3}
\label{subsec:ComparisonOfBestModelsToTransportData}

\begin{figure}
\begin{centering}
\includegraphics[width=0.55\columnwidth]{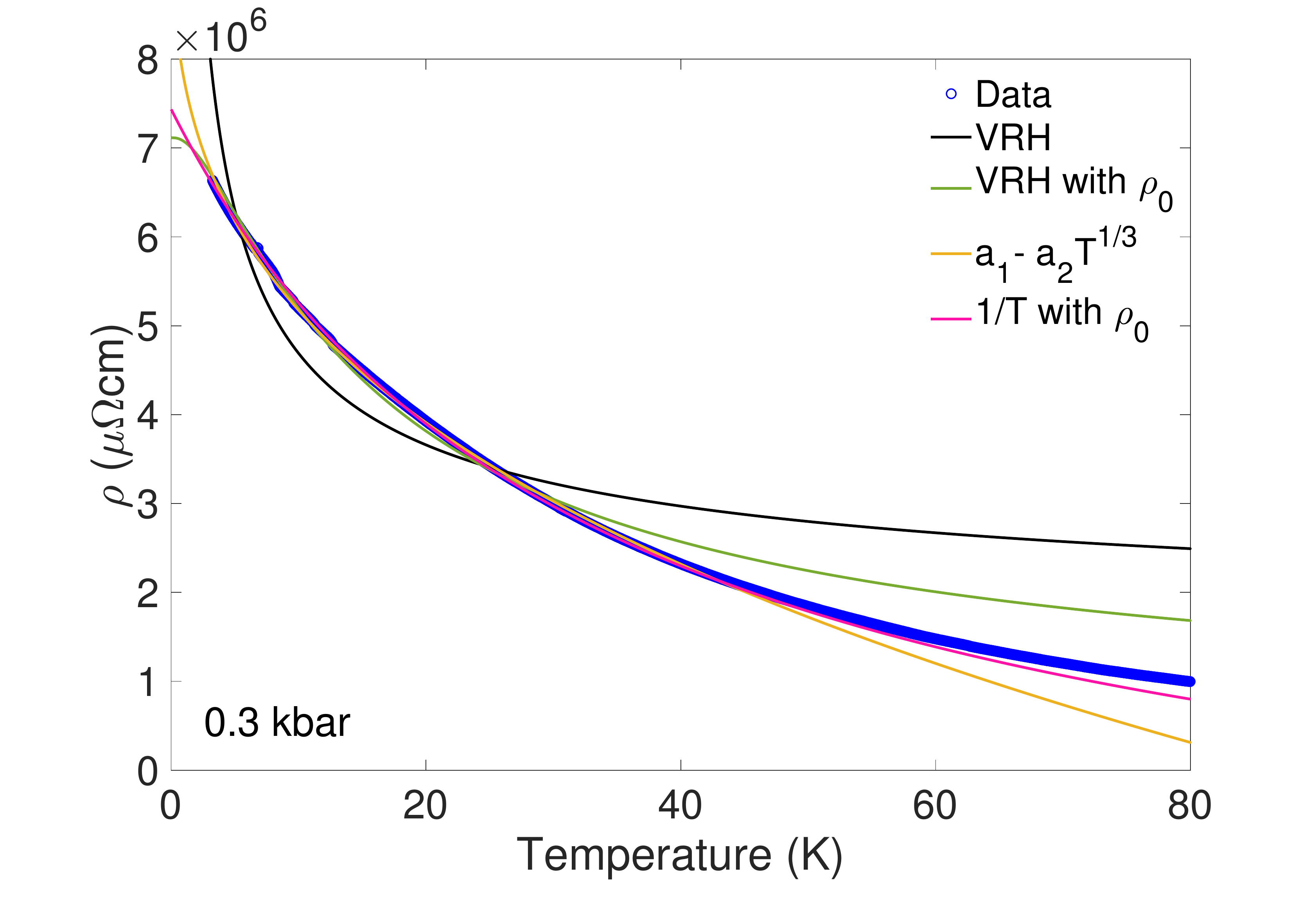}
\par\end{centering}
\begin{centering}
\includegraphics[width=0.55\columnwidth]{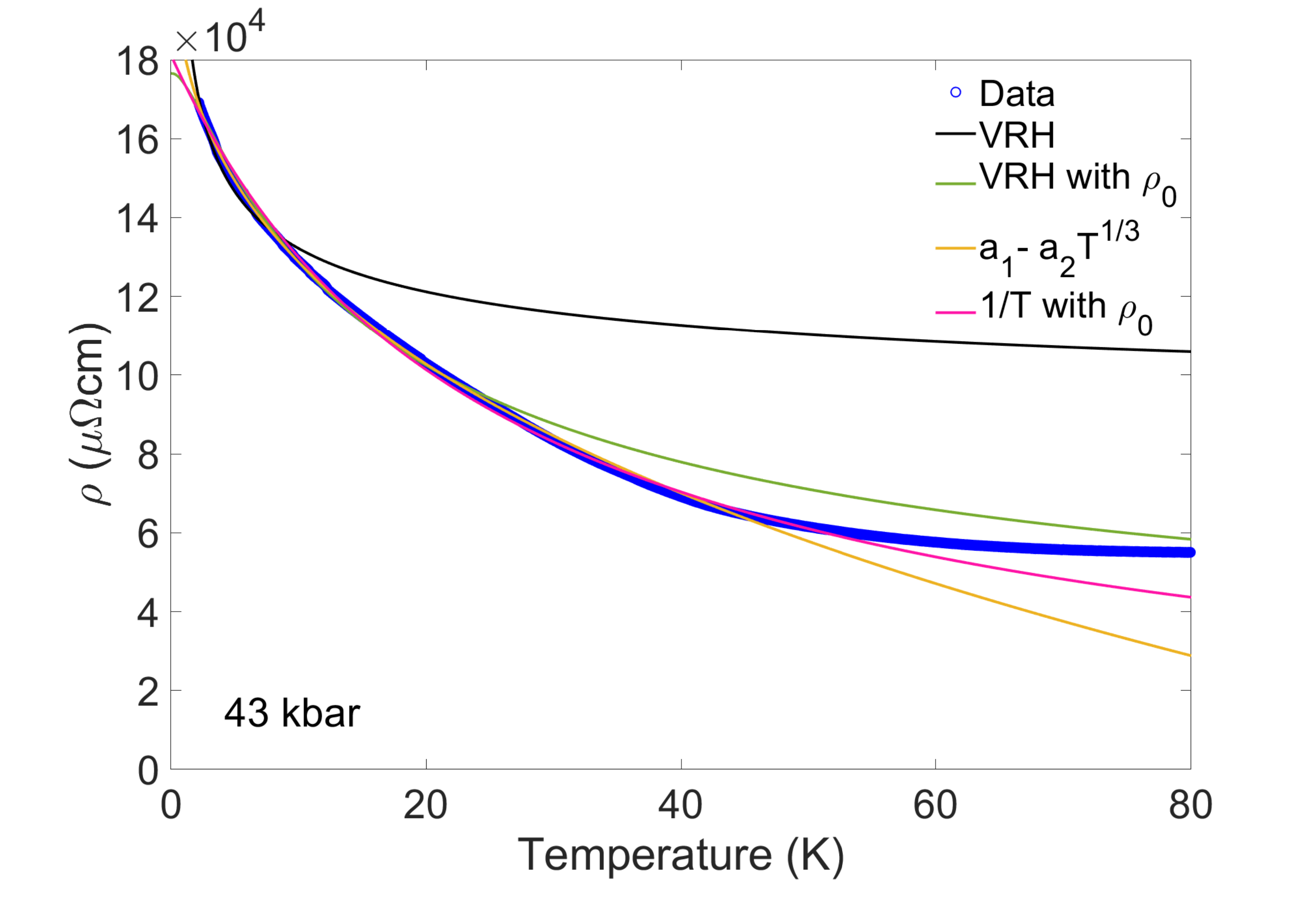}
\par\end{centering}
\centering{}\caption{\label{fig:RvTFitComparisons}Comparison of the best-fitting candidate
transport models at low temperature, at 0.3 and 43~kbar. VRH = variable-range-hopping model.}
\end{figure}

\emph{Comparison of the best fitting transport models to the VRH model} - Supplementary Figure \ref{fig:RvTFitComparisons} shows a comparison of the models found to best fit the low-temperature transport data $\rho(T)$. These were chosen based on and extending those employed on the pyrochlore iridates previously in the literature, including the VRH fits shown in Supplementary Note 2. The models that are described as `with $\rho_{0}$' include a constant resistance channel in parallel to account for the observed non-diverging resistivity, as described in the main text. It can be clearly seen that the VRH model does not give the best description of the data over a reasonable range of temperatures.

The $1/T$ with $\rho_{0}$ model employed in the main text, along with the $T^{1/3}$ empirical model detailed in Supplementary Note 4, give the most convincing fit to the data over a reasonable temperature range. 

\FloatBarrier{}

\newpage{}

\subsection*{Supplementary Note 4}
\label{subsec:fitsToOtherExistingModels}
\begin{figure}
\begin{centering}
\includegraphics[width=0.6\columnwidth]{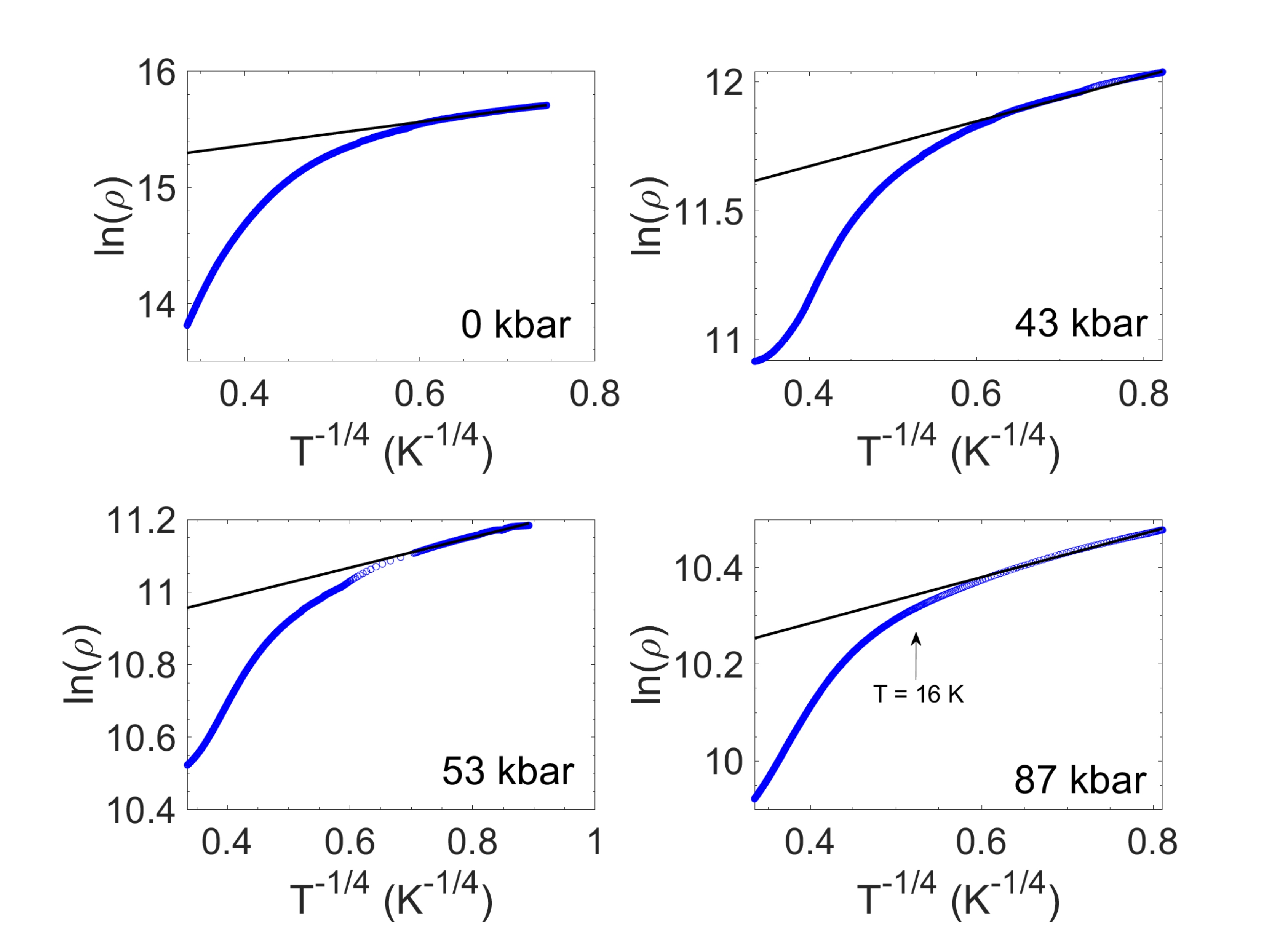}
\par\end{centering}
\centering{}\caption{\label{fig:RvTFits-1}Fits to a variable-range-hopping model $\rho\propto \mathrm{e}{}^{(T_{0}/T)^{1/4}}$: $\ln(p)$ is plotted against $T^{1/4}$ and straight lines (black) are fitted to the data at low temperature.}
\end{figure}

\begin{figure}
\begin{centering}
\includegraphics[width=0.6\columnwidth]{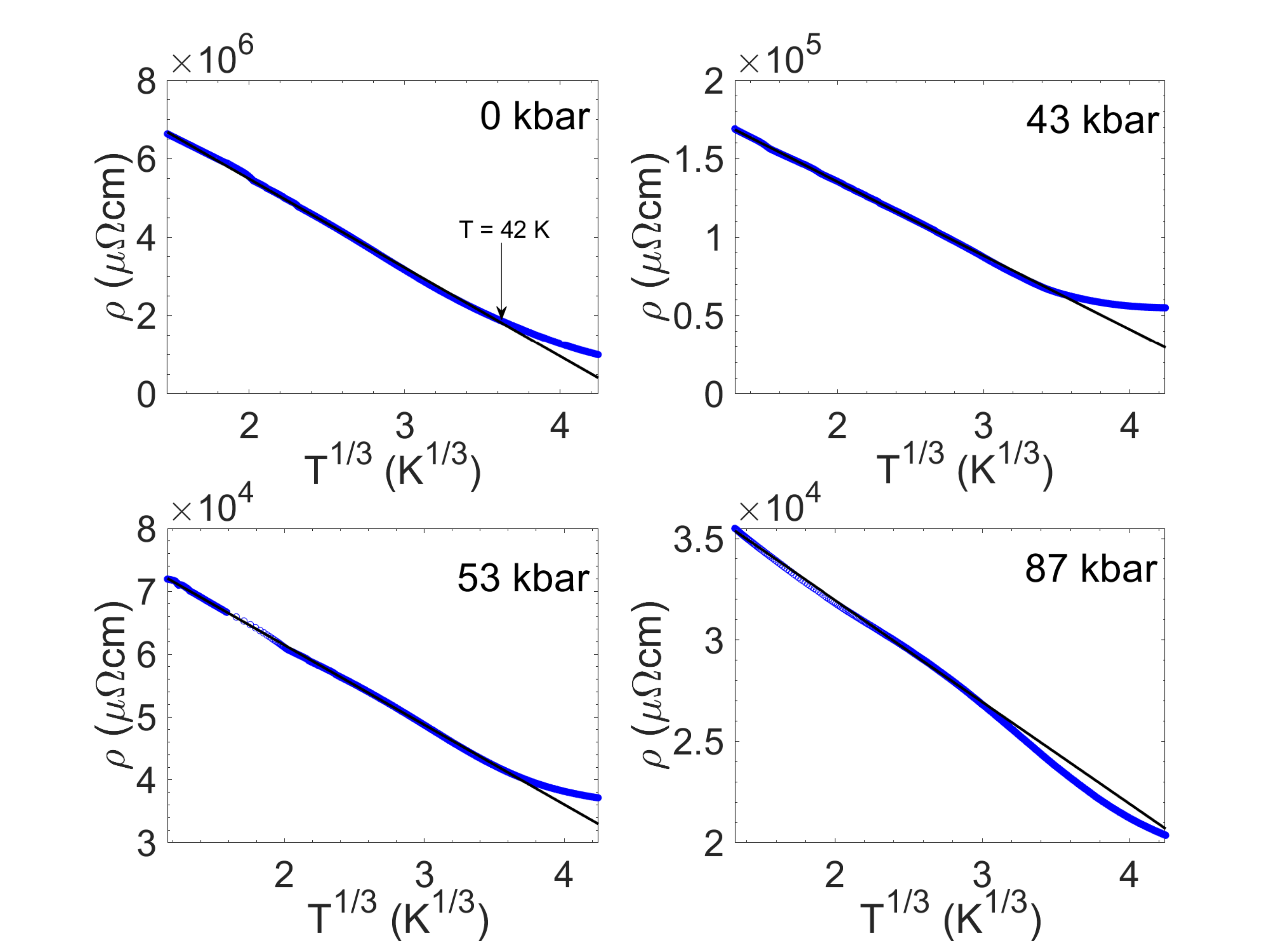}
\par\end{centering}
\centering{}\caption{\label{fig:RvTFits-1-1}Fits to a $\rho\propto T^{1/3}$ model. This temperature exponent was found to best fit the data over the widest range, but with no immediately clear physical meaning. $\rho(T)$ is plotted against $T^{-1/3}$ and straight lines (black) are fitted to the data at low temperature.}
\end{figure}

\begin{figure}
\begin{centering}
\includegraphics[width=0.6\columnwidth]{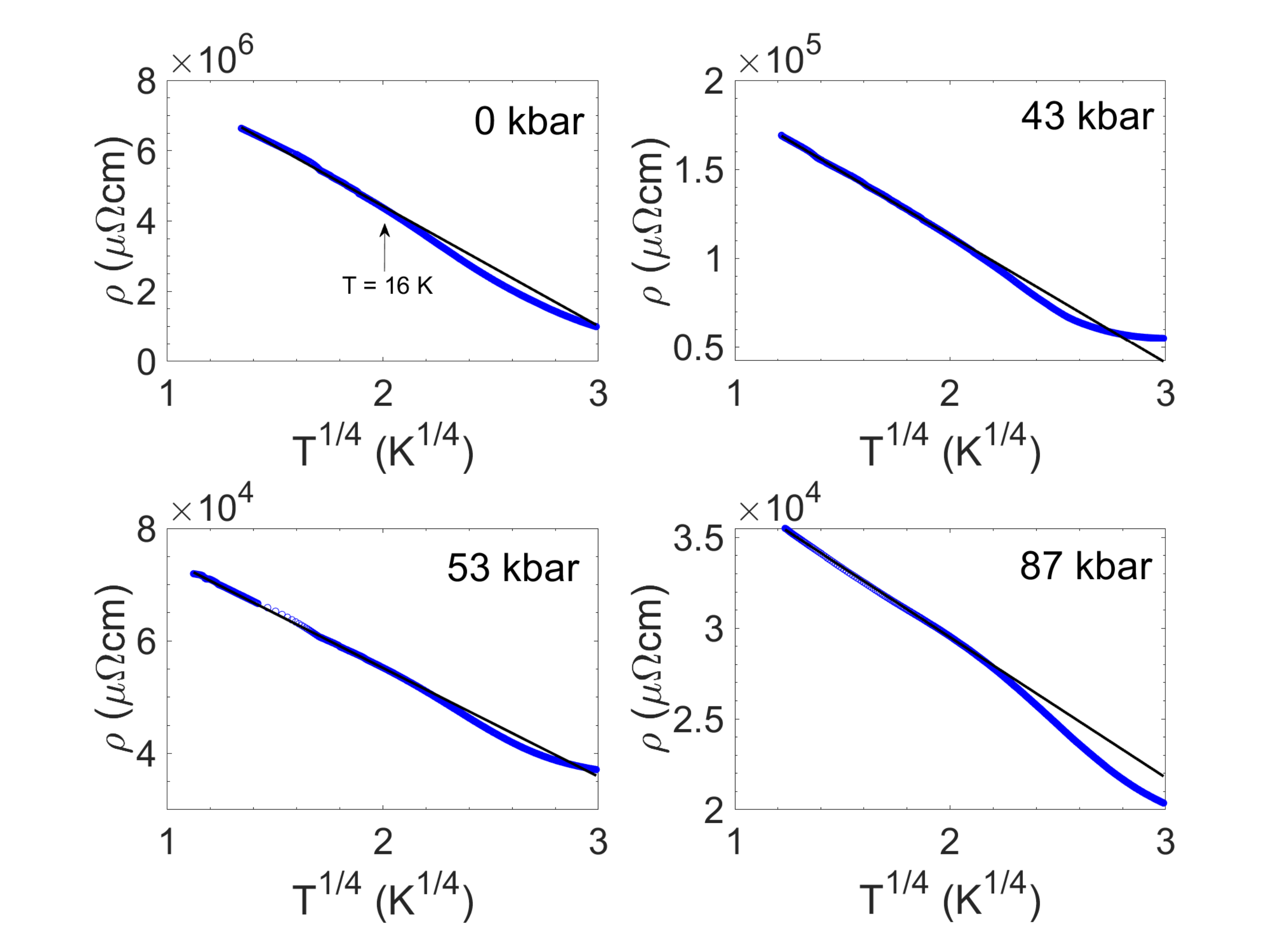}
\par\end{centering}
\centering{}\caption{\label{fig:RvTFits-1-2}Fits to a $\rho\propto T^{1/4}$ quantum critical model put forward in Ref \citep{Telang2021}. $\rho(T)$ is plotted against $T^{-1/4}$ and straight lines (black) are fitted to the data at low temperature.}
\end{figure}

\begin{figure}
\begin{centering}
\includegraphics[width=0.6\columnwidth]{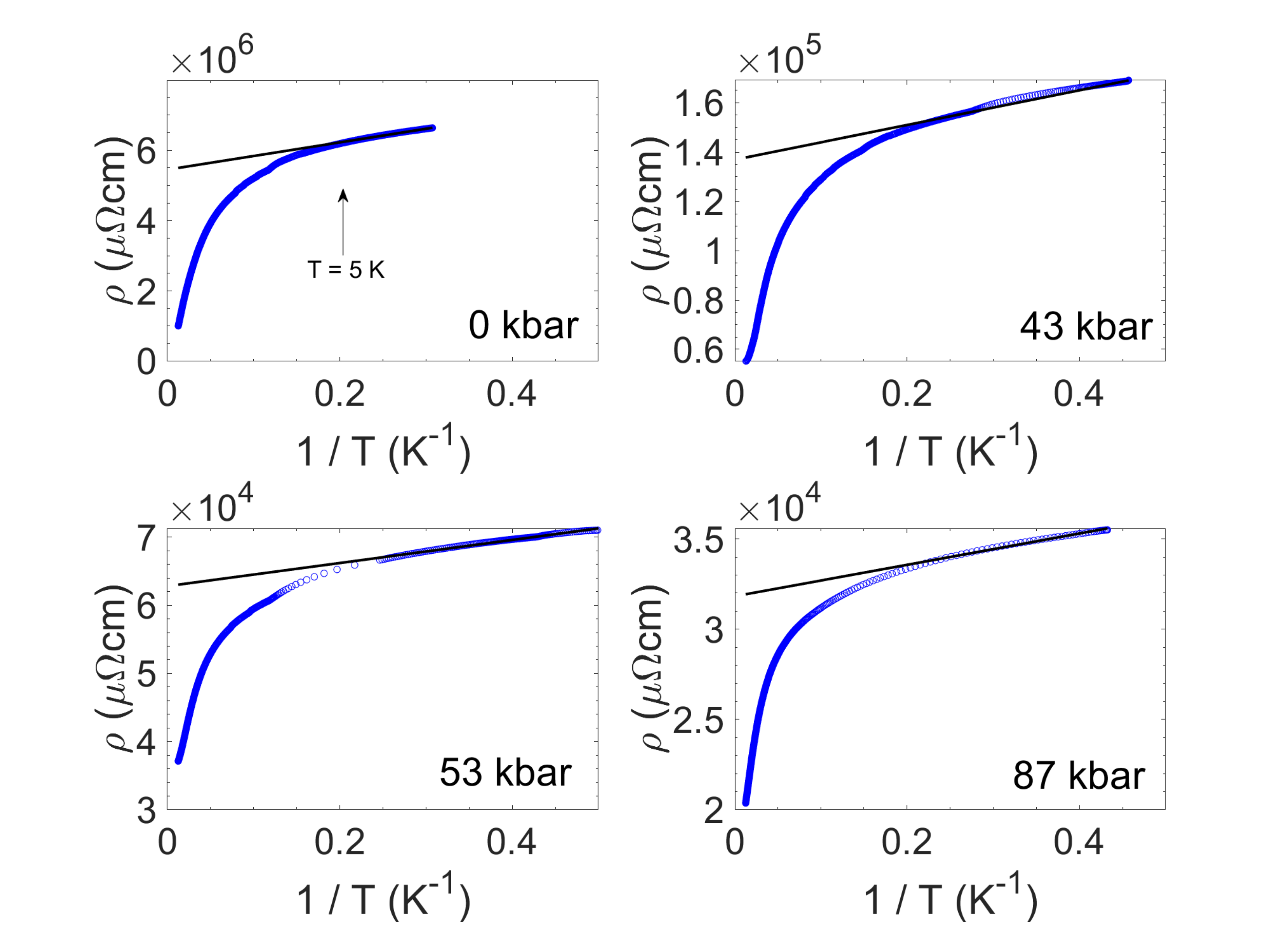}
\par\end{centering}
\centering{}\caption{\label{fig:RvTFits-1-3}Fits to a $\rho\propto T^{-1}$ Weyl semimetal model as employed in Ref \citep{Telang2021}. $\rho(T)$ is plotted against $1/T$ and straight lines (black) are fitted to the data at low temperature.}
\end{figure}

\begin{figure}
\begin{centering}
\includegraphics[width=0.6\columnwidth]{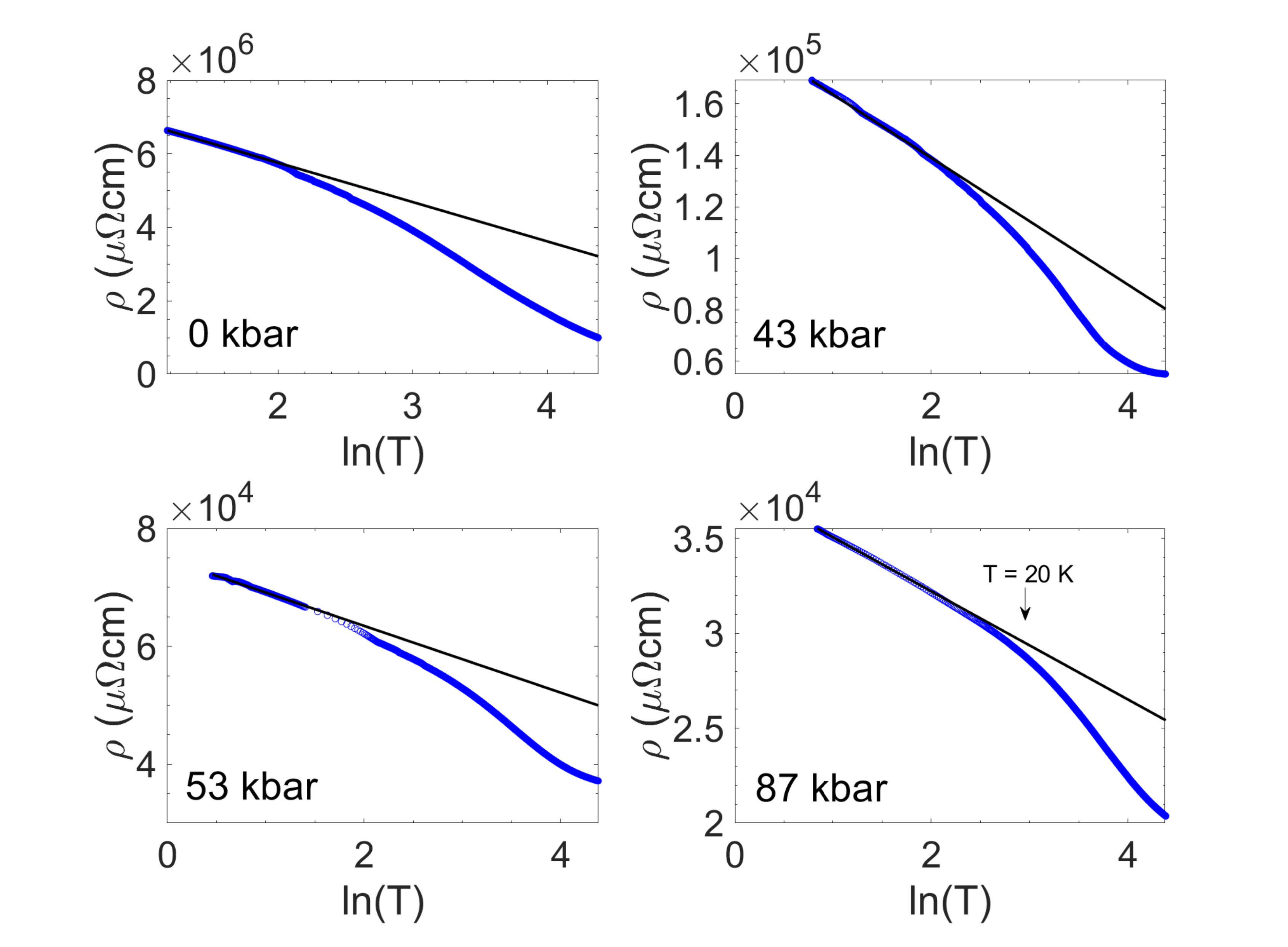}
\par\end{centering}
\centering{}\caption{\label{fig:RvTFits-1-4}Fits to a $\rho\propto \ln(T)$ quadratic-band-touching model as employed in Ref \citep{Telang2021}. $\rho(T)$ is plotted against $\ln(T)$ and straight lines (black) are fitted to the data at low temperature.}
\end{figure}

\begin{figure}
\begin{centering}
\includegraphics[width=0.6\columnwidth]{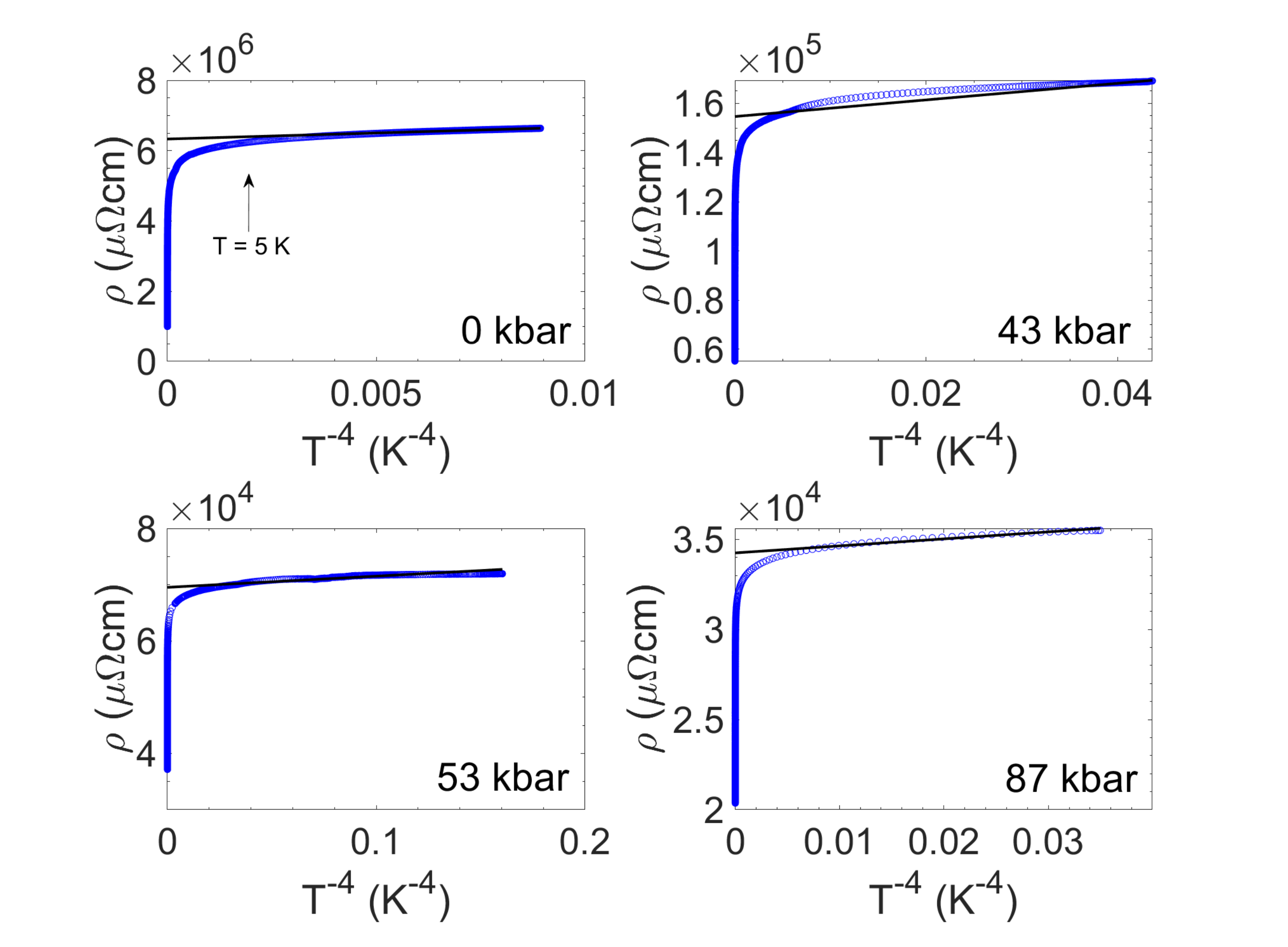}
\par\end{centering}
\centering{}\caption{\label{fig:RvTFits-1-5}Fits to a $\rho\propto T^{-4}$ Weyl semimetal model with thermally screened charged impurities as described in Ref \citep{LaBarre2019}. $\rho(T)$ is plotted against $T^{-4}$ and straight lines (black) are fitted to the data at low temperature.}
\end{figure}

\begin{figure}
\begin{centering}
\includegraphics[width=0.6\columnwidth]{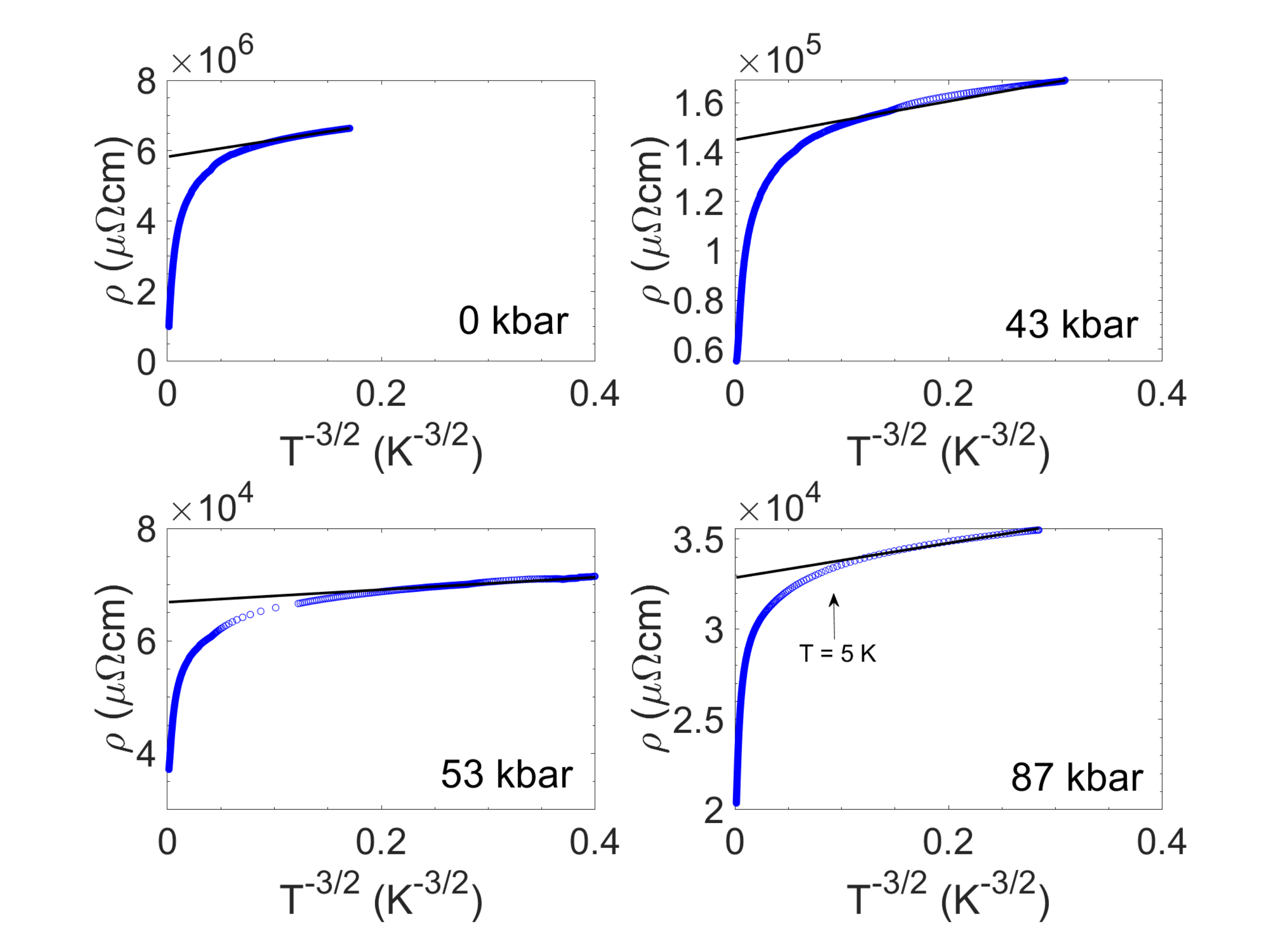}
\par\end{centering}
\centering{}\caption{\label{fig:RvTFits-1-6}Fits to a $\rho\propto T^{-3/2}$ model of a Weyl semimetal in the presence of impurities and electron--electron correlations, as employed in Ref \citep{Liu2020}. $\rho(T)$ is plotted against $T^{-3/2}$ and straight lines (black) are fitted to the data at low temperature.}
\end{figure}

\emph{Fits to other models for comparison} - Following and extending the treatment of Telang and Singh \citep{Telang2021} discussed in the main text, we show here (Supplementary Figures \ref{fig:RvTFits-1}-\ref{fig:RvTFits-1-6}) fits of four representative resistivity datasets to a selection of candidate models presented in a format consistent with Ref.~\citep{Telang2021} for comparison. These models include those used in the preceding two sections, as well as other models used previously in the literature on the pyrochlore iridates. Each is plotted for a temperature range 1.5--80~K.

The variable-range-hopping model $\rho\propto \mathrm{e}{}^{(T_{0}/T)^{1/4}}$ discussed in Supplementary Note 2 and used in much of the literature is shown again first, as well as the best fitting models discussed in Supplementary Note 3. In addition, a direct power-law-exponent fit to the data $\rho\propto T^{N}$ yielded a best value for exponent $N$ of approximately 0.33, so the resulting best fits, $\rho\propto T^{1/3}$ are shown next.

Following these are fits to $\rho\propto T^{1/4}$, $\rho\propto T^{-1}$, $\rho\propto \ln(T)$, $\rho\propto T^{-4}$ and $\rho\propto T^{-3/2}$ models taken from examples used in the literature \citep{Telang2021,Liu2020,LaBarre2019} to demonstrate the comparative applicability of each.

While the temperature ranges over which acceptable agreement with the data is achieved can be seen to differ for each fit, this Section is intended to show that these previously-employed models do not convincingly fit our data. This leads us to conclude that the $1/T$ with $\rho_0$ model employed in the main text and shown in Supplementary Note 3 is the only acceptable description of the observed low temperature data. The $\rho\propto T^{1/3}$ fit also gives convincing agreement, but has no clear physical meaning.

\FloatBarrier{}

\FloatBarrier{}

\clearpage{}

\subsection*{Supplementary Note 5}
\label{subsec:detailOfFitsUsedInMT}
\begin{figure}
\begin{centering}
\includegraphics[width=0.3\columnwidth]{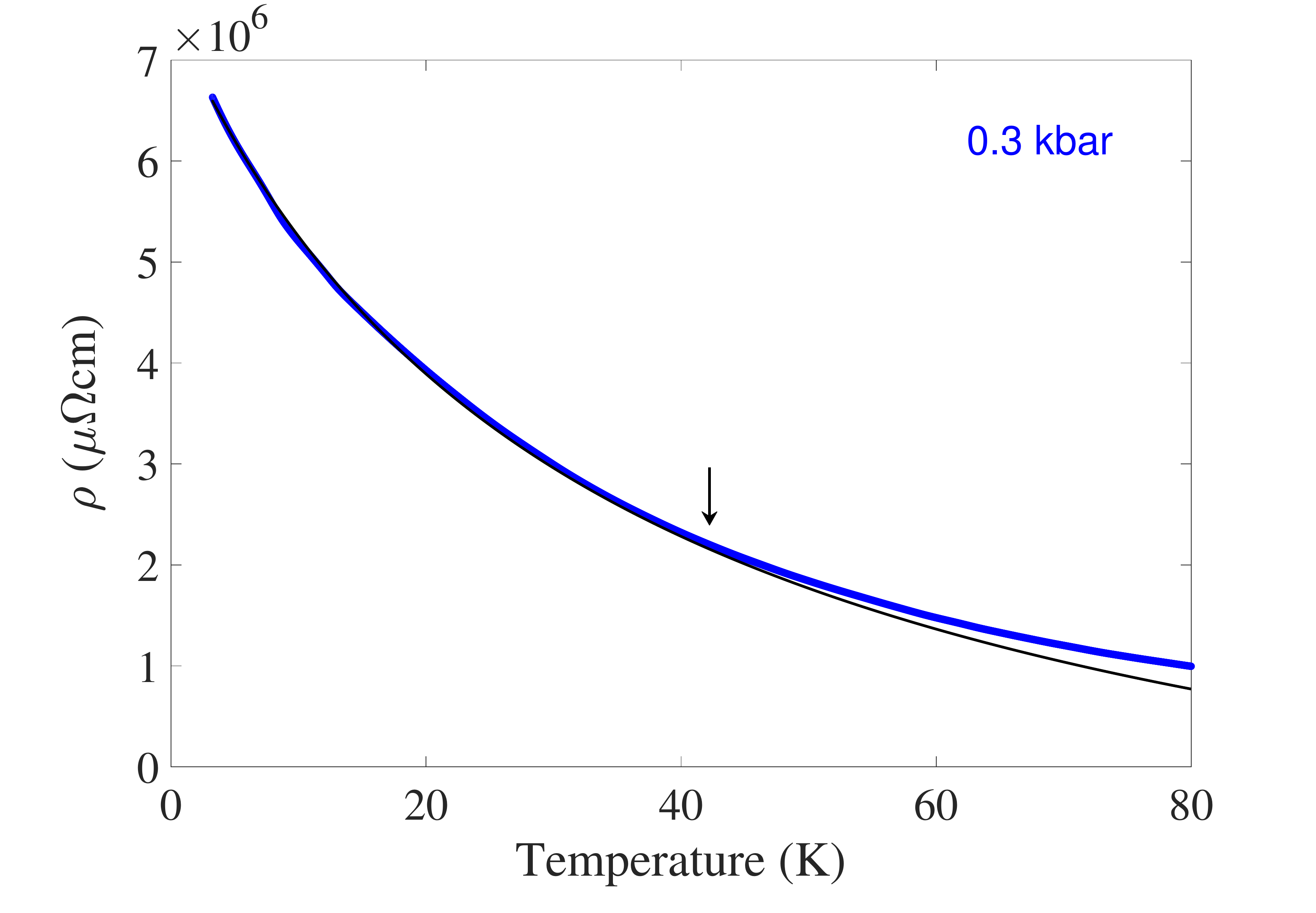}
\includegraphics[width=0.3\columnwidth]{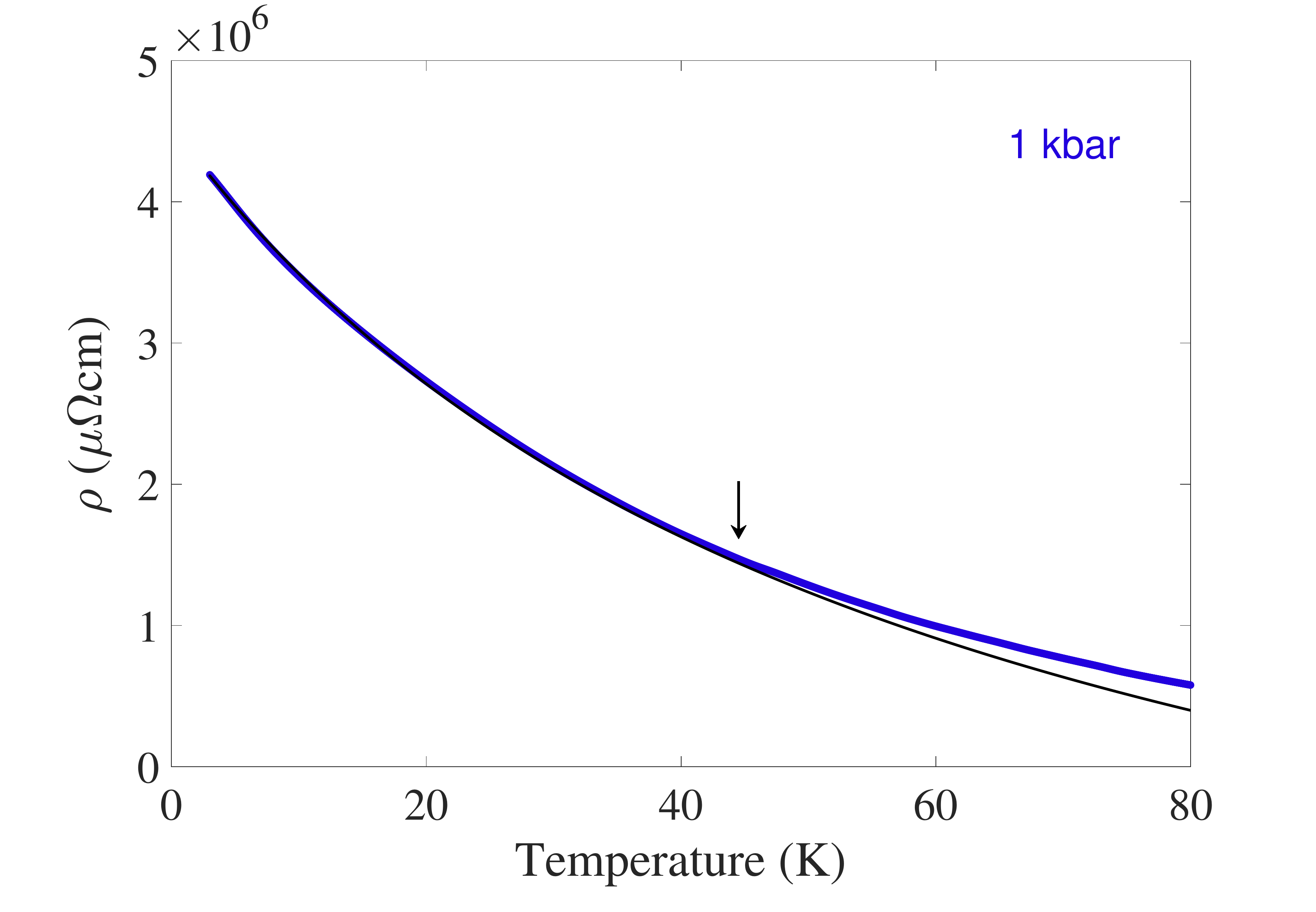}\\
\includegraphics[width=0.3\columnwidth]{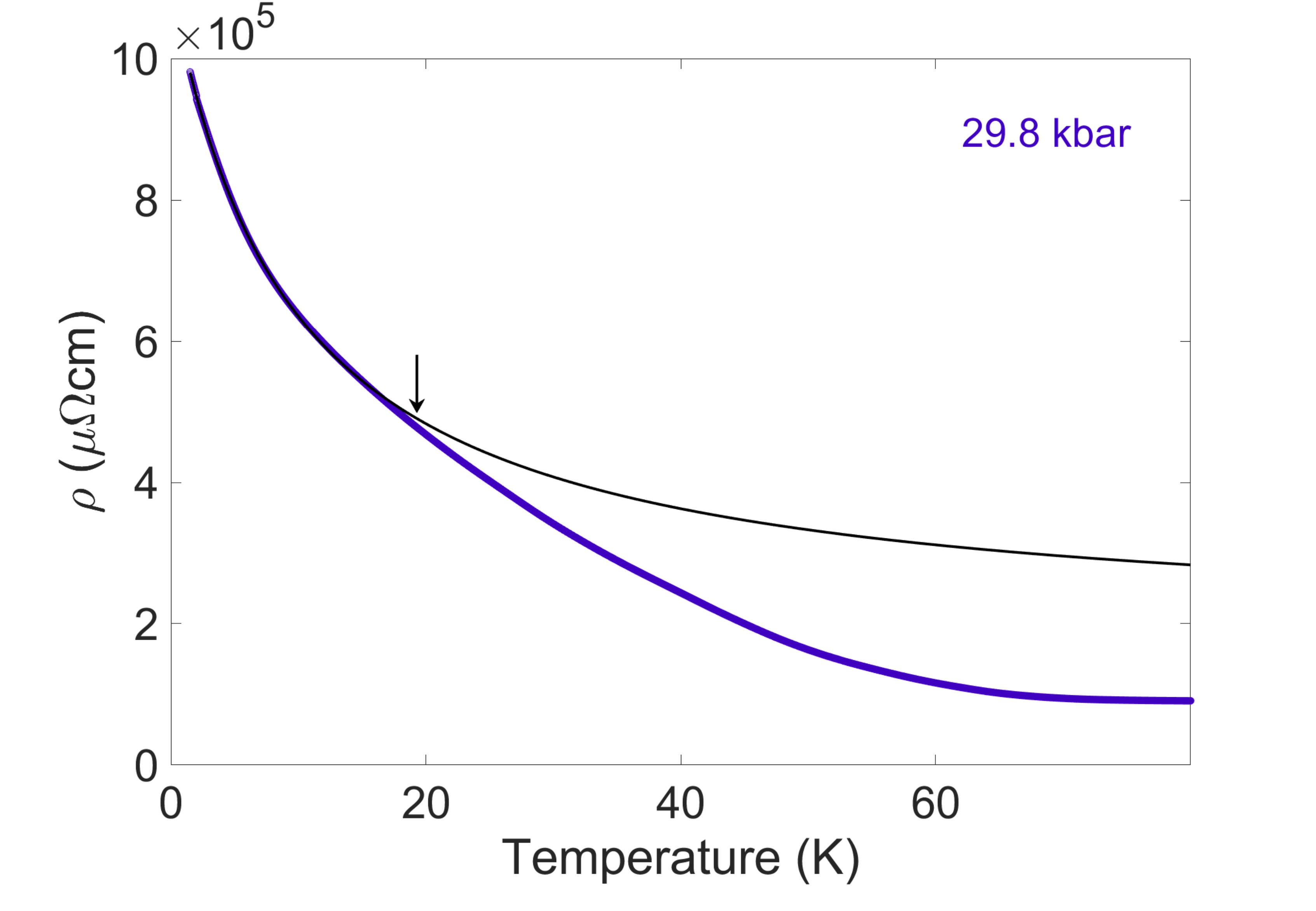}
\includegraphics[width=0.3\columnwidth]{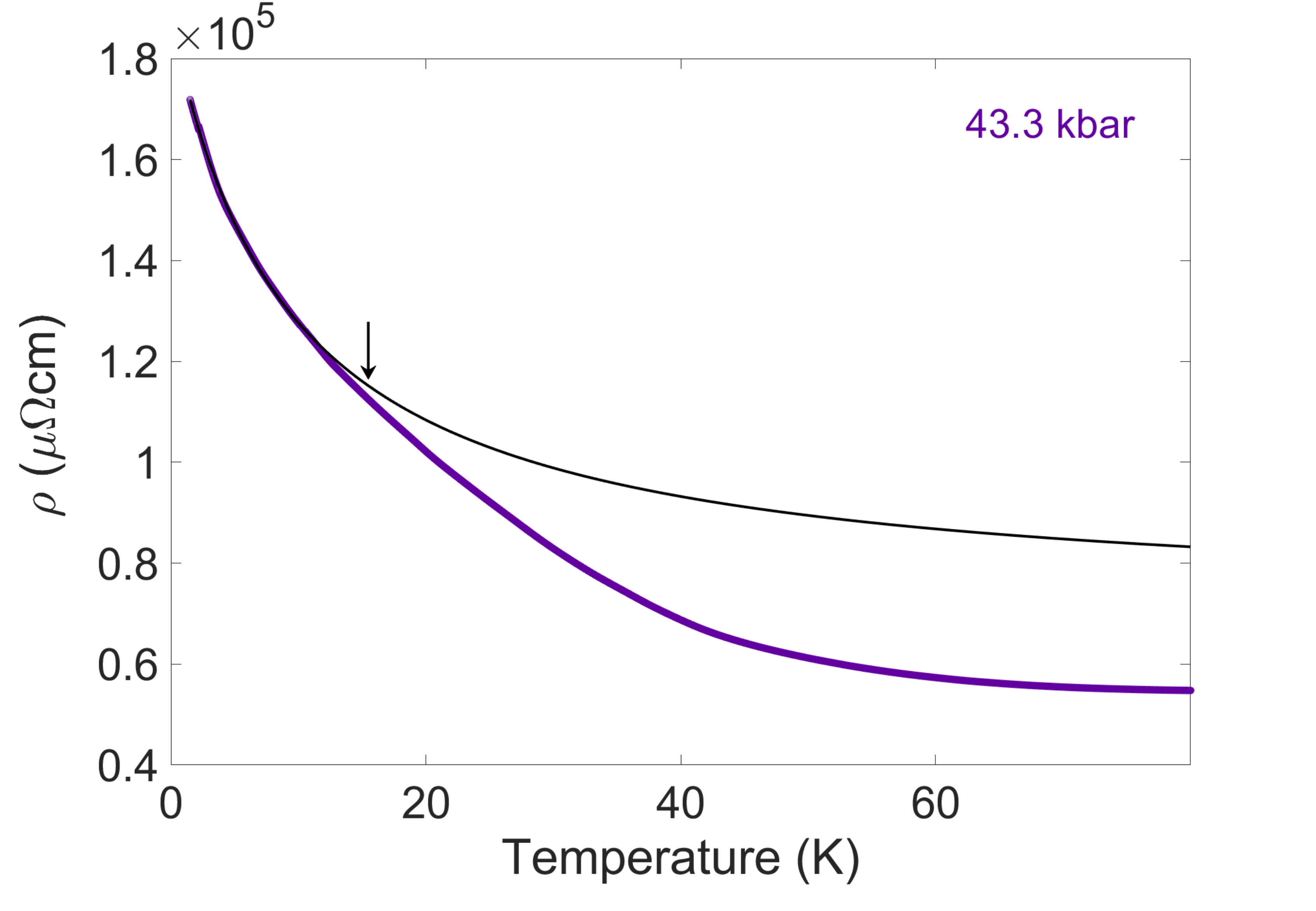}\\
\includegraphics[width=0.3\columnwidth]{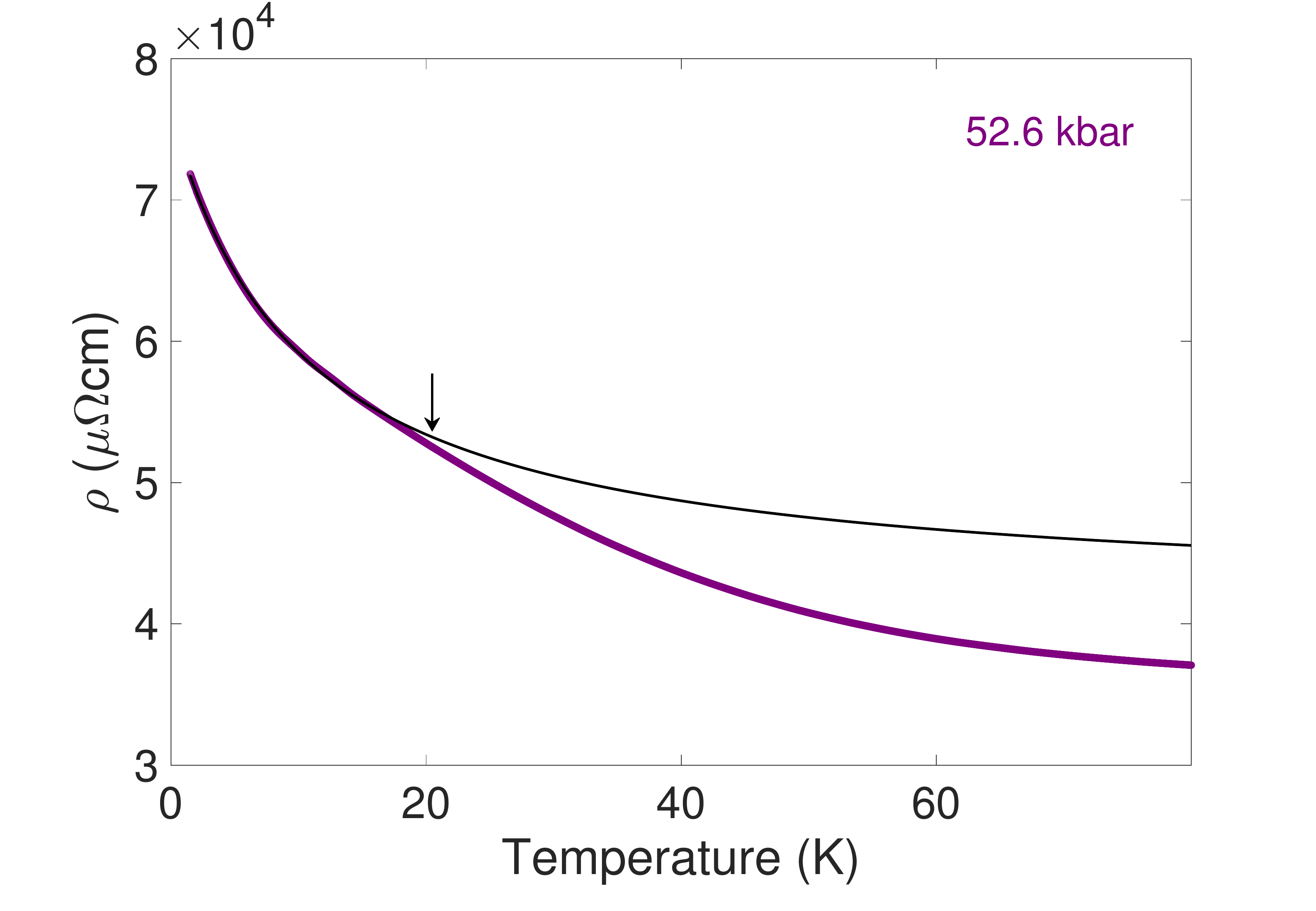}
\includegraphics[width=0.3\columnwidth]{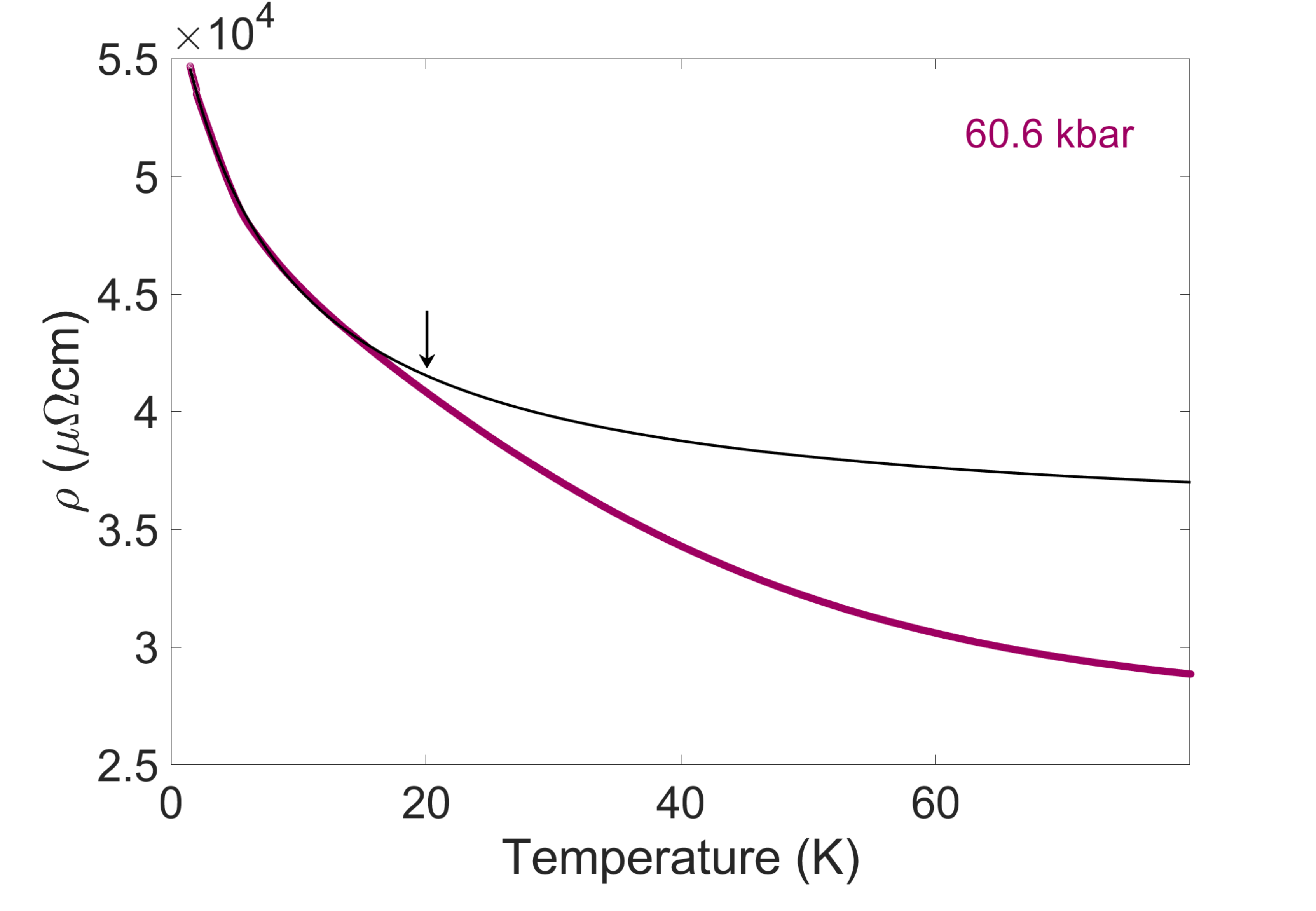}\\
\includegraphics[width=0.3\columnwidth]{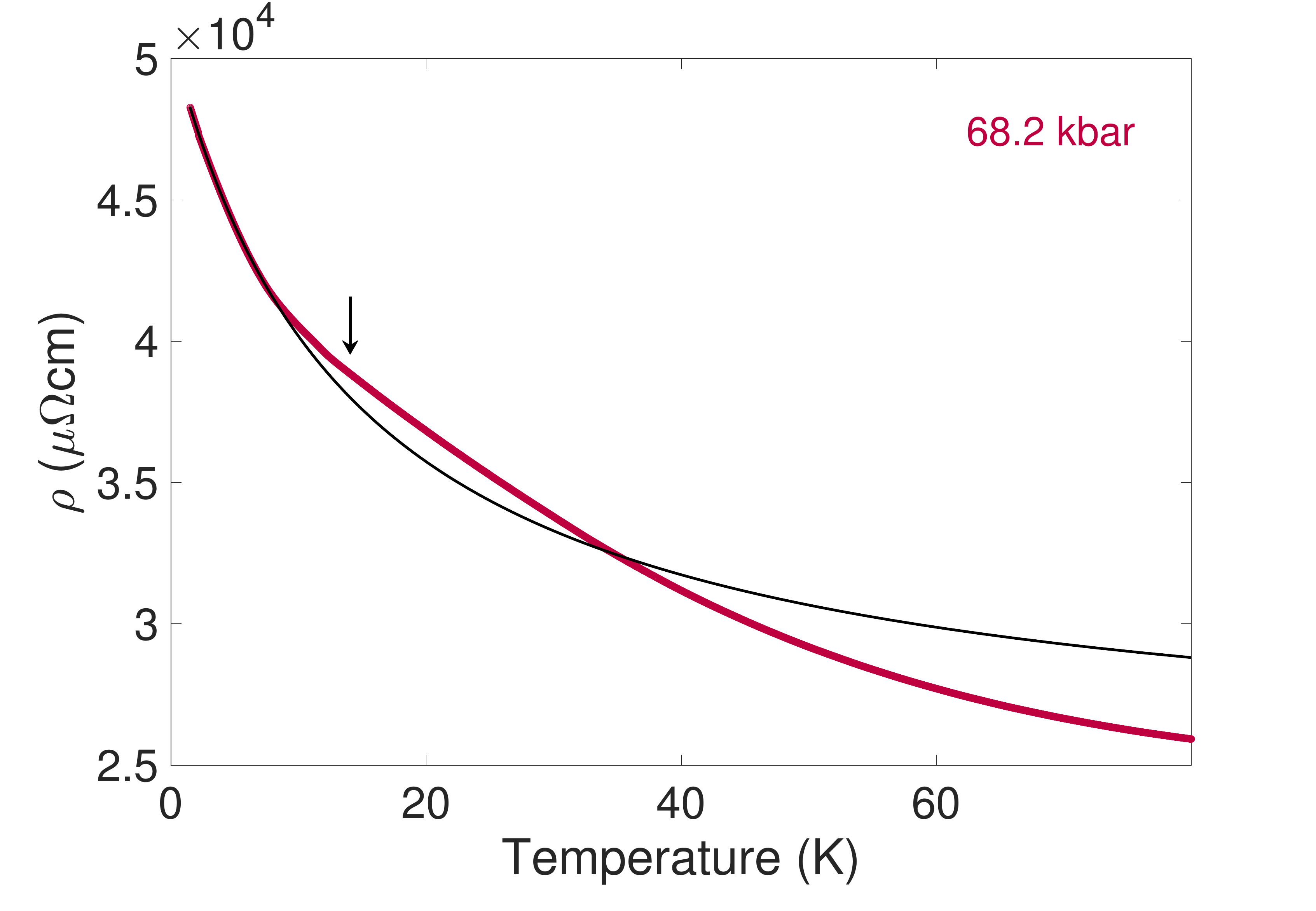}
\includegraphics[width=0.3\columnwidth]{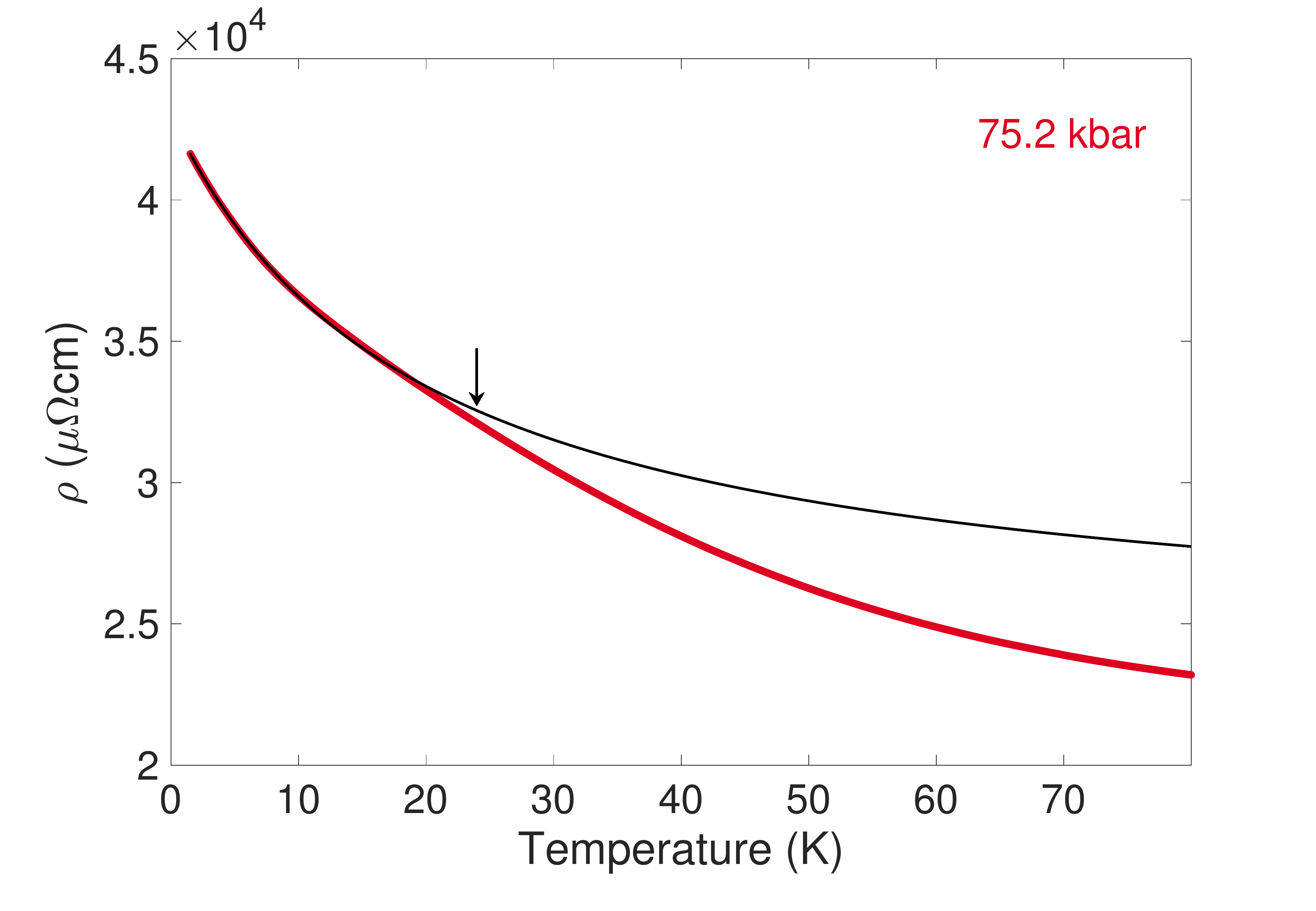}\\
\includegraphics[width=0.3\columnwidth]{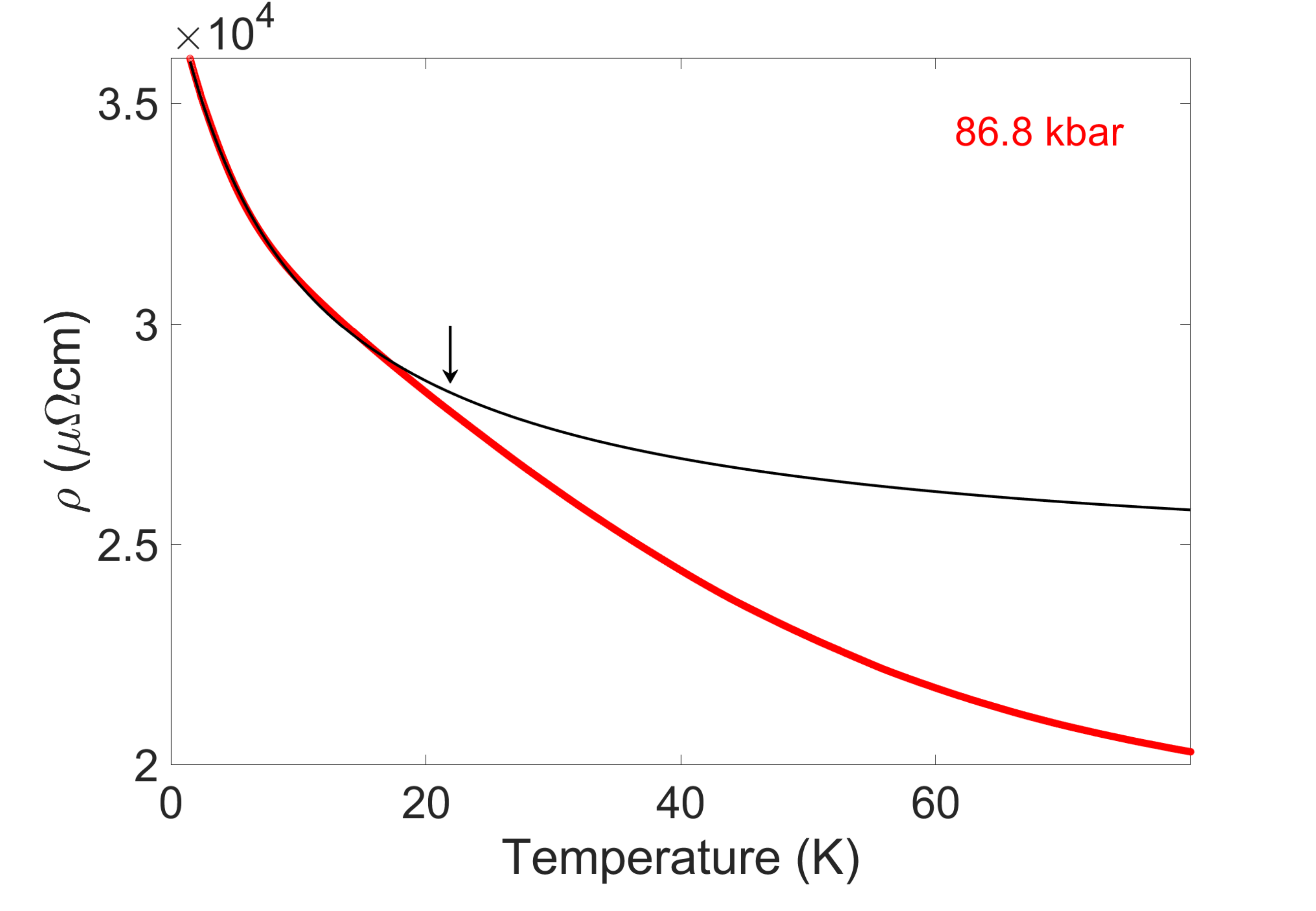}
\par\end{centering}
\centering{}\caption{\label{fig:WSMFits-allP}Data and fit (black lines) for each pressure to the two-channel $1/T$ model employed in the main text and shown in Fig 5. Black arrows indicate the temperatures where the fit deviates from the data by 2\%.}
\end{figure}

\emph{Detail of fits used in the main text} - Supplementary Figure \ref{fig:WSMFits-allP} reproduces the individual fits shown in Fig.~5 of the main text --- the subfigures plot each pressure point separately for ease of comparison of each fit. Black arrows indicate the temperatures where the fit deviates from the data by 2\%. This is chosen as a consistent comparison of the pressure-dependent temperature range of the fitting function's validity.

\FloatBarrier{}

\FloatBarrier{}

\subsection*{Supplementary Note 6}
\label{subsec:treatmentPriorLitData}
\begin{figure}
\begin{centering}
\includegraphics[width=0.6\textwidth]{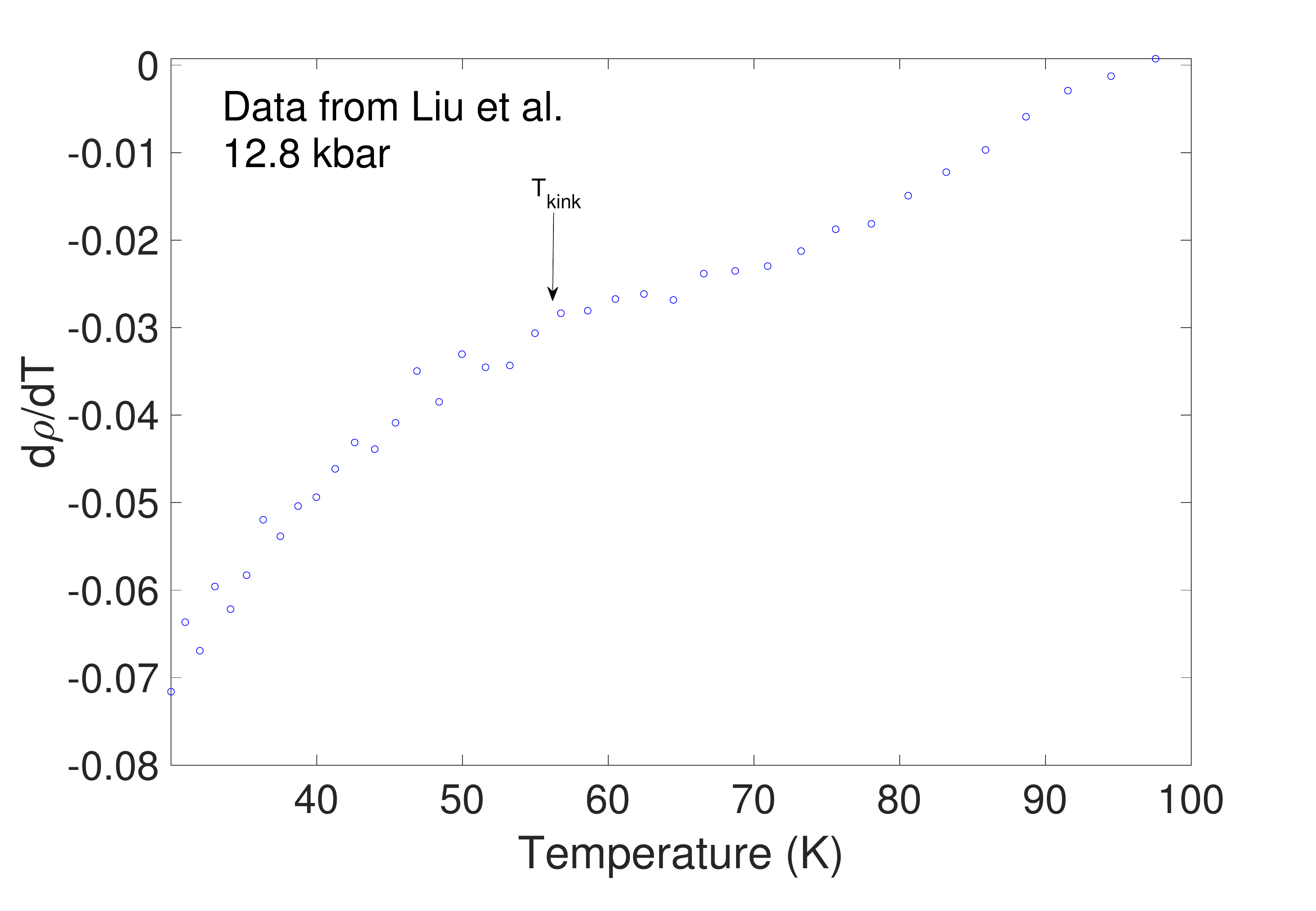}
\par\end{centering}
\begin{centering}
\includegraphics[width=0.6\textwidth]{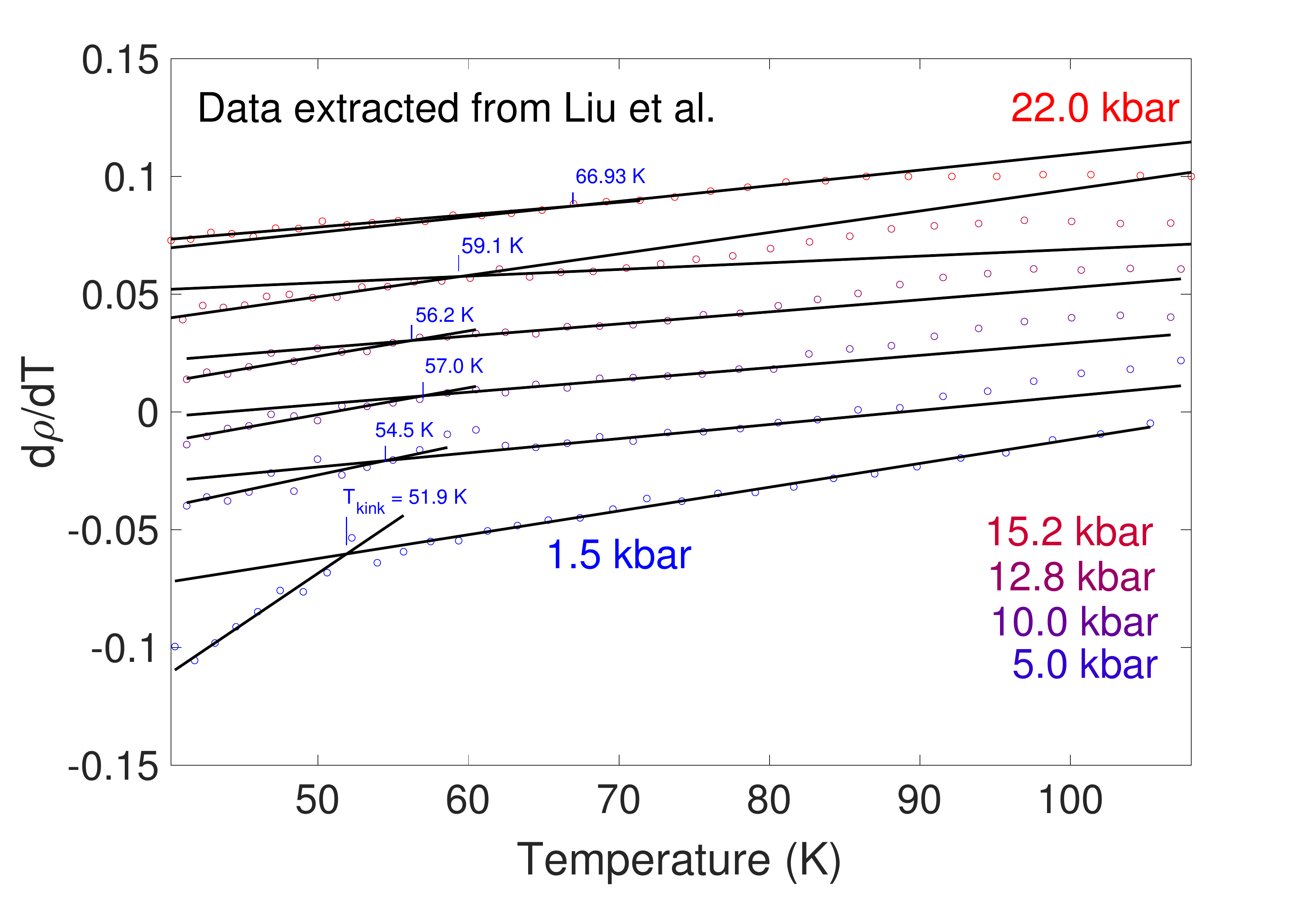}
\par\end{centering}
\centering{}\caption{\label{fig:LiuKinks}Resistivity data extracted from Liu \emph{et al.} and differentiated with respect to temperature. $T_{\mathrm{kink}}$ values are found by fitting straight lines above and below the region of a subtle kink in the gradient, as shown.}
\end{figure}

\emph{Treatment of prior literature transport data} - Supplementary Figure~\ref{fig:LiuKinks} shows the results of digitizing and then differentiating the resistivity data of Liu \emph{et al.} \citep{Liu2018b} with respect to temperature to reveal the subtle feature at $T_{\mathrm{kink}}$ discussed in the main text and plotted in the phase diagram - Figure 9 of the main text.

\FloatBarrier{}
\clearpage{}

\subsection*{Supplementary Note 7}
\emph{Complete magnetotransport data} - For brevity and clarity, the main text of the paper shows only a subset of the complete magnetoresistivity (MR) data collected. For reference, all data, both raw and following symmetrization into background MR, as discussed in the text, are plotted in the following sections. 
Supplementary Note 8 plots the MR at the lowest temperature measured for each pressure.
Supplementary Note 9 shows the MR data at all measured temperatures, for each pressure in turn.
Supplementary Note 10 shows the complete MR dataset symmetrized and binned to extract the single-valued background MR values isolated from the hysteresis effects that accompany them in the raw data.
Supplementary Note 11 shows a comparison of the hysteresis loops reported with those previously seen in other pyrochlore iridate materials.

\subsection*{Supplementary Note 8}
\label{sec:ComparisonBaseTMR}
\begin{figure}
\centering{}\includegraphics[width=0.45\columnwidth]{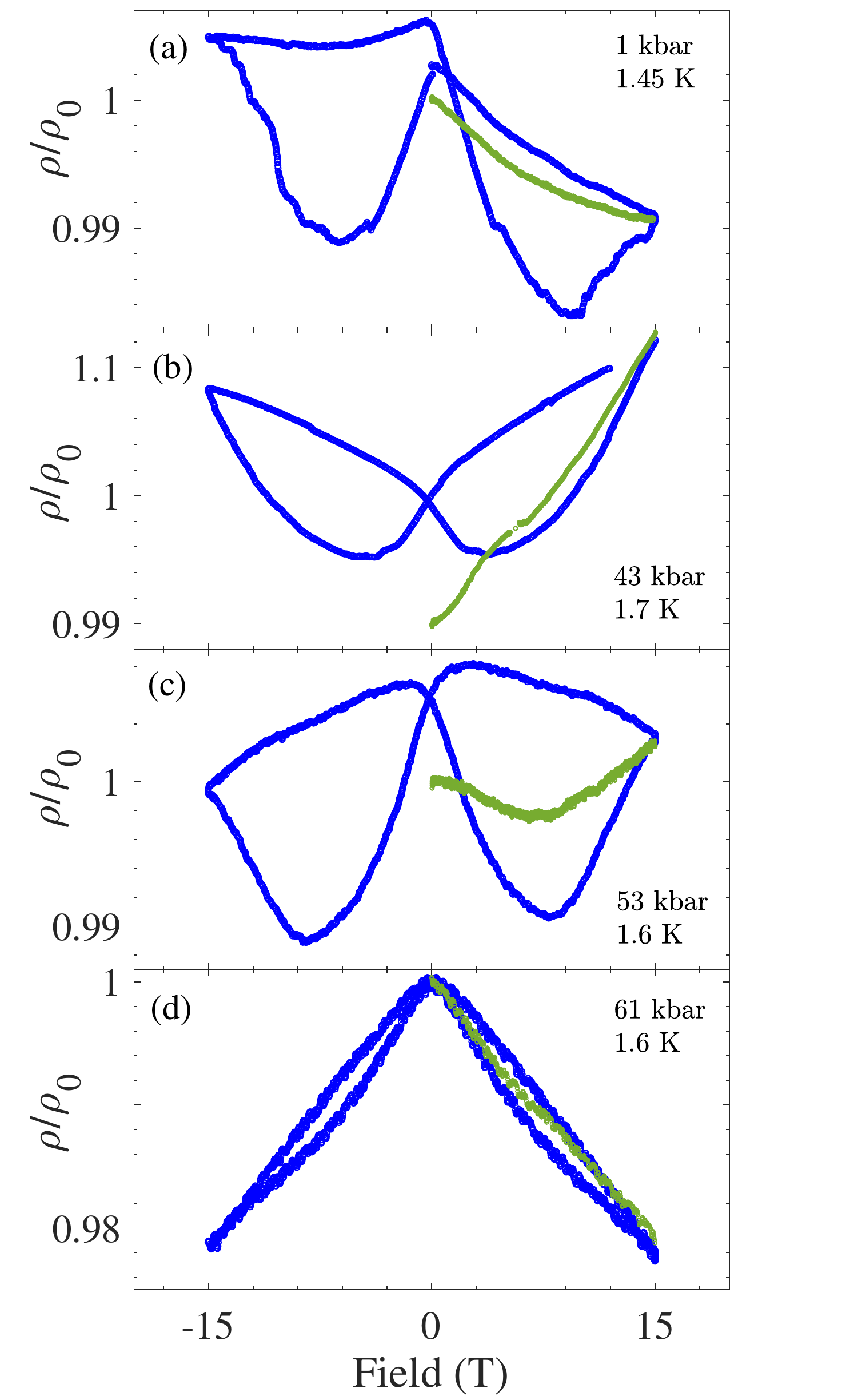}\caption{\label{fig:MR1}Magnetoresistance at fixed 1.6(1)~K temperature,
for increasing applied pressures. Data are normalized to their zero-field
values. In each plot the green points denote the first application
of the field up to +15~T after cooling the sample in zero field.}
\end{figure}

\begin{figure}
\begin{centering}
\includegraphics[width=0.6\columnwidth]{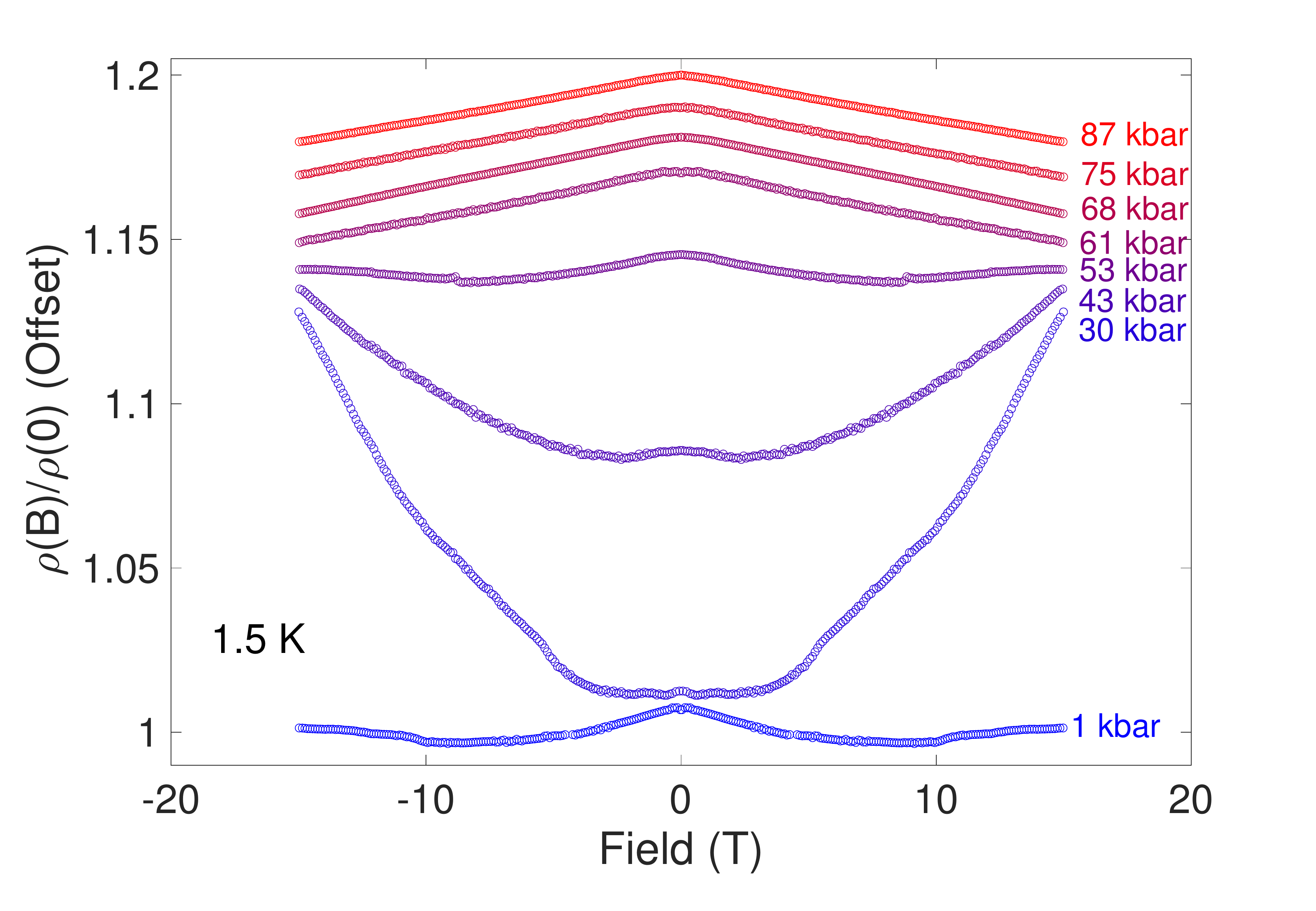}
\par\end{centering}
\centering{}\caption{\label{fig:MR-allP_BaseT}Relative change in effective single-valued
resistivity against magnetic field at each pressure, offset for clarity,
at 1.5(1)~K. Data have been binned and symmetrized as described in the
main text.}
\end{figure}

\emph{Comparison of magnetoresistance data at base temperature for each pressure} - Further to the data presented in the main text, this section plots the raw data (Supplementary Figure \ref{fig:MR1}), and then the symmetrized and binned single-valued MR data (Supplementary Figure \ref{fig:MR-allP_BaseT}), at the lowest temperature measured for various pressures.

\FloatBarrier{}

\FloatBarrier{}
\clearpage{}

\subsection*{Supplementary Note 9}
\label{sec:AllMR}
\begin{figure}
\begin{centering}
\includegraphics[width=0.35\columnwidth]{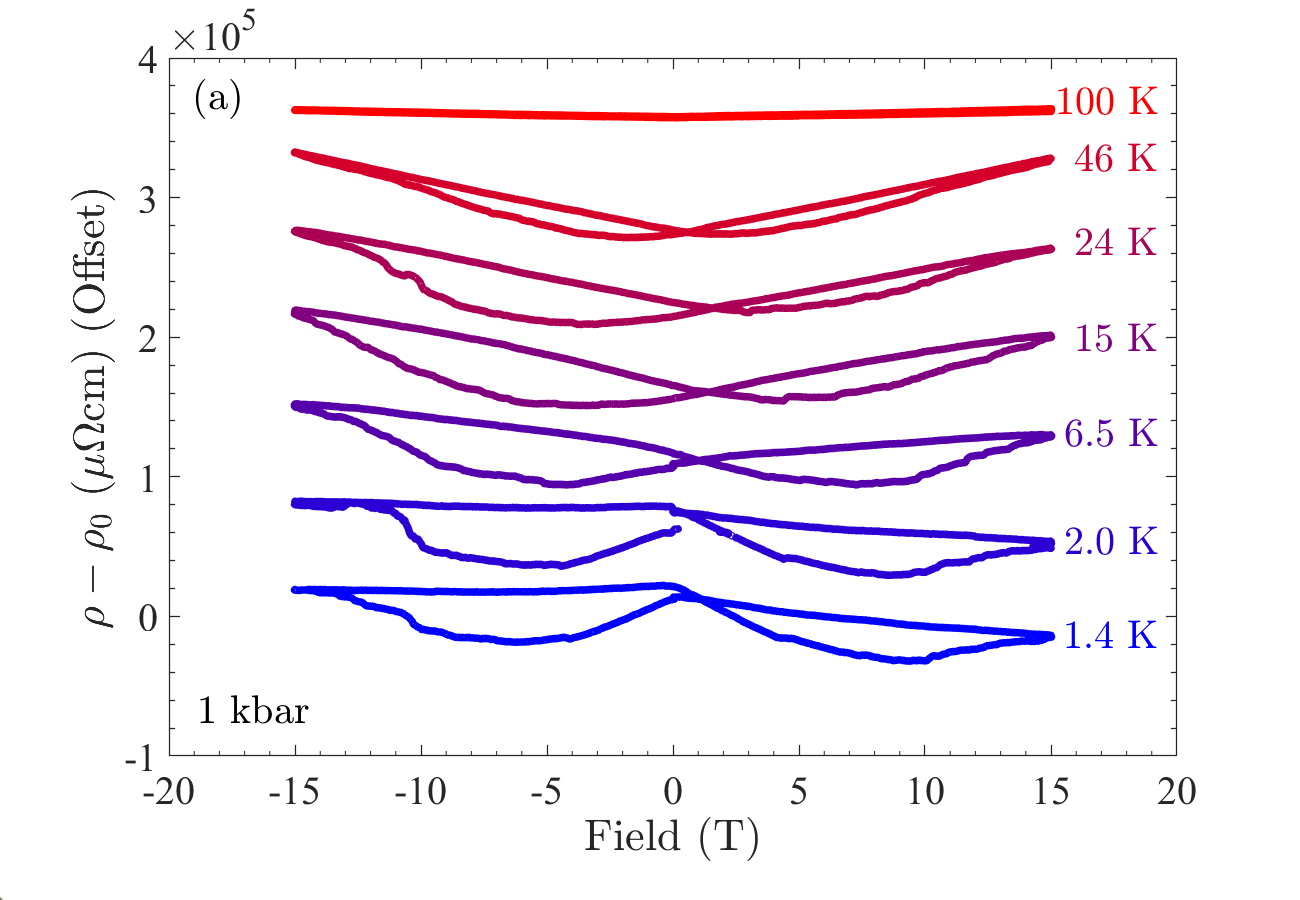}
\includegraphics[width=0.35\columnwidth]{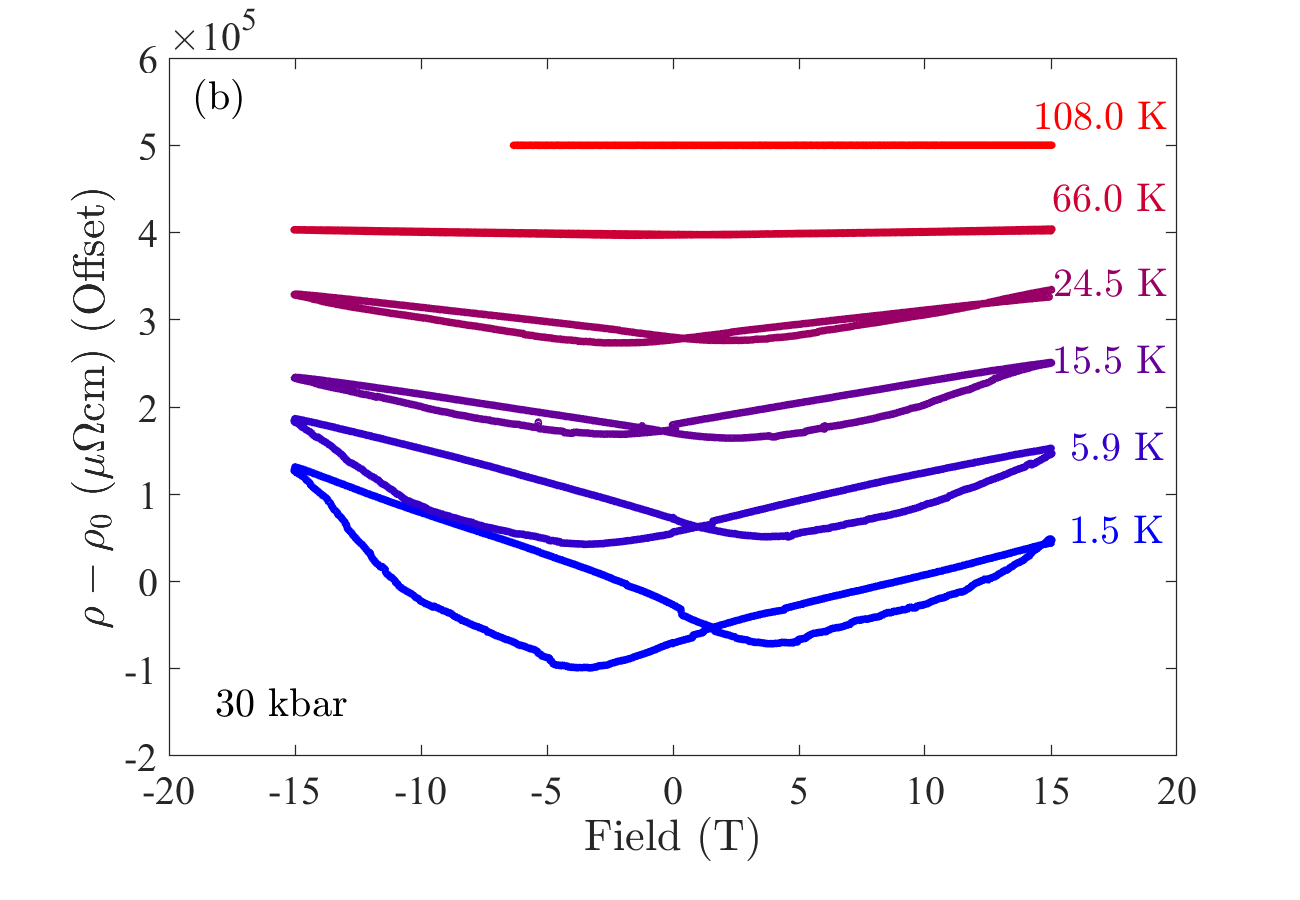}
\includegraphics[width=0.35\columnwidth]{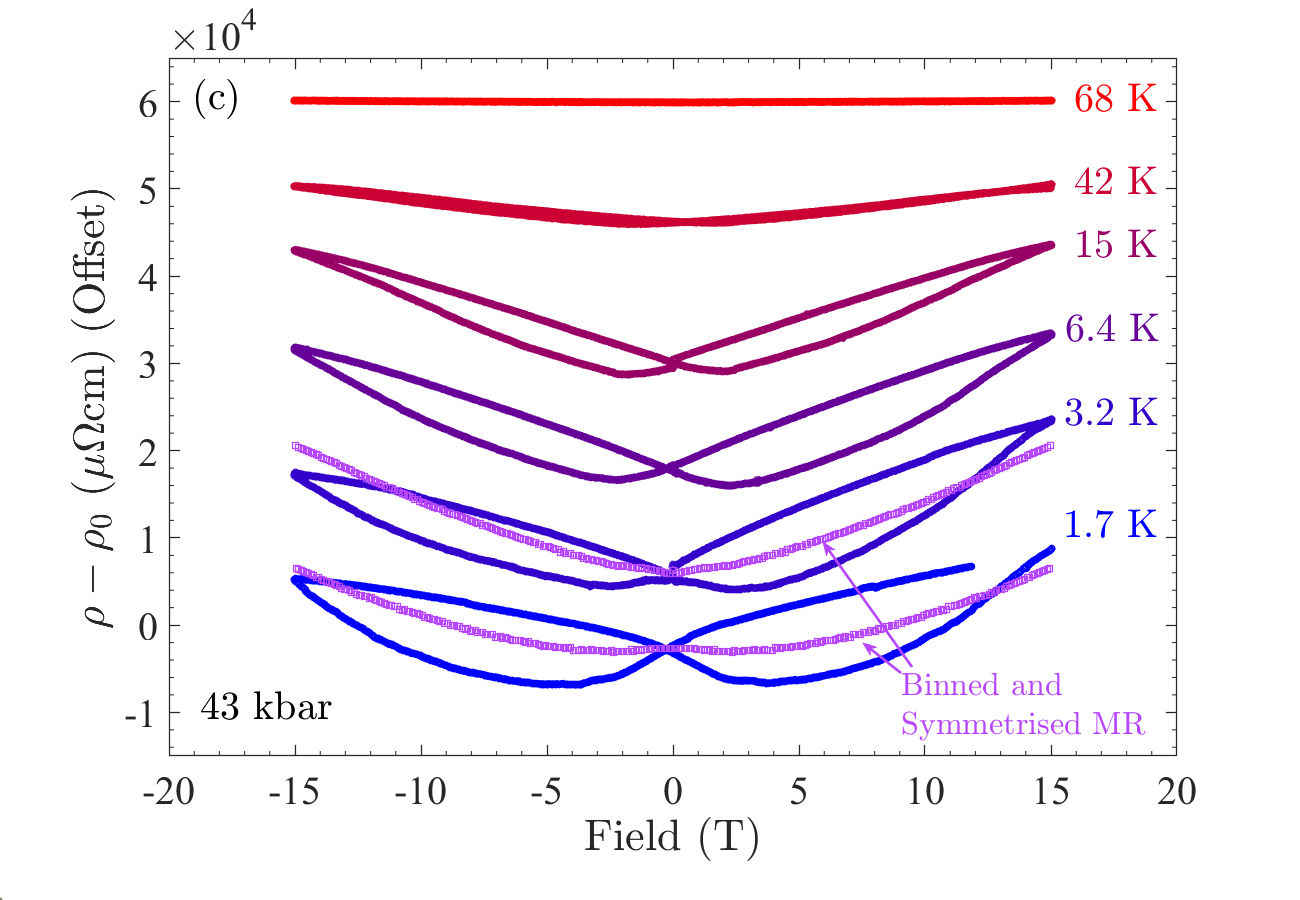}
\includegraphics[width=0.35\columnwidth]{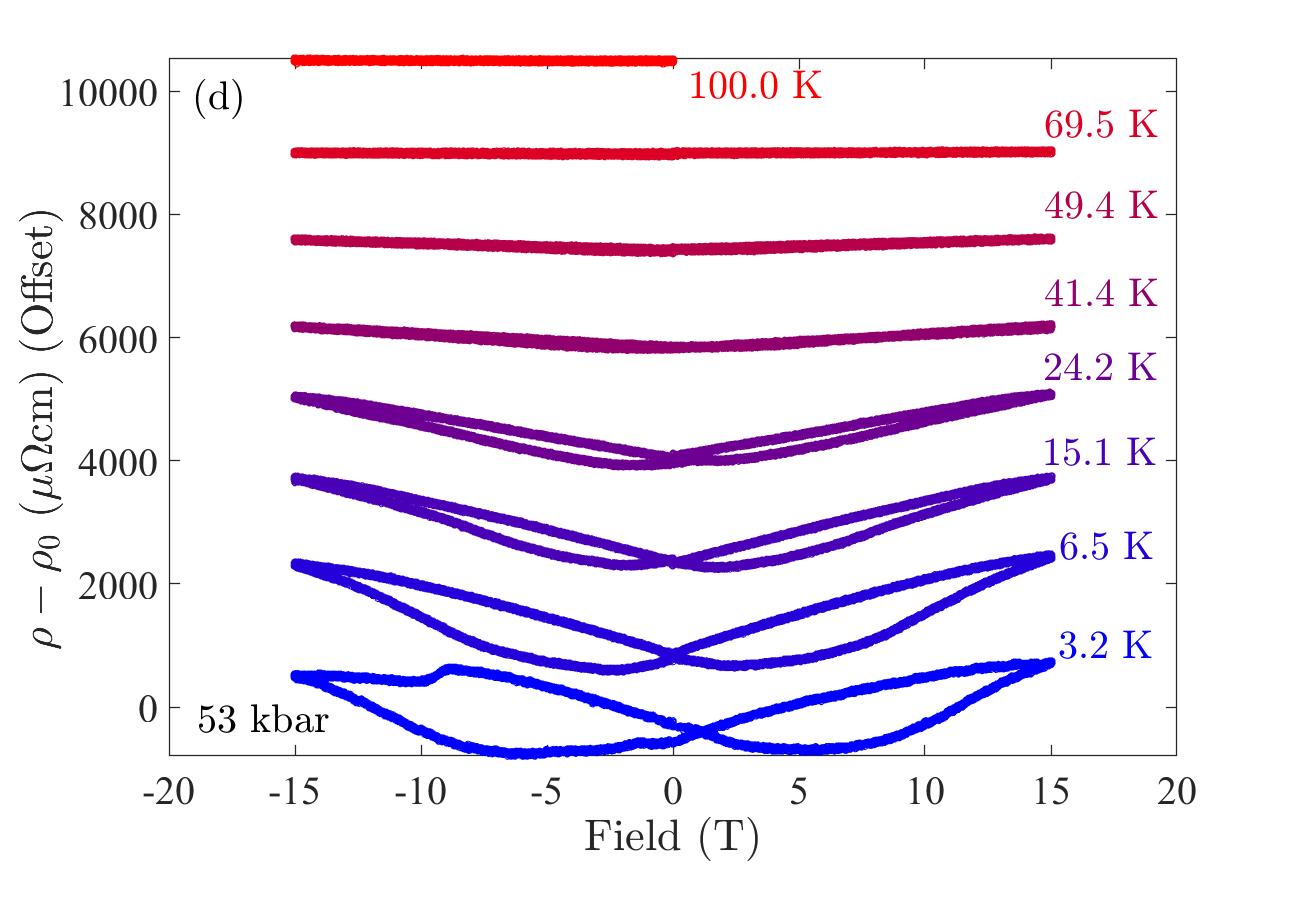}
\includegraphics[width=0.35\columnwidth]{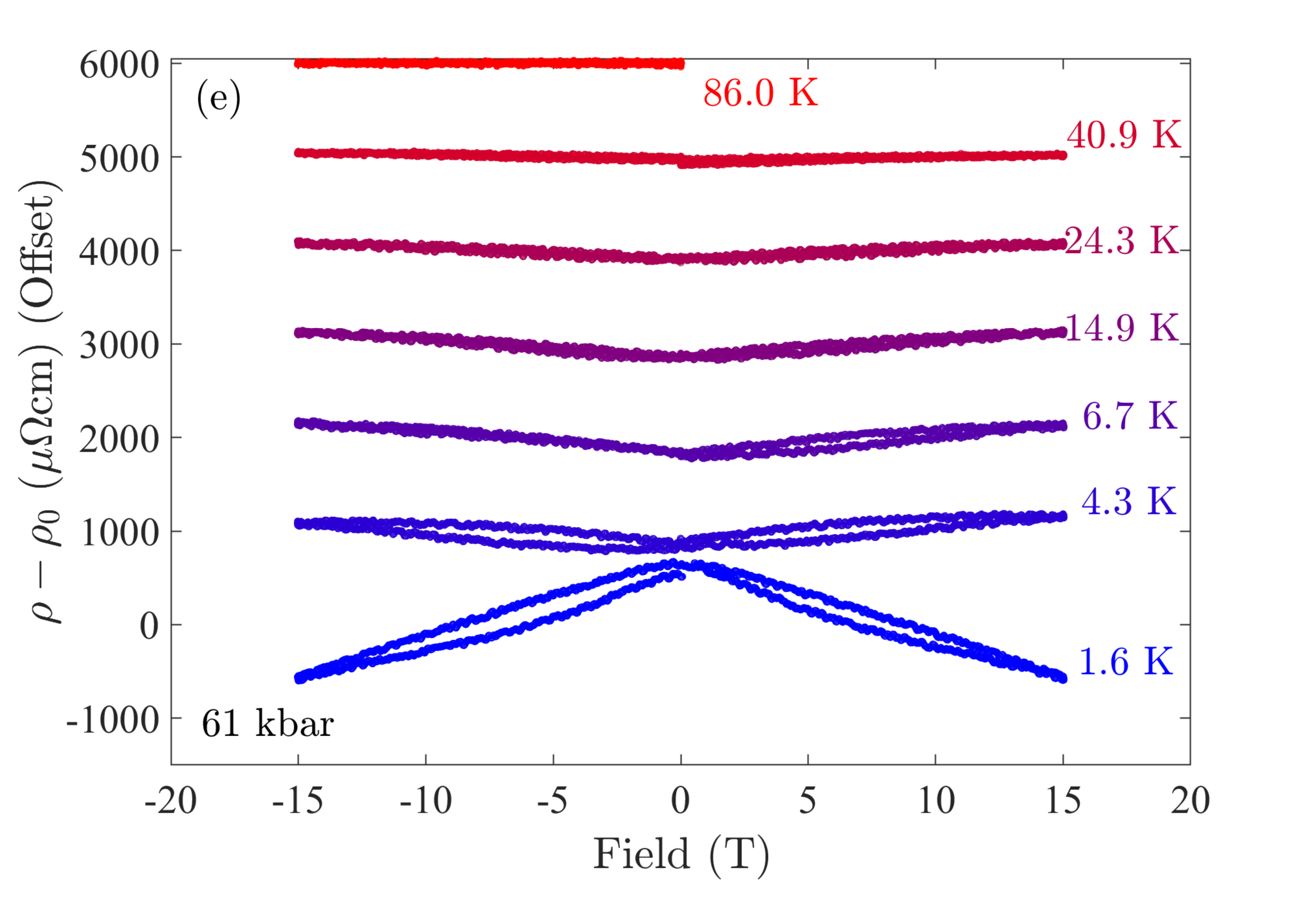}
\includegraphics[width=0.35\columnwidth]{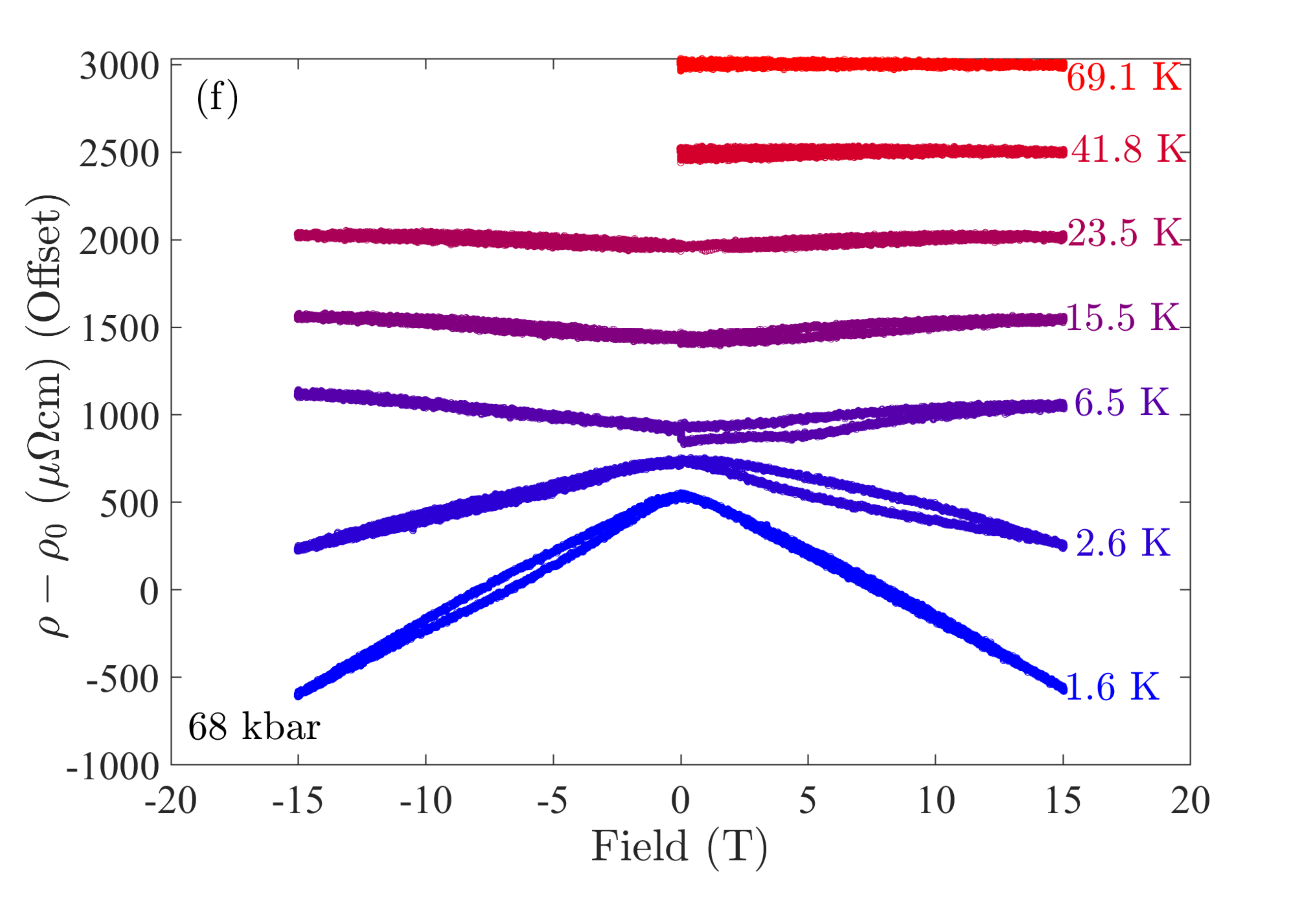}
\includegraphics[width=0.35\columnwidth]{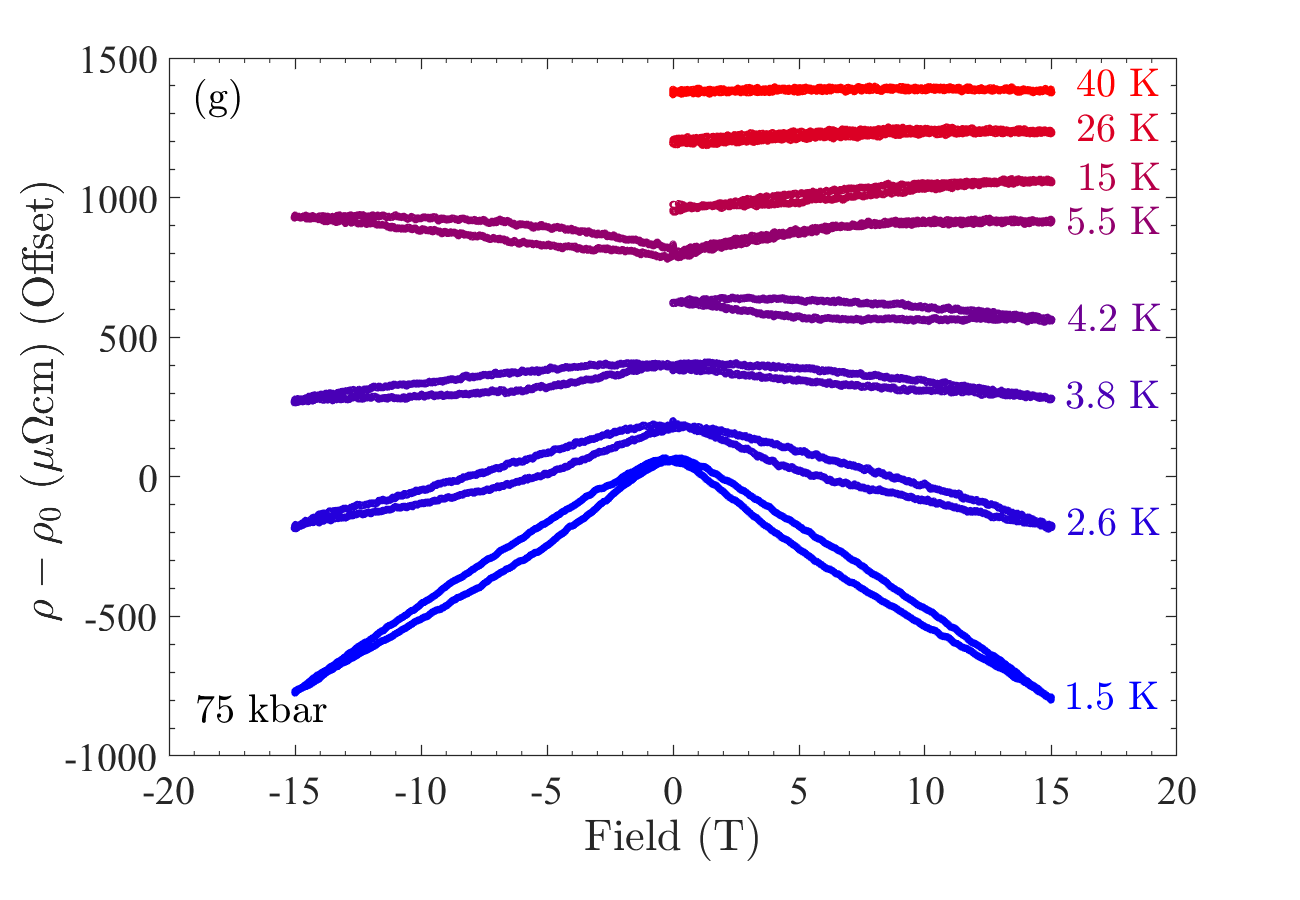}
\includegraphics[width=0.35\columnwidth]{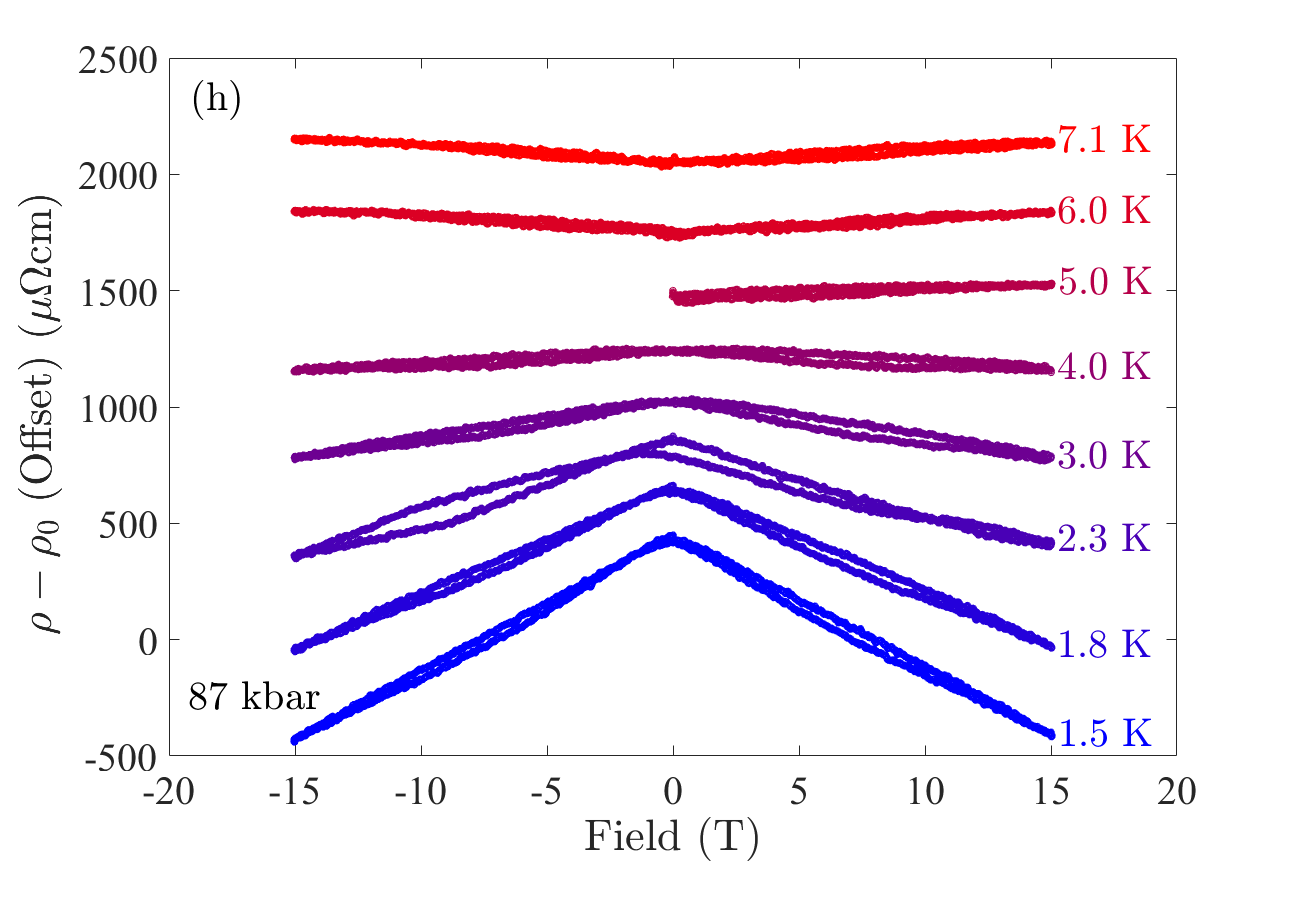}
\par\end{centering}
\centering{}\caption{\label{fig:RvB-allT_allPPanel}Change in resistivity against magnetic field
at each measured temperature, plotted for each pressure measured.}
\end{figure}

\emph{Full magnetoresistance data at each pressure and temperature} - In this section we plot (Supplementary Figure \ref{fig:RvB-allT_allPPanel}) the MR curves at each temperature measured, for each pressure in turn. The data are presented as relative change in MR, $\rho(B) - \rho(B=0)$, and curves are offset for clarity of viewing.

\FloatBarrier{}

\FloatBarrier{}
\clearpage{}

\FloatBarrier{}
\subsection*{Supplementary Note 10} 
\label{Sec:sym_MR_each_pressure}
\begin{figure}
\begin{centering}
\includegraphics[width=0.35\columnwidth]{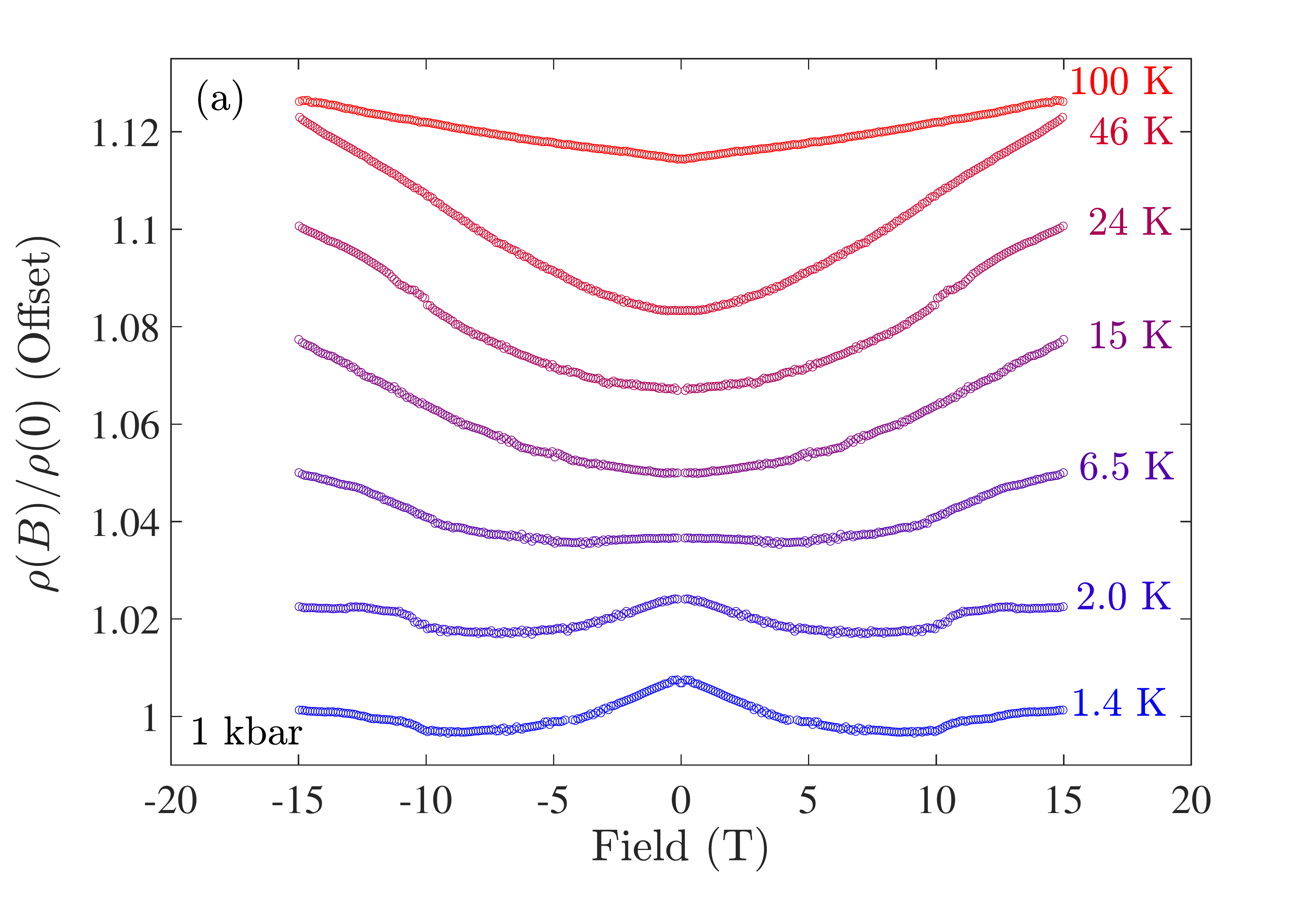}
\includegraphics[width=0.35\columnwidth]{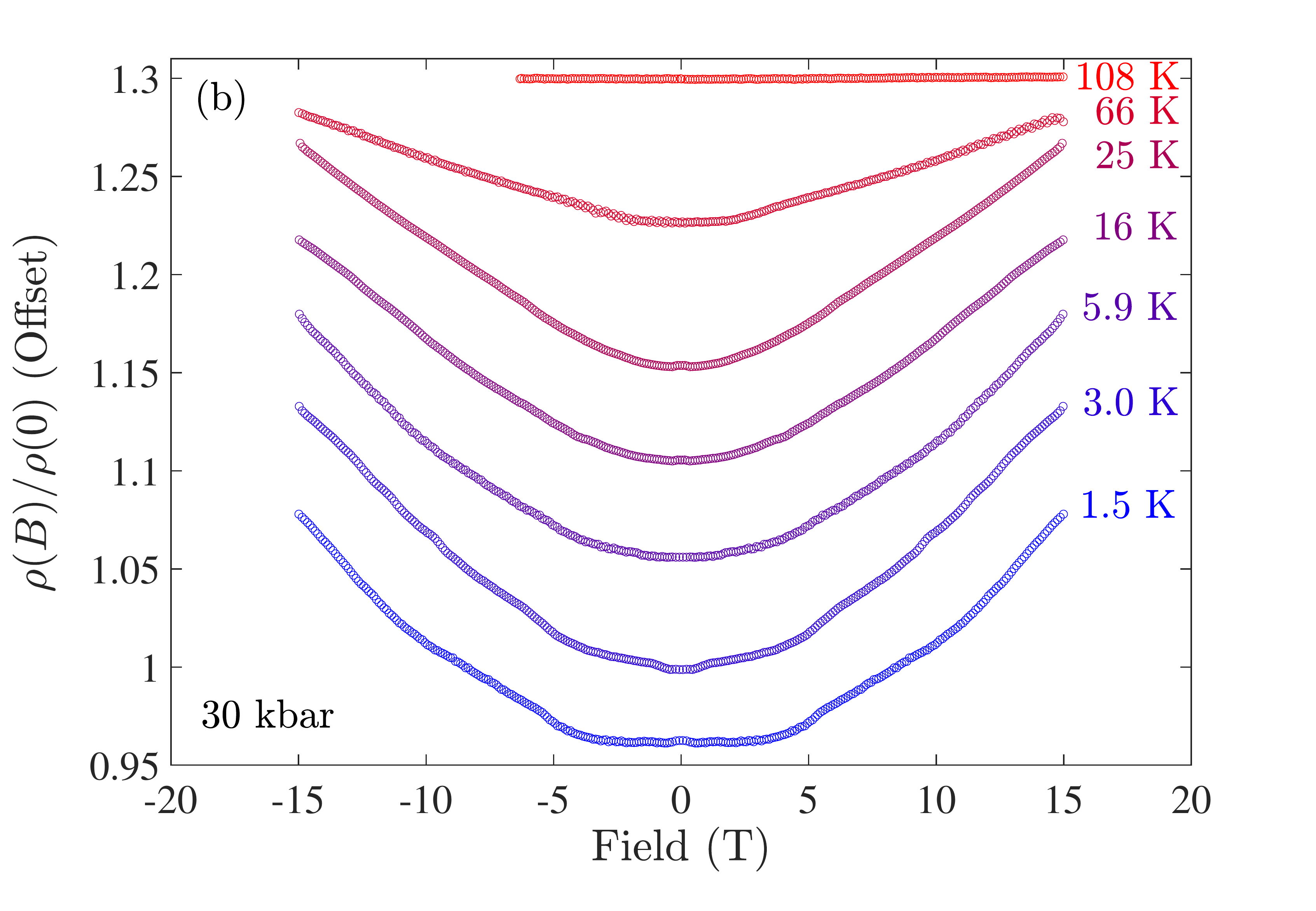}
\includegraphics[width=0.35\columnwidth]{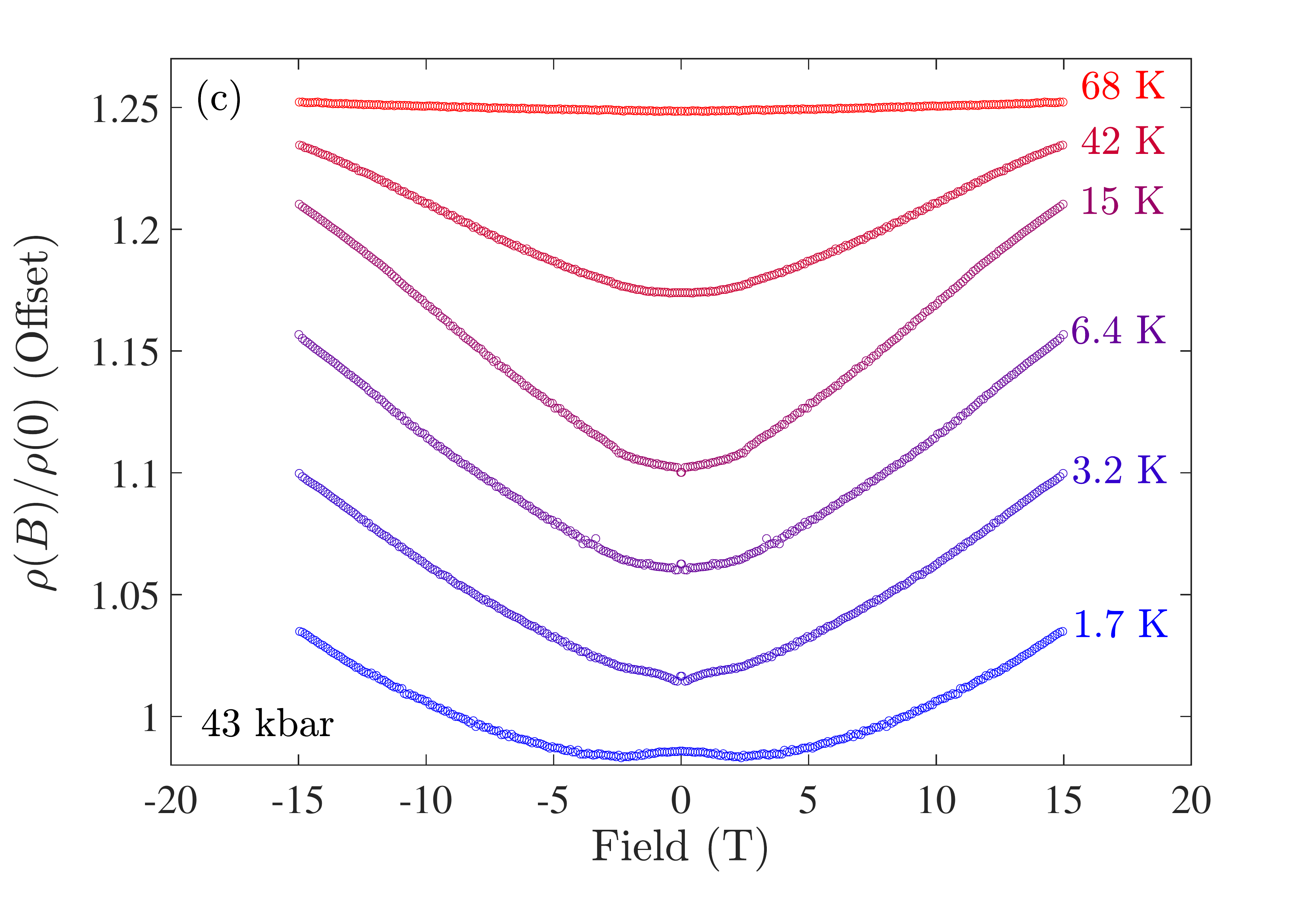}
\includegraphics[width=0.35\columnwidth]{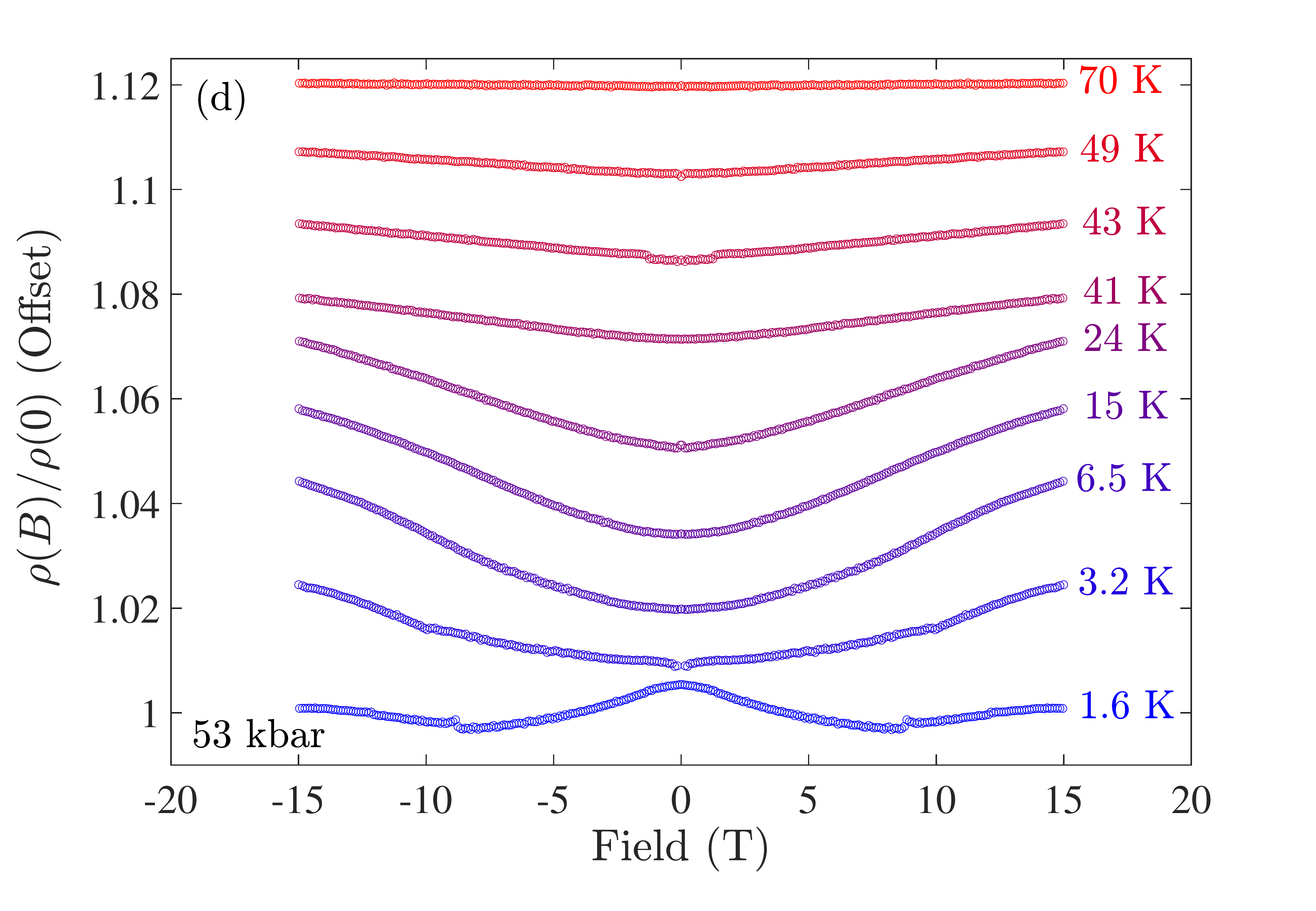}
\includegraphics[width=0.35\columnwidth]{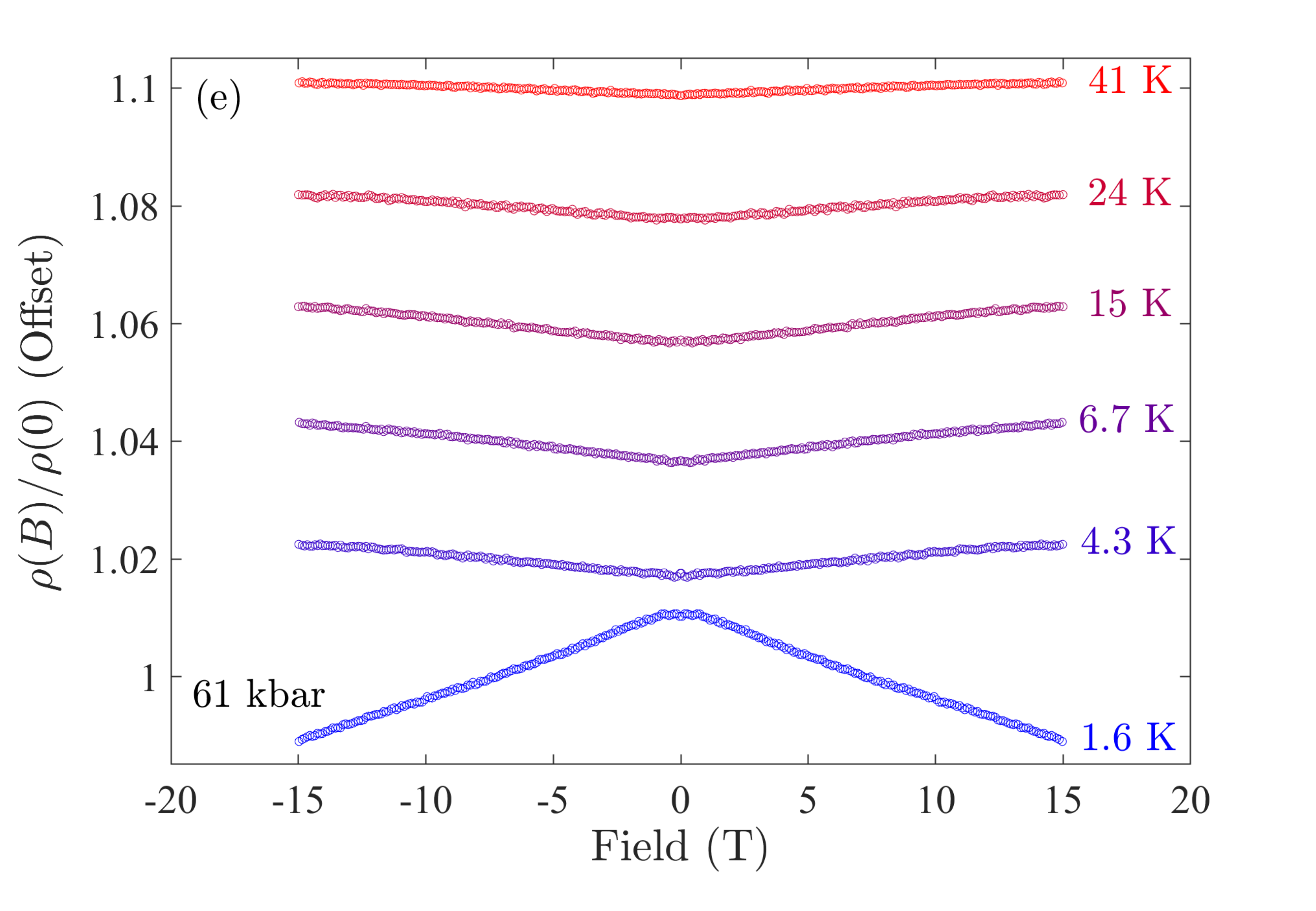}
\includegraphics[width=0.35\columnwidth]{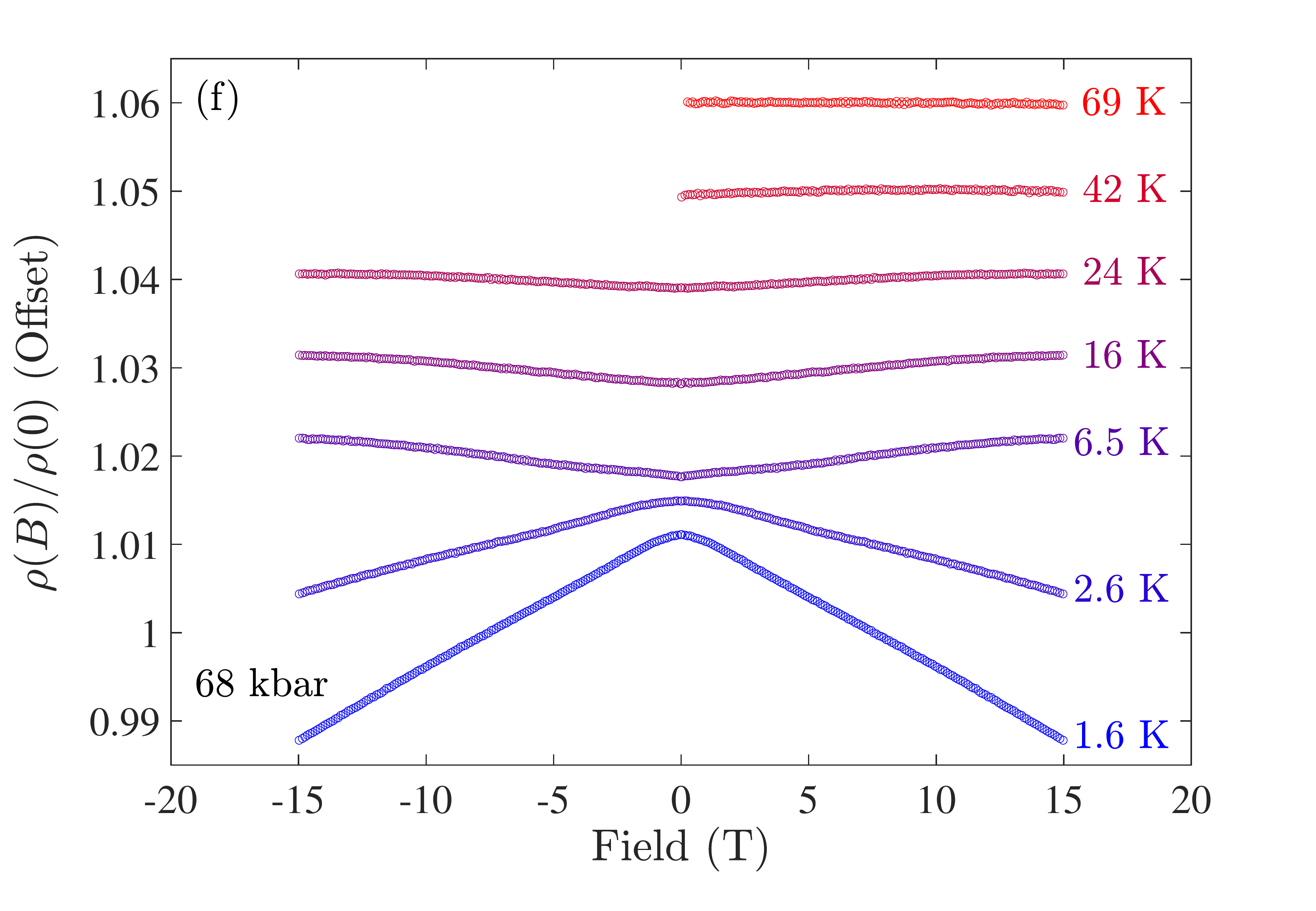}
\includegraphics[width=0.35\columnwidth]{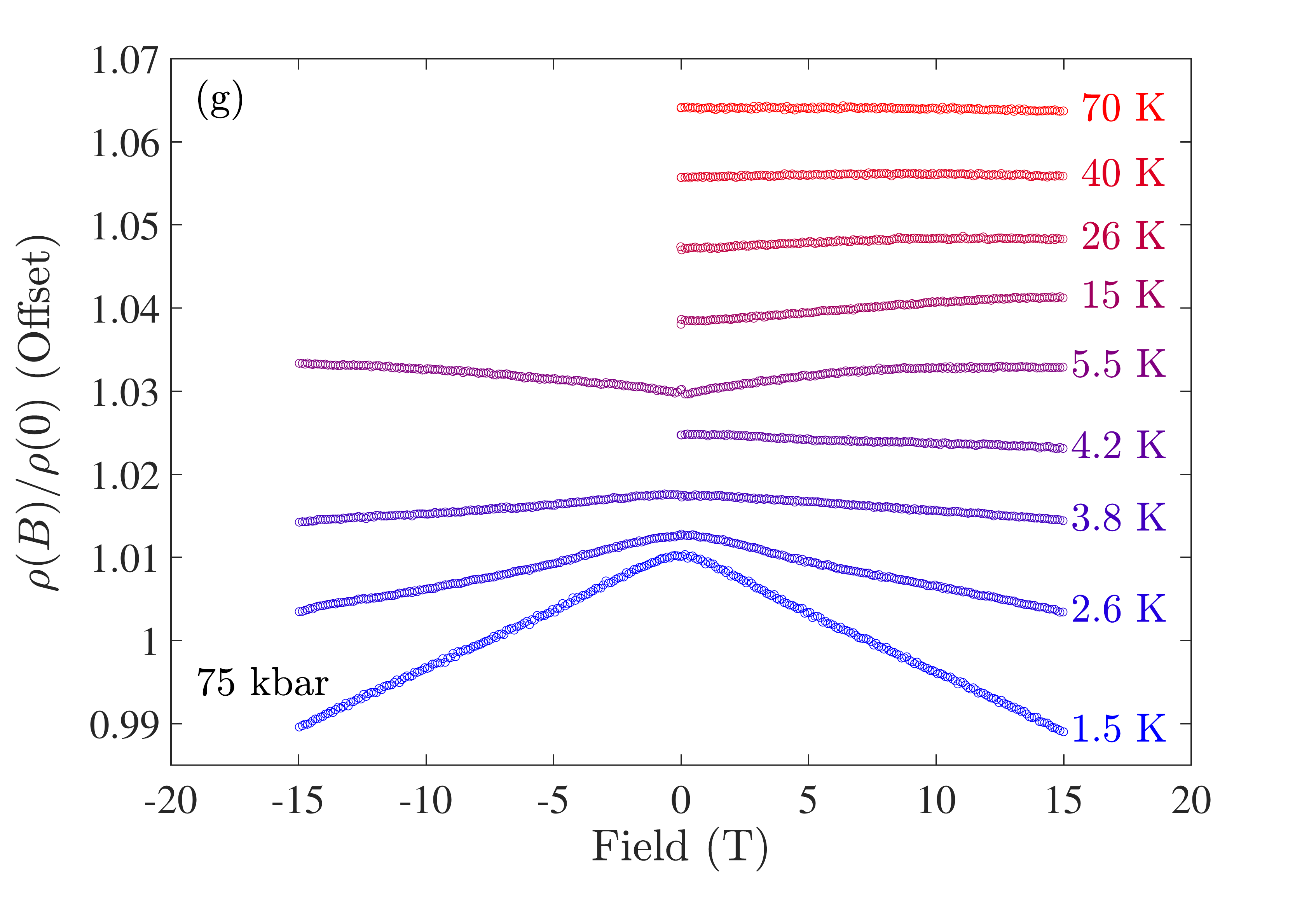}
\includegraphics[width=0.35\columnwidth]{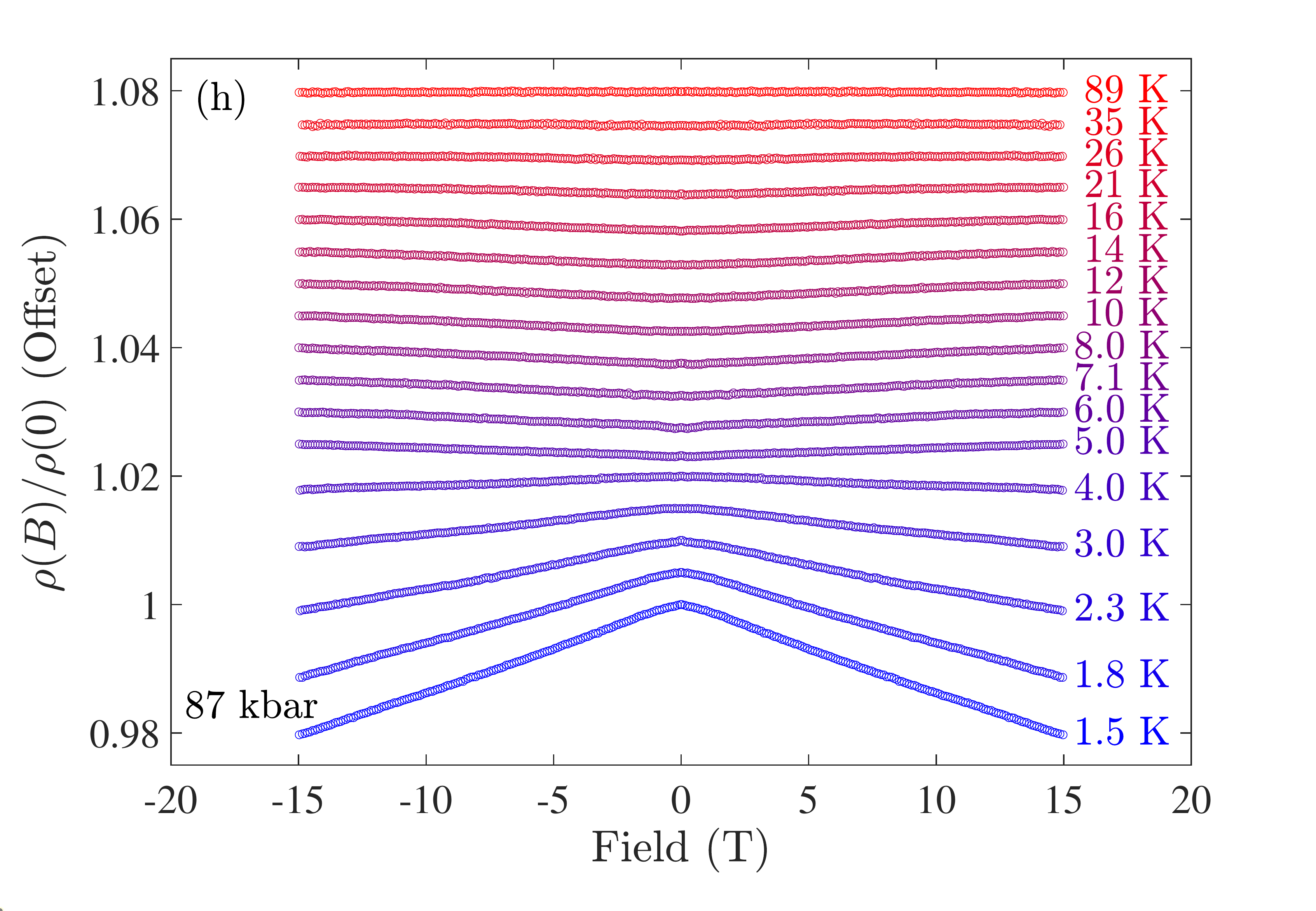}
\par\end{centering}
\centering{}\caption{\label{fig:MR-allT_allP}Relative change in effective single-valued resistivity
against magnetic field at various fixed temperatures, offset for clarity, at increasing pressures for each panel. Data have been binned and symmetrized as described in the main text.}
\end{figure}

\emph{Symmetrized magnetoresistance curves at each pressure} - In this section we plot (Supplementary Figure \ref{fig:MR-allT_allP}) the symmetrized and binned single-value background MR curves at each temperature measured (as discussed in the main text), for each pressure in turn. The data are presented as relative fractional change of MR, $\rho(B) / \rho(B=0)$, and are offset from their $B=0$ values of 1, for clarity of viewing.

\clearpage{}

\subsection*{Supplementary Note 11}
\emph{Comparison of magnetoresistance hysteresis with other pyrochlore iridates} - For ease of comparison with existing results in the literature, we plot below in Supplementary Figure \ref{fig:RvB-NdHoSm} equivalent magnetoresistance hysteresis loops for Sm$_2$Ir$_2$O$_7$, Nd$_2$Ir$_2$O$_7$, and Ho$_2$Ir$_2$O$_7$, for equivalent temperatures and orientations.
\FloatBarrier{}

\begin{figure}
\begin{centering}
\includegraphics[width=0.5\columnwidth]{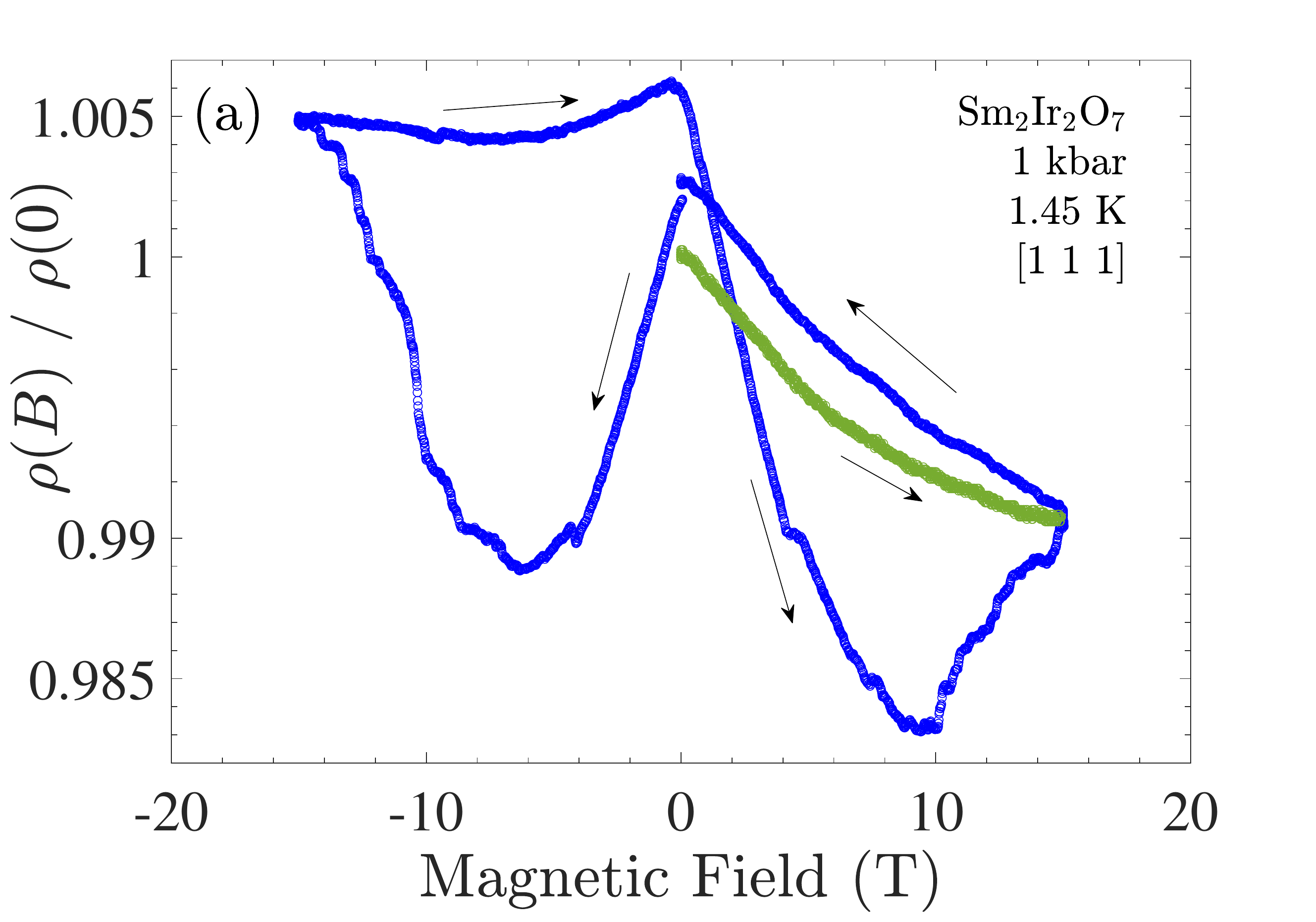}
\includegraphics[width=0.5\columnwidth]{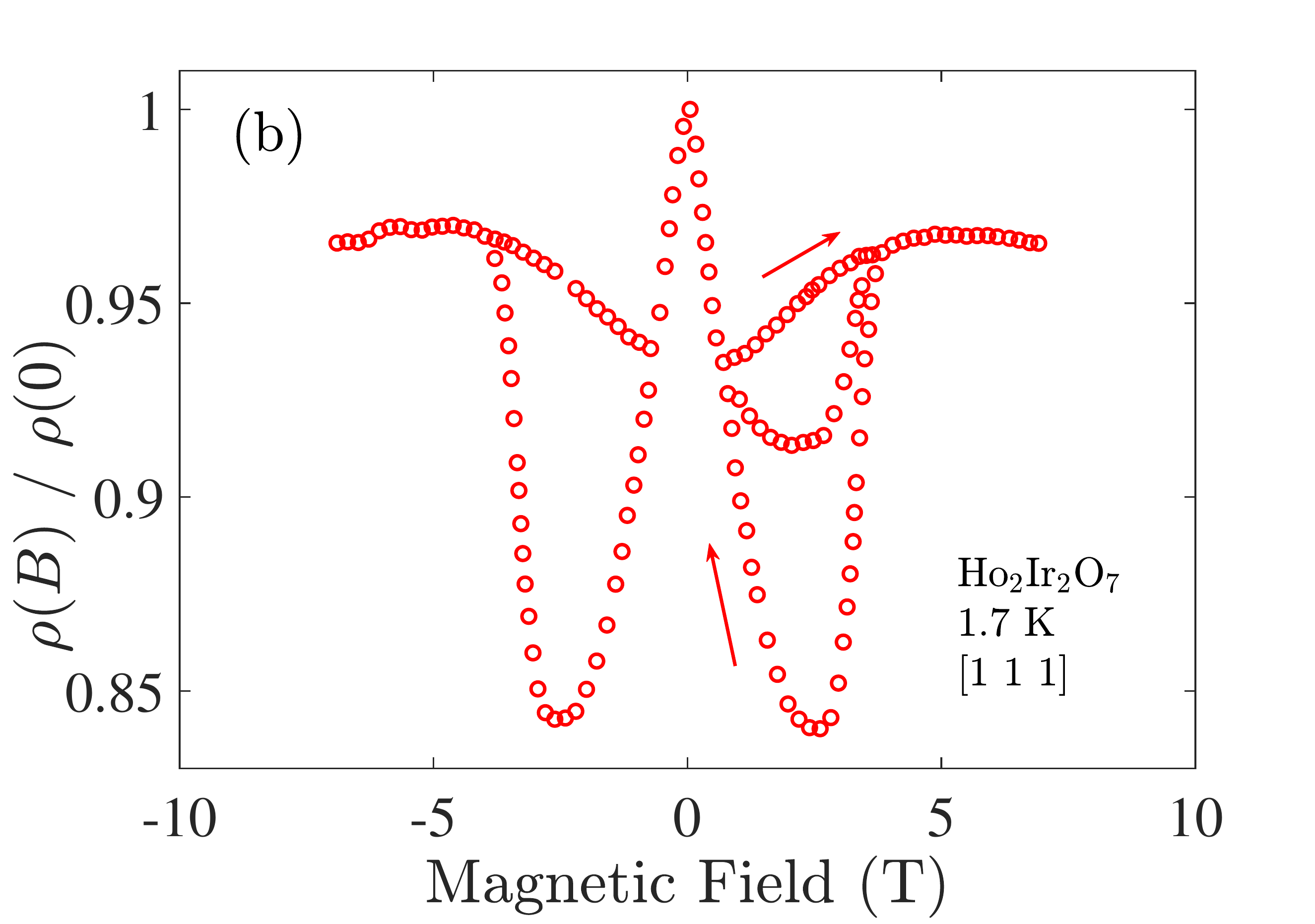}
\includegraphics[width=0.5\columnwidth]{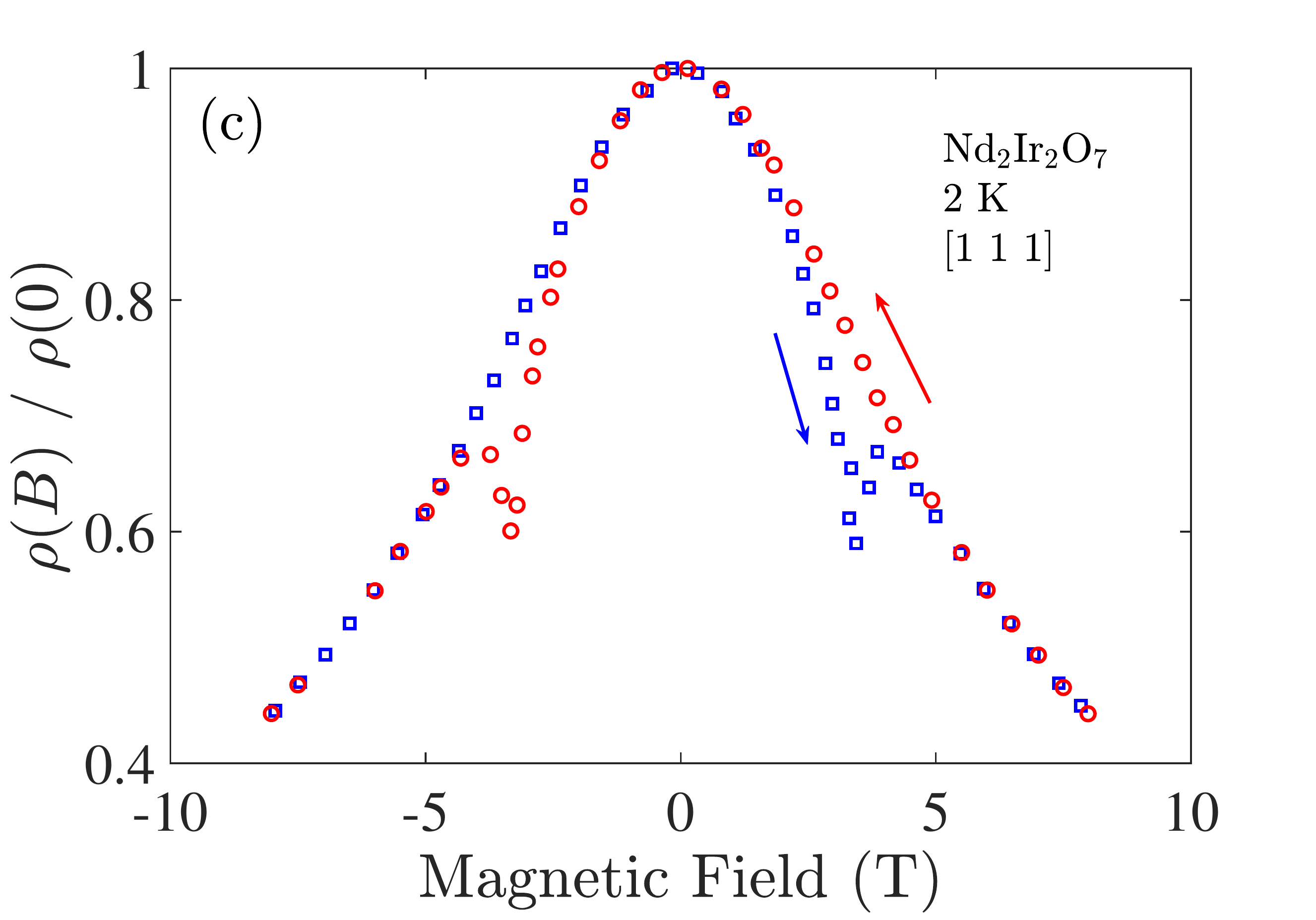}
\par\end{centering}
\centering{}\caption{\label{fig:RvB-NdHoSm}Relative change in resistivity
against magnetic field at each base temperature measured (approx 2 K) for (a) Sm$_2$Ir$_2$O$_7$ (this work), (b) Ho$_2$Ir$_2$O$_7$ \citep{Pearce2022} and (c)  Nd$_2$Ir$_2$O$_7$ \citep{Ueda2015a}.}
\end{figure}

\FloatBarrier{}
\clearpage{}

\subsection*{Supplementary Note 12}
\begin{figure*}
\begin{centering}
\includegraphics[width=0.9\columnwidth]{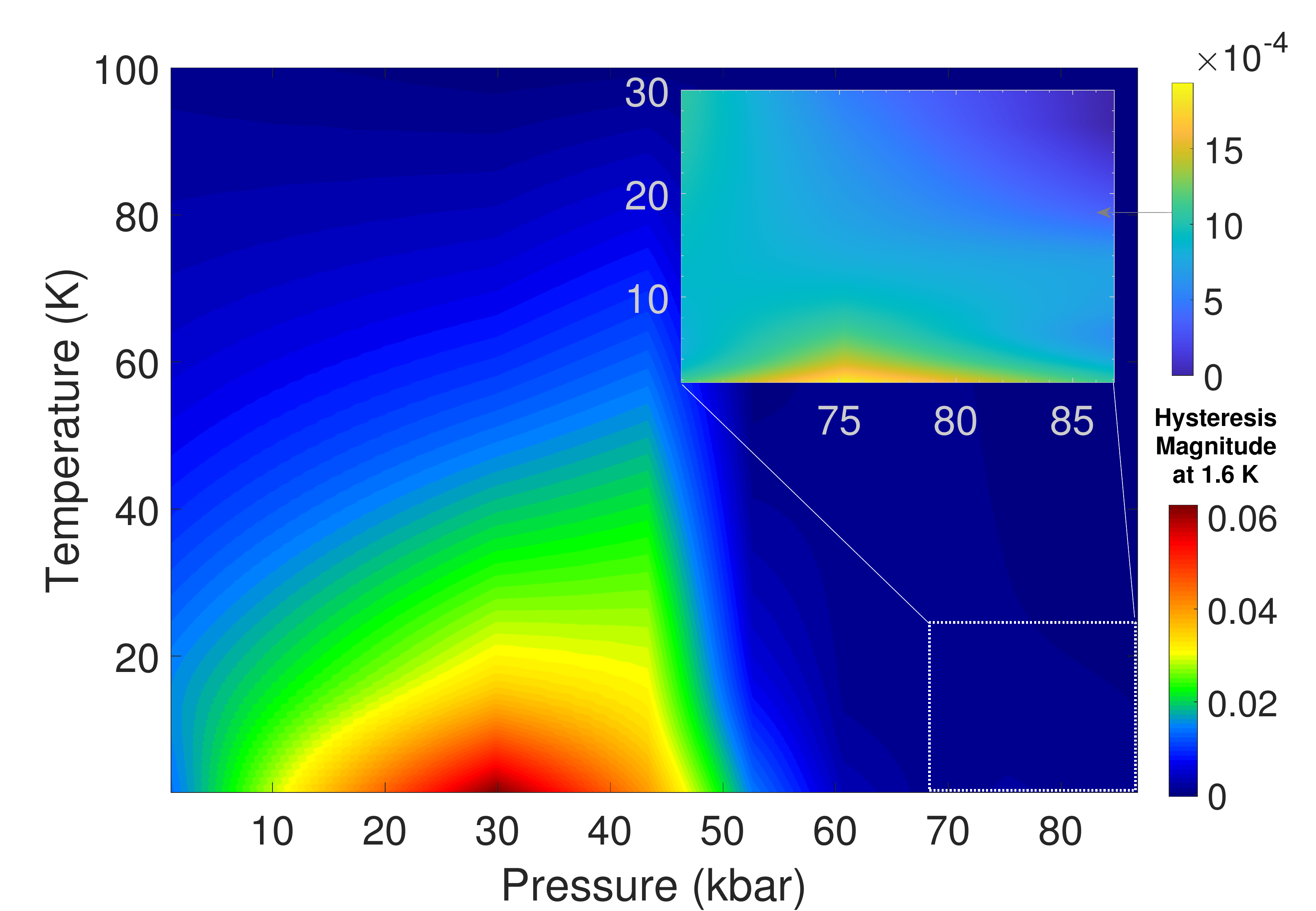}
\par\end{centering}
\centering{}\caption{\label{fig:colormap1-4}Color plot showing the evolution of the magnitude
of the hysteresis in magnetoresistance with pressure and temperature
in Sm$_2$Ir$_2$O$_7$. The color scale indicates the maximum difference
between `up' and `down' field sweeps, divided by the $\rho(B=0)$ zero-field
resistivity at that point. The inset is an enlargement of the high-pressure -
low-temperature region, where a different, weaker form of hysteresis appears.}
\end{figure*}

\begin{figure}
\begin{centering}
\includegraphics[width=0.45\columnwidth]{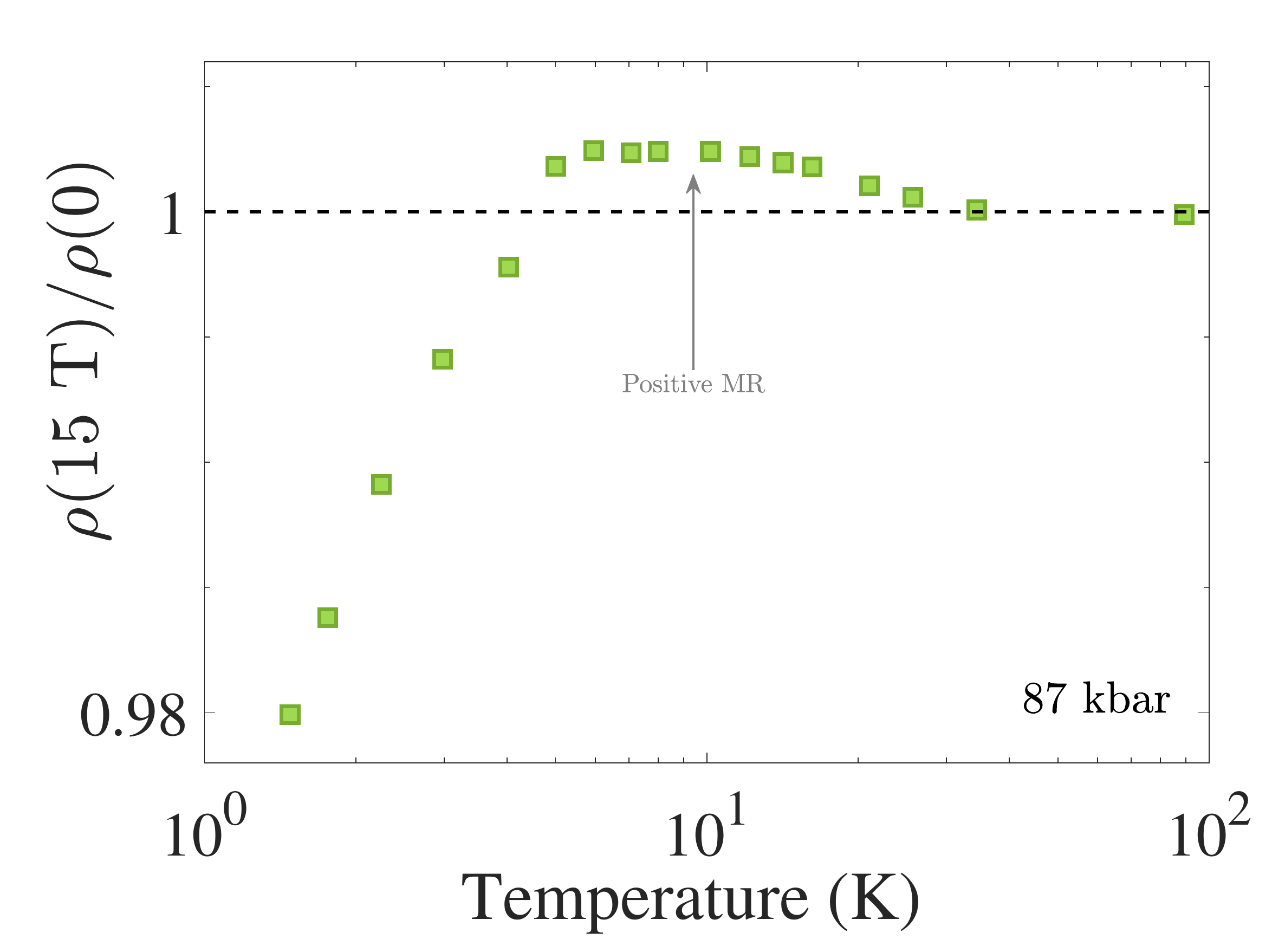}
\par\end{centering}
\centering{}\caption{\label{fig:MR-highpTDep}Maximal MR ratio as a function of temperature at the highest pressure, 87(2)~kbar (See Supplementary Figure \ref{fig:MR-allT_allP}(h) for source data). A crossover from negative to positive MR can be seen between 4 and 5~K.}
\end{figure}

\emph{Magnetotransport phase diagram details and additional data} - This section extends the data shown in the phase diagram, Fig. 9, in the main body of the paper. Supplementary Figure~\ref{fig:colormap1-4} charts the magnitude of the hysteresis in the MR as it varies with temperature and pressure. The `hysteresis closes' points shown in Fig. 9 in the main text are taken from the values in this figure. To create a uniform colormap of MR values spanning the 2--125 K temperature range shown in Fig.~9, interpolation and extrapolation of the discrete data points collected was performed. Each data point represents a sweep of field from 0 to 15 T at a fixed measurement temperature. These were not carried out at the same values of temperatures at every pressure, so for each pressure point a spline was generated to interpolate data values at intermediate temperatures. A Bezier spline was adjusted to pass through all the individual data points of $\rho(15~\mathrm{T})$ / $\rho(0)$ values plotted against temperature, with smooth variation.  Once each spline was generated, they were sampled in evenly separated intervals to comprise the temperature axis. For the horizontal pressure axis, the discrete pressure values of the experiment were used, with interpolated shading selected to give the coloring between them. Raw data used in the generation of these splines are reflected in Supplementary Figure~\ref{fig:MR-allT_allP} and in the example Supplementary Figure~\ref{fig:MR-highpTDep}.

\FloatBarrier{}

Supplementary Figure~\ref{fig:MR-highpTDep} makes clear the change in sign of overall background (i.e. symmetrized and binned) magnetoresistance, by taking the values of the data shown in Supplementary Figure~\ref{fig:MR-allT_allP}(h), and plotting their values at the highest field, 15 T, against temperature. The region of weak positive magnetoresistance at the highest pressures in Fig. 9 of the main paper becomes even clearer here. The MR starts negative at the lowest temperatures, crosses over into positive (greater than 1 in MR ratio) between 4 and 5 K, then gradually dies away into a flat MR ($\rho(15~\mathrm{T})$ / $\rho(0)$  = 1) at higher temperatures. This trend is easily seen in the high-pressure MR data presented in Supplementary Figure~\ref{fig:MR-allT_allP}(e-h).

\FloatBarrier{}
\clearpage

\subsection*{Supplementary Note 13}

\emph{Crystal electric field calculations} - In order to begin to model and understand the magnetization properties of Sm$_2$Ir$_2$O$_7$, we looked at the single-ion behavior. Using crystallographic data and a point charge description, we assembled a crystal electric field (CEF) Hamiltonian for a Sm atom in its local environment. The procedure outlined in the following paragraphs allowed us to write the CEF Hamiltonian as
 \begin{equation}\label{H_CEF}
 \begin{aligned}
 \mathcal{H} &= B^{0}_{2} \widehat{O}^{0}_{2} + B^{0}_{4} \widehat{O}^{0}_{4} + B^{3}_{4} \widehat{O}^{3}_{4} + B^{0}_{6} \widehat{O}^{0}_{6} + B^{3}_{6} \widehat{O}^{3}_{6} + B^{6}_{6} \widehat{O}^{6}_{6} \text{.}
 \end{aligned}
 \end{equation}
 \noindent 

 The Hamiltonian, energy levels and eigenstates in the local reference frame customary for the pyrochlore lattice are given in the section below. 
 We find a ground state (GS) doublet largely spanned by the vectors $\vert J, m_J \rangle = \vert 5/2, \pm 3/2 \rangle$, with an energy gap $\Delta_{\rm CEF}$ to the next levels of approximately $11.73$~meV (namely, $136$~K). 
 These results are in agreement with earlier work on Sm$_2$Ti$_2$O$_7$~\cite{Mauws2018}.

 For temperatures and Zeeman energies that are small compared to the gap $\Delta_{\rm CEF}$, the maximum magnetic moment expected from the GS $m_J=\pm3/2$ doublet is $\mu = g_J J_z = 3/7 \simeq 0.43$~$\mu_B$, using the Land\'{e} factor $g=2/7$. However, one ought to account for possible anisotropic effects due to the structure of the CEF Hamiltonian. 
 These are most effectively illustrated by plotting the magnetization vectors computed as thermal expectation values at $0.1$~K from the CEF Hamiltonian after applying a Zeeman contribution, $H_{\rm Zeeman} = \mu_B g_J \vec{J}\cdot\vec{h}_{\rm ext}$, due to an external magnetic field $\vec{h}_{\rm ext}$ of $0.5$~T, of random orientation chosen uniformly on the unit sphere (see Supplementary Figure~\ref{fig:anisotropy}, left panel). 

 We observe a strong easy-axis anisotropy, whereby Sm develops a magnetization almost purely along the local $[1\,1\,1]$ direction, and we find that this behavior persists in applied fields up to $\sim 20$~T and temperatures up to $\sim 10$~K. 
 At correspondingly low temperatures and small applied fields, we therefore expect to be able to approximate reasonably well the single-ion behavior by considering only the ground state doublet and modelling it as a classical Ising degree of freedom along the local $[111]$ axis, of magnitude $0.43$~$\mu_B$. 

 As the applied field is increased, the off-easy-axis components become more relevant, as illustrated in $50$~T in the right panel of Supplementary Figure~\ref{fig:anisotropy}. 
 Given that the magnetic moment in the CEF GS doublet is not the maximal allowed for a $J=5/2$ atom, when a strong applied field tilts the moment away from the easy axis, it also increases its magnitude above $0.43$~$\mu_B$ (towards the maximal allowed value for $J_z=5/2$: $\mu = g_J J_z = 5/7 \simeq 0.71$~$\mu_B$). Indeed, we find that the $M(H)$ behavior of this material in very high magnetic fields, shown in Fig~2 of the main text, cannot be captured by an effective model of Ising moments pointing along the local $[1\,1\,1]$ axes and limited to 0.43$\mu_B$, nor by a model of isotropic Heisenberg spins of fixed magnitude. Our findings are consistent with related pyrochlore systems, such as Sm$_2$Ti$_2$O$_7$~\citep{Malkin2010,Singh2008,Mauws2018, PecanhaAntonio2019}, where the Sm spins are constrained to lie along their local $\left[1\,1\,1\right]$ axes due to CEF splitting, adopting a doublet ground state with mostly $m_J=\pm3/2$ and four non-colinear axes in each tetrahedron.

 We note that in our crystal electric field calculations, we adopted a point charge model (PCM) \cite{Hutchings1964} using Steven's formalism \cite{Stevens1952}, which allowed us to write the CEF Hamiltonian as an expansion of CEF parameters $B^{m}_{n}$ and Stevens operators $\widehat{O}^{m}_{n}$, with $n$ and $m$ based on the symmetry of the environment that surrounds the magnetic ion. In the PCM formalism, the electrostatic potential is approximated by a sum over the Coulomb potentials from the surrounding ions. This allows us to calculate the CEF parameters 
 \begin{figure}
 \centering
 \includegraphics[width=0.4\textwidth]{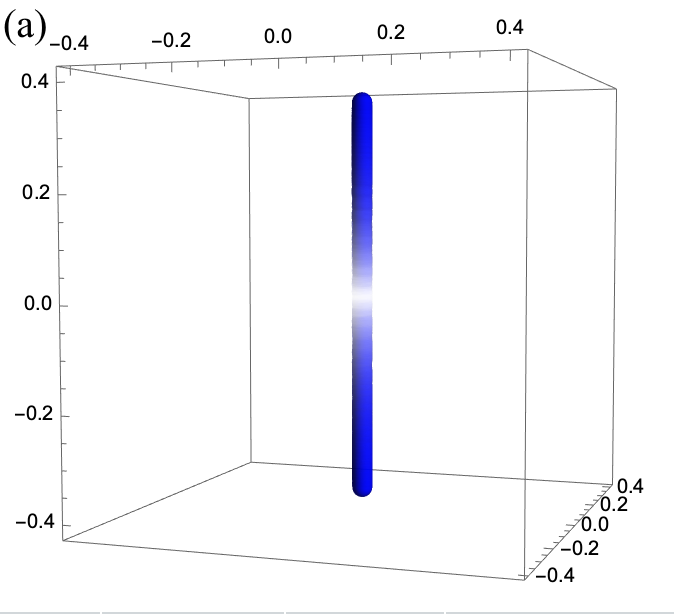}
 \includegraphics[width=0.4\textwidth]{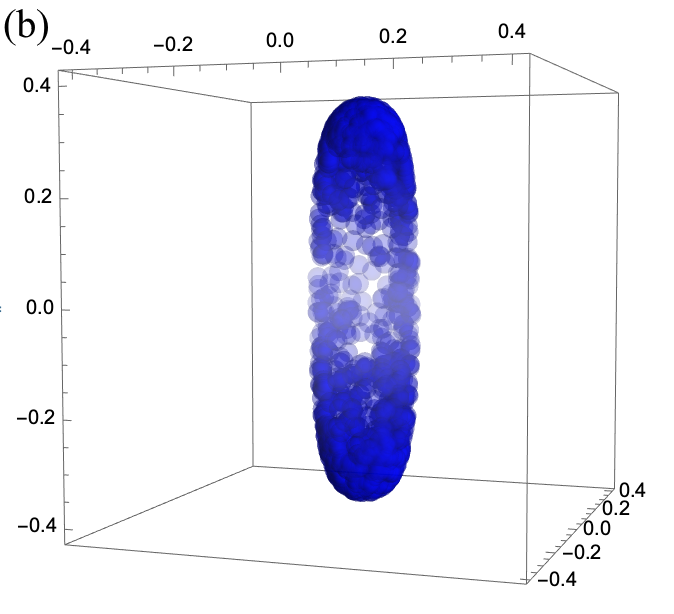}
 \caption{\label{fig:anisotropy} 
 Illustration of the magnetization of a Sm atom in Sm$_2$Ir$_2$O$_7$, in response to an applied field of random orientation on the unit sphere. Each blue dot corresponds to the tip of the magnetization vector drawn from the position of the Sm atom. The vertical axis corresponds to the local $\left[1\,1\,1\right]$ direction. The left panel shows the behavior for $0.1$~K and $0.5$~T; the right panel shows $10$~K and $50$~T. The axes are in units of $\mu_B$. 
 } 
 \end{figure}
 \begin{equation}\label{B_CEF}
 B^{m}_{n}=1.44\cross10^{4}(0.5292)^{n}Z\left( \gamma_{m,n}C_{m,n}\langle r^{n}\rangle \theta_{n}\right)
 =-eV(\boldsymbol{r})
 \end{equation}
 \noindent
 where $\langle r^{n}\rangle$ is the expectation value of the radial part of the wave function, $\theta_{n}$ is a numerical factor that depends on the magnetic ion, $C_{m,n}$ is a normalization factor and $Z$ is the proton number of the magnetic ion. $C_{m,n}$, $\theta_{n}$ and $\langle r^{n}\rangle$ are tabulated values that can be found in \cite{Hutchings1984}. $\gamma_{m,n}$ is a linear summation of the Tesseral Harmonics, with each coefficient in the summation given by the distance to the neighboring atoms, their charges and $n$. In our sample, the reference system was rotated to align the local $\langle 111 \rangle$ direction along $\hat{z}$, resulting in the Sm ions environment having a $D_{3d}$ symmetry. Only the closest oxygens were considered, as customary for pyrochlore oxides.
 
  \subsubsection*{\label{app:CEF} Single-ion crystal electric field Hamiltonian}

 Ignoring terms of order $10^{-12}$ or smaller for convenience, we find that the CEF Hamiltonian for a Sm atom in Sm$_2$Ir$_2$O$_7$ in the conventional local reference frame for pyrochlore oxides~\cite{Onoda_2011} takes the form: 
 \begin{equation}
   \left( 
     \begin{array}{cccccc}
     23.83 & 0 & 0 & -6 & 0 & 0 \\
     0 & -18.37 & 0 & 0 & 0 & 0 \\
     0 & 0 & -5.46 & 0 & 0 & 6 \\
     -6 & 0 & 0 & -5.46 & 0 & 0 \\
     0 & 0 & 0 & 0 & -18./37 & 0 \\
     0 & 0 & 6 & 0 & 0 & 23.83 
     \end{array}
   \right)
 \, , 
 \end{equation}
 with eigenvalues and eigenvectors: 
 \begin{equation}
     \begin{array}{cl}
     0 \ {\rm meV} & \quad (0,0,0,0,1,0) \\
     0 \ {\rm meV} & \quad (0,1,0,0,0,0) \\
     11.73 \ {\rm meV} & \quad (0.19,0,0,0.98,0,0) \\
     11.73 \ {\rm meV} & \quad (0,0,-0.98,0,0,0.19) \\
     43.39 \ {\rm meV} & \quad (0.98,0,0,-0.19,0,0) \\
     43.39 \ {\rm meV} & \quad (0,0,0.19,0,0,0.98) 
 \, , 
     \end{array}
 \end{equation}
 where the eigenstate amplitudes are given in the $m_J=-5/2,-3/2,-1/2,1/2,3/2,5/2$ basis, from left to right. We point out the characteristic ground state doublet spanned by the $m_J=\pm 3/2$ states~\cite{Mauws2018}, separated by a large gap from the excited states, $\Delta_{\rm CEF} = 11.73$~meV~$= 136$~K times $k_B$.

\FloatBarrier